\documentclass[aps,pra,twocolumn,groupedaddress,longbibliography]{revtex4-1}

\usepackage{graphicx}

\usepackage{amssymb}
\newcommand{\vort}{\mbox{\boldmath$\omega$}} 
\newcommand{\vel}{\mbox{\boldmath$v$}}

\newcommand{\nrm}{\mbox{\boldmath$n$}}

\newcommand{\Avec}{\mbox{\boldmath$A$}}
\newcommand{\Bvec}{\mbox{\boldmath$B$}}
\newcommand{\Evec}{\mbox{\boldmath$E$}}

\newcommand{\cubr}[1]{\left(#1\right)}

\newcommand*\diff{\mathop{}\!\mathrm{d}}

\usepackage{color}
\newcommand{\tb}[1]{{\color{black} #1}}

\begin{document}


\title{Reconnection of Vortex Tubes with Axial Flow}


\author{P.~McGavin and D.~I.~Pontin}
\affiliation{Division of Mathematics, University of Dundee, Dundee, DD1 4HN, UK}


\date{\today}

\begin{abstract}
This paper addresses the interaction of initially anti-parallel vortex tubes containing an axial flow that induces a twisting of the vortex lines around the tube axes, using numerical simulations. Vortex tube configurations with both the same and opposite senses of twist -- corresponding to same and opposite signs of kinetic helicity density -- are considered. It is found that the topology of the reconnection process is very different between the two cases. For tubes with the same sense of twist, the reconnection is fully three-dimensional (3D): vortex lines reconnect at a finite angle, and 3D vortex null points may be created. Following reconnection the vortex line topology in both bridge and thread structures exhibits a high degree of complexity. For oppositely-twisted tubes the reconnection is locally two-dimensional, occurring along vorticity null lines, that in contrast to the untwisted case are not perpendicular to the tube axes. This leads to a break in the symmetry between the two vortex bridges generated during reconnection. For all cases studied, increasing the twist in the vortex tubes leads to a later, faster, and more complete reconnection process.
\end{abstract}


\maketitle
\section{Introduction}
Vortex reconnection in classical, high Reynolds number fluids has been under investigation for the last 30 years or so, motivated by its proposed role in the breakdown of wakes from engines and turbines \citep{1970AIAAJ...8.2172C}, leading to generation of turbulence and noise \citep{1983PhFl...26.2816H,2011PhFl...23b1701H}. Further motivation stems from the proposal that an intimate understanding of vortex dynamics can lead to insights into the classical problem of turbulent flows \citep[e.g.][]{saffman1993}. Vortex reconnection has been studied in various different configurations, notably pairs of anti-parallel vortex tubes \citep[e.g.][]{1987PhRvL..58.1636P,1989PhFl....1..633M,iutam1989melander,2011PhFl...23b1701H,2012PhFl...24g5105V}, vortex rings \citep{1987PhRvL..58.1632A,1991JFM...230..583K}, and knotted vortex tubes \citep{kida1987,2013NatPh...9..253K}.

Traditionally, most studies of anti-parallel vortex reconnection used vortex tubes in which the vortex lines are untwisted, being oriented parallel to the tube axes. However, it is also of physical interest to study the interaction of vortex tubes that individually have non-zero helicity -- since for example wing-tip vortices are known to exhibit non-zero axial flow \citep{Moore491,2016JFM...803..556F}. Introducing twist to the vortex tubes in the anti-parallel configuration, there are two options we can consider: both tubes having the same sign of twist, or twists of opposite signs. 
Interaction of helical vortex tubes is relevant for understanding processes in helical turbulent flows. A further motivation to study reconnecting twisted vortex tubes is that reconnection between untwisted vortex tubes naturally leads to helical vortex structures, that themselves go on to reconnect in secondary processes. Previous studies including the effect of non-zero helicity, corresponding to twisting of the vortex lines around the tube axes, include that of \cite{2012PhFl...24g5105V}, who considered only the case where both tubes have the same helicity (twist). Here we make a systematic study where we vary the levels of twist, and consider both same and opposite senses of twist in the tubes.

We begin by defining what we mean by vortex reconnection. As pointed out by \cite{kida1991,takaoka1996} the reconnection of vortex lines is critically different from the `reconnection' of vorticity isosurfaces: it is only the former that is prohibited in an inviscid fluid. Here we define vortex reconnection as a change in the topology of the vorticity field as in the \emph{general magnetic reconnection} framework of \cite{1988JGR....93.5547S}.
Note that two vorticity fields are topologically equivalent if and only if one can be transformed into the other by means of a smooth (continuously differentiable) deformation. Such an evolution between topologically equivalent fields preserves all linkages or knottedness of field lines within the volume as well as all connections of vortex lines between co-moving boundary points. Note that for a topological change to be defined as reconnection, it must be due to a local non-ideal evolution (as opposed to e.g.~a global diffusion). The relationship between  topology and reconnection in this sense is discussed in more detail for the magnetic case by \cite{hornig1996}.

In a companion study, we have addressed the reconnection between untwisted vortex tubes, focussing particularly on the topology of the vortex lines \citep[][]{mcgavin2018a}.  In this paper we analyse the topological and geometrical structure of reconnection during the interaction of vortex tubes with axial flow (in which the vorticity field lined are twisted), to understand the characteristics of this interaction.
This topological analysis is facilitated by the advances in recent years of the theory of magnetic reconnection in highly conducting plasmas. The mathematical parallels between the vorticity evolution in a barotropic fluid and the magnetic field in a plasma  are summarised in the Appendix -- the reader is also directed to \cite{1993PhFlB...5.2355G,2001LNP...571..373H}. Locally two-dimensional (2D) reconnection occurs at null lines (in 3D space) of $\vort$, when a non-zero component of $\nabla\times\vort$ along the null line exists. By contrast, topological change of the vortex lines in a fully 3D vorticity field occurs in localised regions where $(\nabla\times\vort)\cdot\vort\neq 0$. Reconnection between untwisted, anti-parallel vortex tubes turns out to be locally two-dimensional, and instabilities in the double-vortex sheet between the tubes may result in multiple reconnection X-lines being created, leading to the formation of secondary vortex rings \citep[][]{mcgavin2018a}. 

Our aim here is to investigate the topological properties of reconnecting twisted anti-parallel vortex tubes, and explore the insights this can give into other aspects of the reconnection process.
The paper is organised as follows. In Section \ref{sec:setup} we describe the computational setup. In Sections \ref{sec:nh} and \ref{sec:zh} we describe the results of the simulations with finite net helicity and zero net helicity, respectively, and in Section \ref{sec:comp} their quantitate properties are compared. Finally in Section \ref{sec:conc} we present our conclusions.

\section{Simulation setup}\label{sec:setup}

The initial condition for our simulation comprises two twisted vortex tubes, constructed as follows. There is an azimuthal flow -- cylindrical coordinates -- that produces a vorticity component along the axis of the vortex tube. The presence in addition of a radially-varying axial flow introduces a twist to the vortex lines, and we choose this radial dependence such that the twist is constant for all radii. Specifically, in cylindrical co-ordinates ($r$, $\phi$, $z'$) centred on the axis of the vortex tube ($r=0$) the flow is
\begin{equation}\label{vfield}
v_{\phi}=\frac{1}{16r}\tanh\cubr{8r^2}, \quad v_{z'}=\pm v_0\frac{\pi}{48}\cubr{1-\tanh{\cubr{8r^2}}},
\end{equation}
$v_0$ constant. At $t=0$ a pair of such vortex tubes is initiated with axes located at $y=\pm 1$, $z=-9$. 
Our simulation domain is 6 units long in the direction of the tube axis ($x$), such that for $v_0=1$ vortex lines twist around the central axis exactly once within the simulation domain. \tb{To quantify the strength of the axial flow we define the `swirl number' for our simulations as the ratio of the spatial maxima of $v_{z'}$ and $v_\phi$ at $t=0$, specifically $q=||v_{z'}(t=0)||_\infty/||v_{\phi}(t=0)||_\infty$.}

\begin{figure}
\centering
(a)\includegraphics[width=0.4\textwidth]{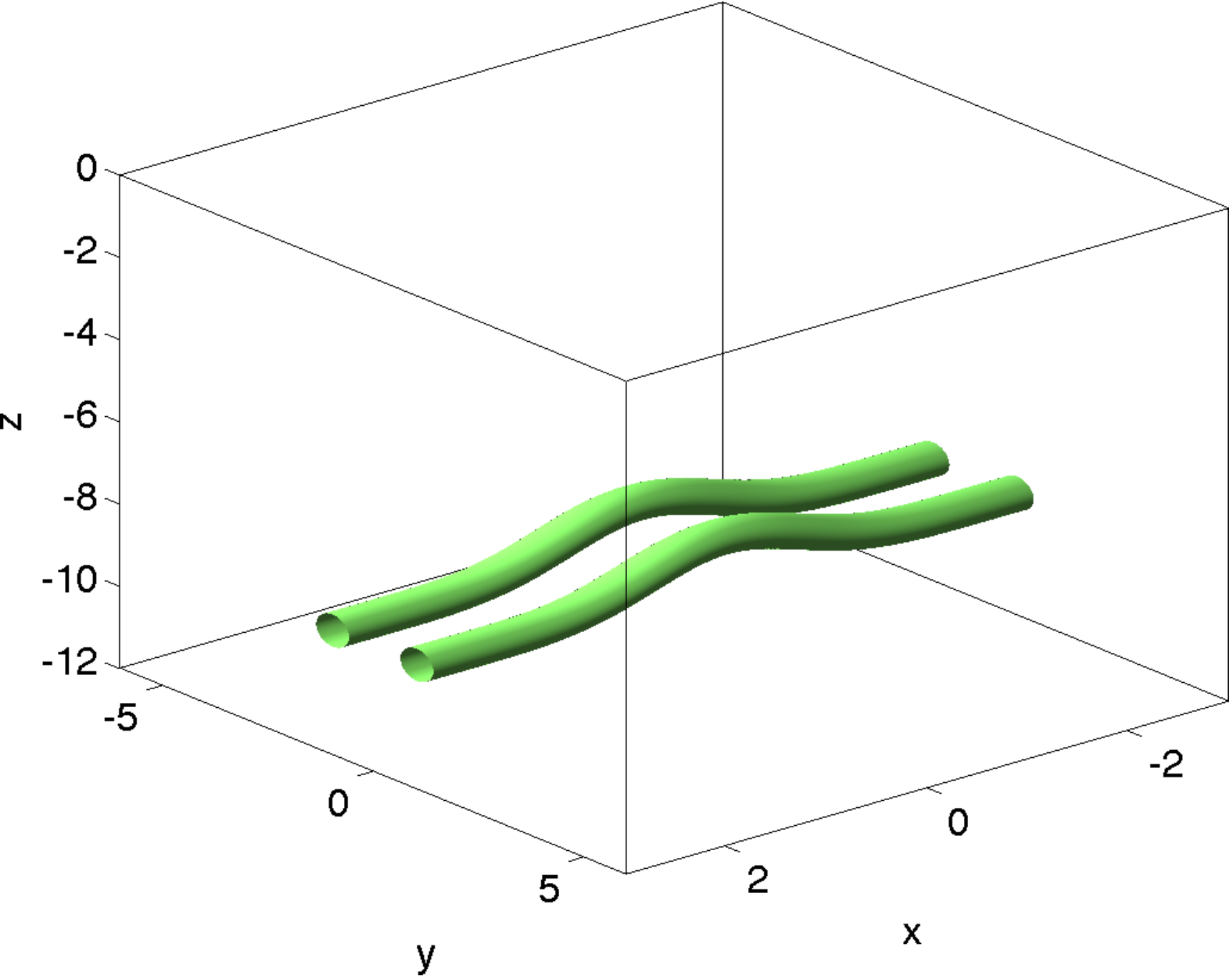}
(b)\includegraphics[width=0.4\textwidth]{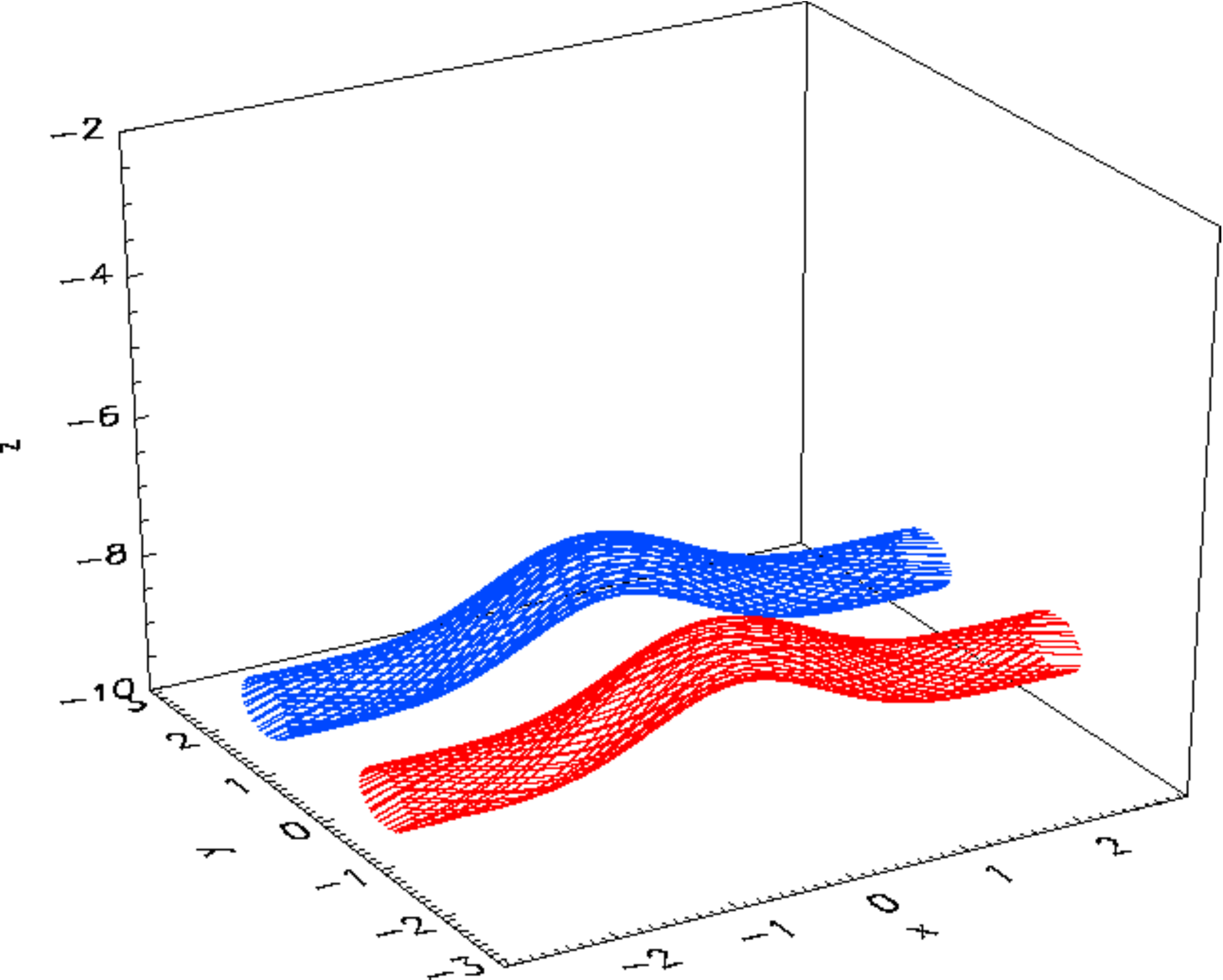}
\caption{(a) Vorticity isosurface and (b) vorticity field lines for the initial condition with non-zero net helicity.}
\label{fig:afinitcond}
\end{figure}

In line with the previous studies mentioned above, a perturbation is applied that deforms the vortex tubes in the positive-$z$ direction such that they will impinge on one another. 
The perturbation is achieved by applying a deformation of the form  $z\to z + \cos^6\!\!\left({\pi x}/{6} \right), ~x\in[-3,3]$, and then calculating the pull-back on the 1-form ${\bf v}$ \citep{frankel2004geometry}. This ensures that the vorticity field lines are also deformed as per the perturbation (though note that the velocity field is no longer divergence-free).
Generating the vorticity distribution and perturbation in this way allows us to preserve the exact divergence-free nature of the vorticity field, avoiding the complications described by, e.g., \cite{1989PhFl....1..633M}, as well as the issues of numerical noise generation that hampered previous studies that initialised the vortex tubes with compact support \citep{2008PhyD..237.1912B}.
The initial condition is shown in Figure \ref{fig:afinitcond} for $\tb{q=0.485}$. 

The simulations are conducted using a 3D code developed and thoroughly tested for hydrodynamic and magnetohydrodynamic problems \citep{nordlund19953d,galsgaard1996}. This is a high-order finite difference code using staggered grids to maintain conservation of physical quantities. The derivative operators are sixth-order in space -- meaning that numerical diffusion is minimised -- while the interpolation operators are fifth-order. The solution is advanced in time using a third-order explicit predictor-corrector method. We solve the compressible Navier-Stokes equations in the form 
\begin{eqnarray}
\frac{\partial(\rho \mathbf{v})}{\partial t} &=& -\mathbf{\nabla}\cdot(\rho \mathbf{v}\mathbf{v}) -{\nabla}p + \mu 
\left( \nabla^2 {\bf v} + \frac{1}{3} \nabla (\nabla \cdot {\bf v}) \right) \label{momentum}\\
\frac{\partial \rho}{\partial t}&=& -{\nabla} \cdot (\rho \mathbf{v}) \label{mass}\\
\frac{\partial e}{\partial t}&=& -{\nabla}\cdot(e\mathbf{v})-p\mathbf{\nabla}\cdot\mathbf{v} \nonumber\\
&&+ \mu\left(  \frac{{\partial v_i}}{\partial x_j}\frac{{\partial v_i}}{\partial x_j}+\frac{{\partial v_j}}{\partial x_i}\frac{{\partial v_i}}{\partial x_j}-\frac{2}{3}(\nabla\cdot{\bf v})^2   \right) \label{energy}
\end{eqnarray}
where $\mathbf{v}$ is the fluid velocity, $\rho$ the density, $e$ the thermal energy, $p=(\gamma-1)e=2e/3$ the gas pressure, $\mu$ the viscosity, and summation over repeated indices is assumed. The viscosity is set explicitly to a constant value throughout the volume. 
{Note also that while the simulation is compressible, we find that in practice density fluctuations are small (maximum of 2-3\%), and so the compressibility has a minimal effect on the dynamics.}

The above equations are solved over a periodic domain $x=\pm 3$, $y=\pm6$, $z=\pm12$. This is facilitated by the inclusion of a pair of `image vortices'  with axes at $y=\pm 1$, $z=+9$, but we focus on the evolution within the sub-domain $z<0$.
At $t=0$ we set the density $\rho=0.1$, and the thermal energy $e=0.09$, uniform in the domain, giving a sound speed of 0.77 in non-dimensional code units. 
A grid resolution of $[120,240,240]$ is chosen for the sub-domain $x\in[-3,3]$, $y\in[-6,6]$, $z\in[-12,0]$. The grid is uniformly-spaced in $x$ and $z$, but is stretched along $y$ so as to increase the density of points in the vicinity of $y=0$, in order to better resolve the double-vortex sheet that forms. 
In line with previous studies we define the Reynolds number to be $\Gamma/(\mu/\rho)$, where $\Gamma$ is the tube circulation at $t=0$. We present here the results of simulations with $Re=800$. This value is chosen since it offers a balance between minimising diffusion and avoiding the additional complications involved with generation of secondary vortex rings at higher $Re$ \citep{mcgavin2018a}. \tb{We follow previous studies by reporting the results in terms of a normalised time $t^*=t/(2\pi b^2/\Gamma)=t/64$, where $b$ is the separation of the unperturbed tube axes.}

We run a series of simulations with \tb{$q=\{0.0485, 0.243, 0.485, 0.970, 1.455 \}$, corresponding to $v_0 = \{0.1, 0.5, 1, 2, 3\}$} meaning that we cover a range of configurations from a weakly twisted to a relatively strongly twisted case where the azimuthal vorticity is comparable to the axial vorticity. {Note that the most strongly twisted case is likely close to the stability limit for a straight vortex tube \citep[e.g.][]{mayer1992}, and so we do not consider \tb{$q>1.455$}.}
Depending on the relative signs chosen in Equation (\ref{vfield}), the two tubes may have the same or opposite twists, and we distinguish these two situations by the net helicity in the domain, that may  be zero (when the helicities of the individual tubes are equal but opposite) or non-zero (when they are equal): we refer to the two corresponding sets of simulations as ``zero net helicity" and ``non-zero net helicity" in what follows.

\section{Results: tube pair with net helicity}\label{sec:nh}

\subsection{Internal topology change at early times: loss of twist}\label{subsec:afngtwist}

\begin{figure}
\centering
(a)\includegraphics[width=0.45\textwidth]{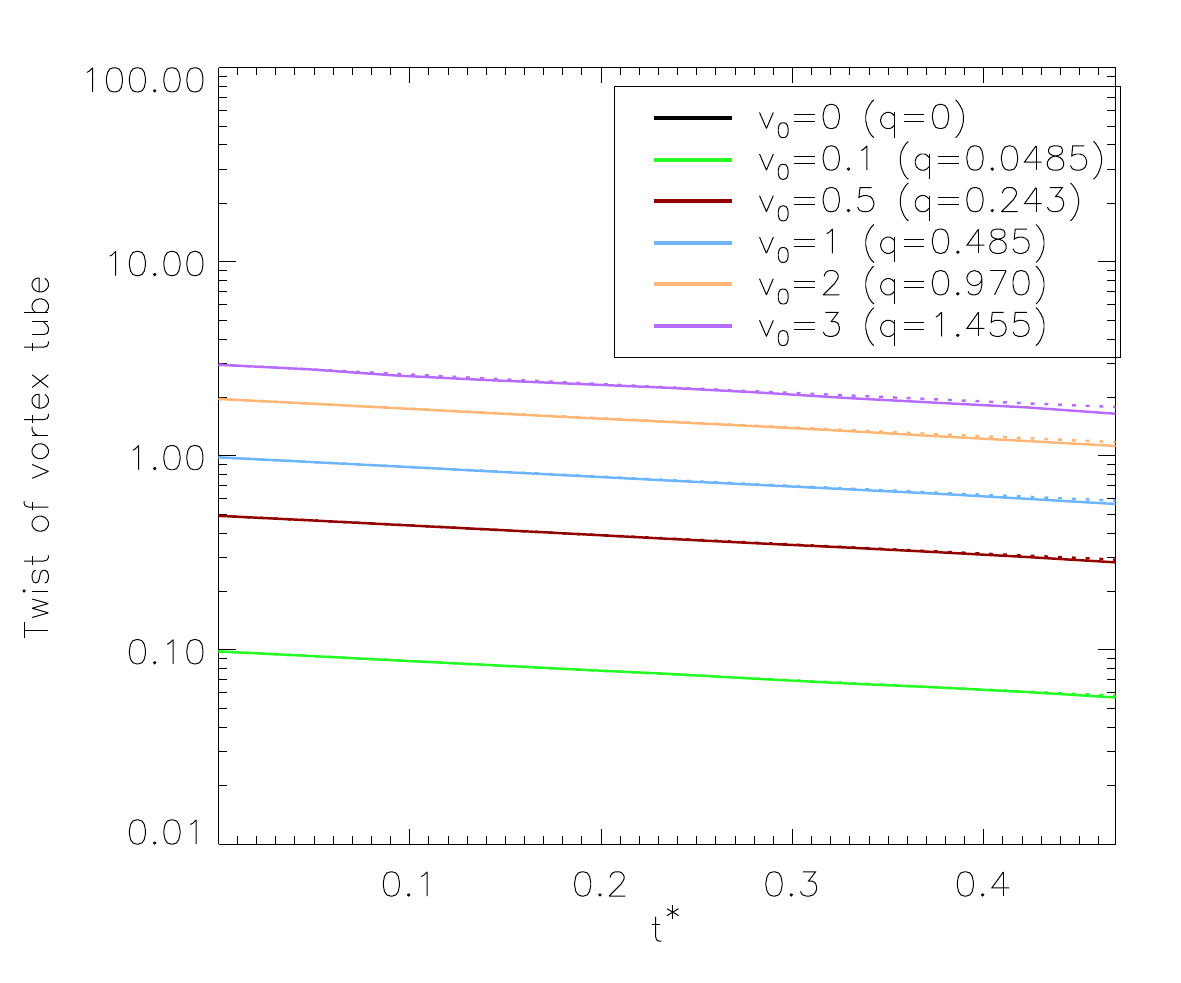}
(b)\includegraphics[width=0.45\textwidth]{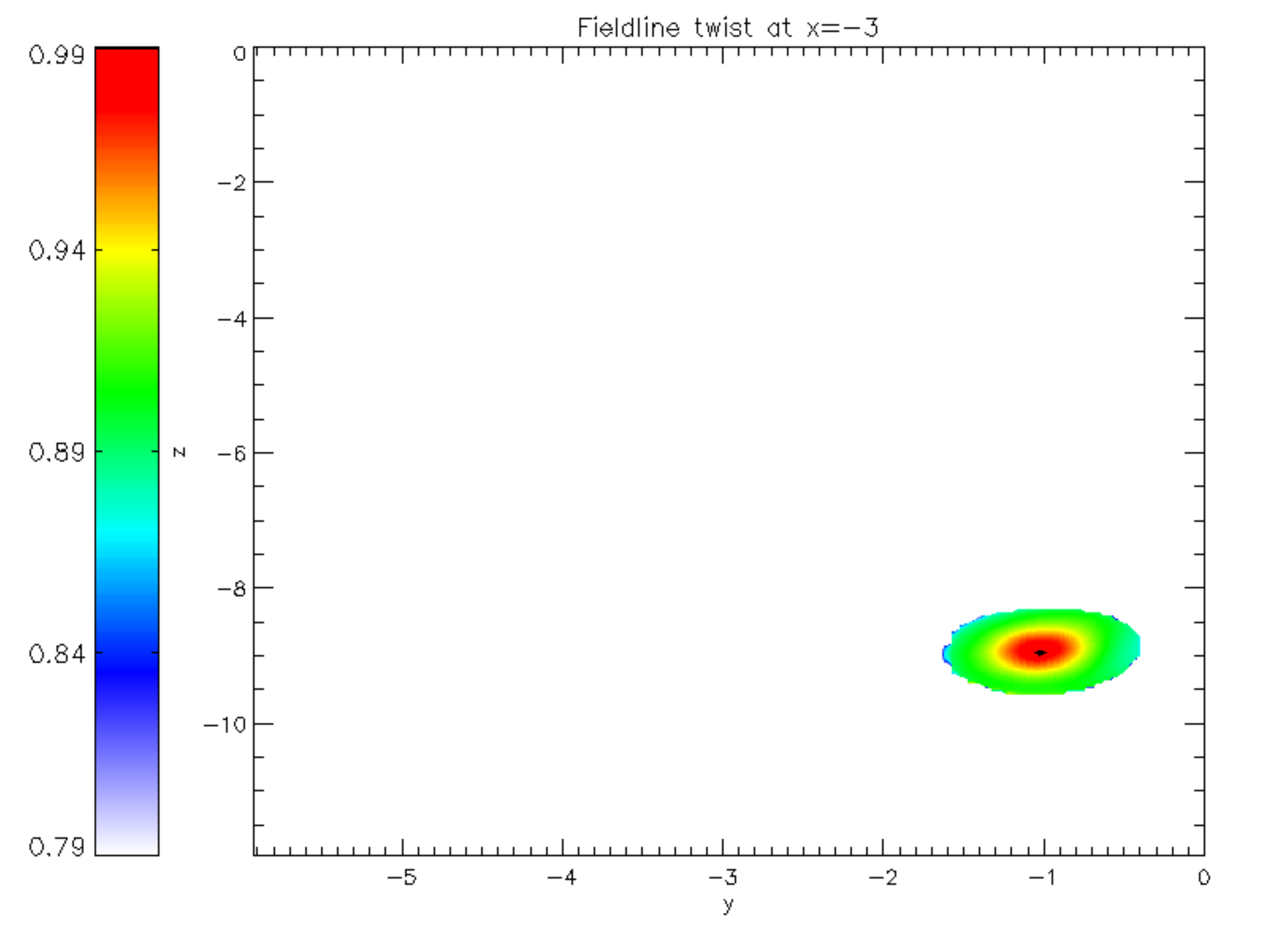}
\caption{(a) Measured change in twist of each vortex tube at early times (solid) compared to the change in twist predicted by evaluating $\nu\left\{\int(\nabla\times\vort)_\parallel\diff l\right\}_{max}$ (dashed). (b) Twist of field lines around the tube axis plotted on the $x=-3$ boundary at \tb{$t^*=0.05$} for \tb{$q=0.485$}. {Note that the twist is defined as the number of turns that a vortex line makes around the tube axis between $x=-3$ and $x=3$ (equivalently the angle of rotation around the tube axis in this interval, divided by $2\pi$).}}
\label{fig:aftwistloss}
\end{figure}

For non-zero values of $\tb{q}$ -- axial flow -- there is a non-zero $(\nabla\times\vort)_\parallel$ distribution within the vortex tubes, and this -- coupled with the non-zero viscosity -- leads to a loss of helicity and correspondingly twist in each tube. In Figure~\ref{fig:aftwistloss} we plot the average twist of the vortex lines around the central axis (solid) and compare it to the predicted loss of twist obtained by evaluating the vortex reconnection rate given by $\nu\int(\nabla\times\vort)_\parallel\diff l$ along the central axis of the tube (see the Appendix). The loss of twist is exponential with time, with the same exponent for all initial twists, since $(\nabla\times\vort)_\parallel$ increases linearly with the twist of the vortex tubes. We note that the loss of twist occurs independent of whether or not the initial perturbation is applied to the vortex tubes, while the twist is conserved better at higher Reynolds numbers. 
We calculate the change in twist only up until \tb{$t^*=0.5$}, when the tubes begin reconnecting with each other. In Figure~\ref{fig:aftwistloss}(b) we see that greater degrees of twist are preserved in the core of the tube at later times, {which is expected due to the profile of $(\nabla\times\vort)_\parallel$ in the tube, as shown by \cite{mcgavin2017}}.

\subsection{Qualitative description of the reconnection process}

\begin{figure*}
\centering
(a)\includegraphics[width=0.4\textwidth]{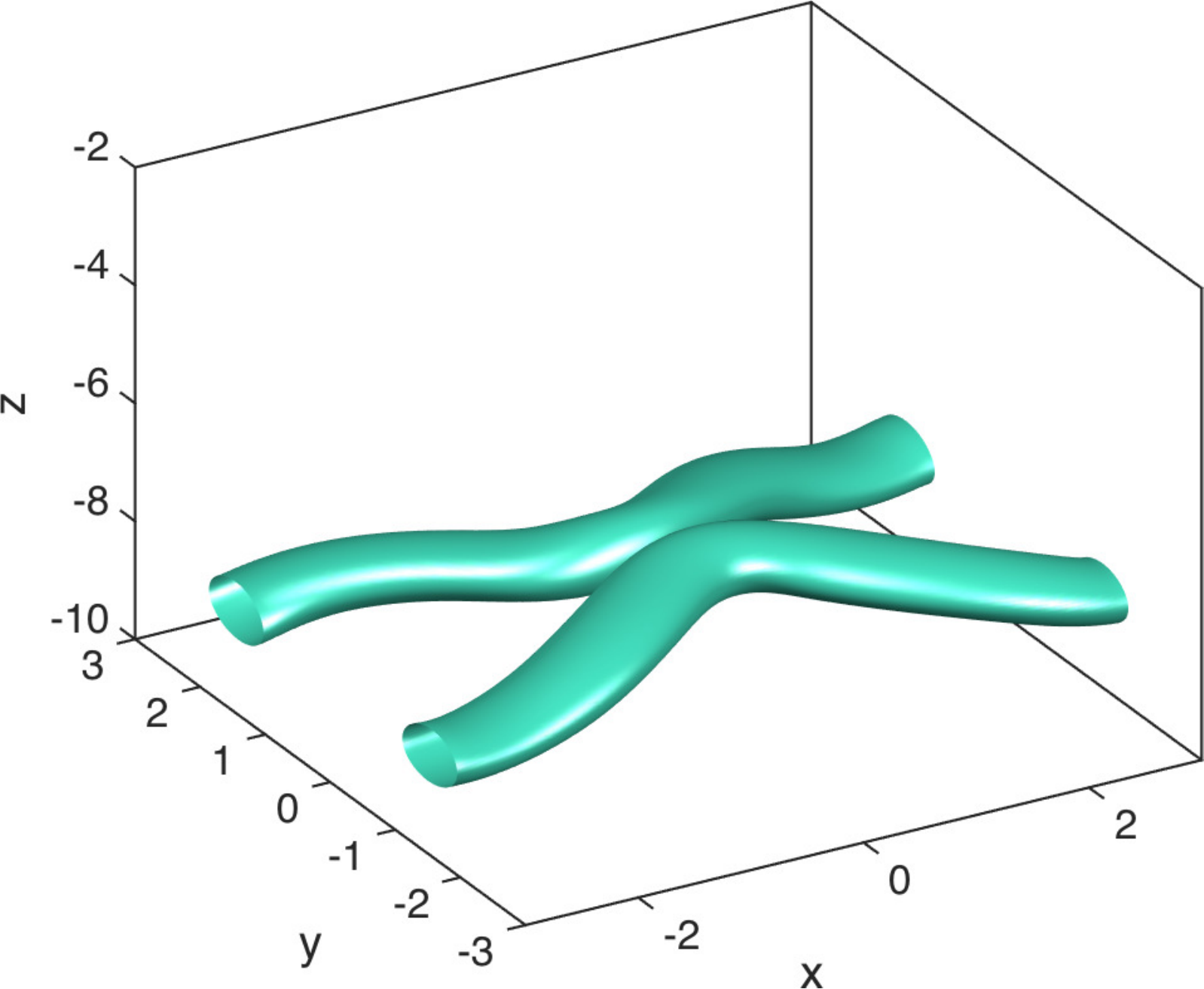}
\includegraphics[width=0.4\textwidth]{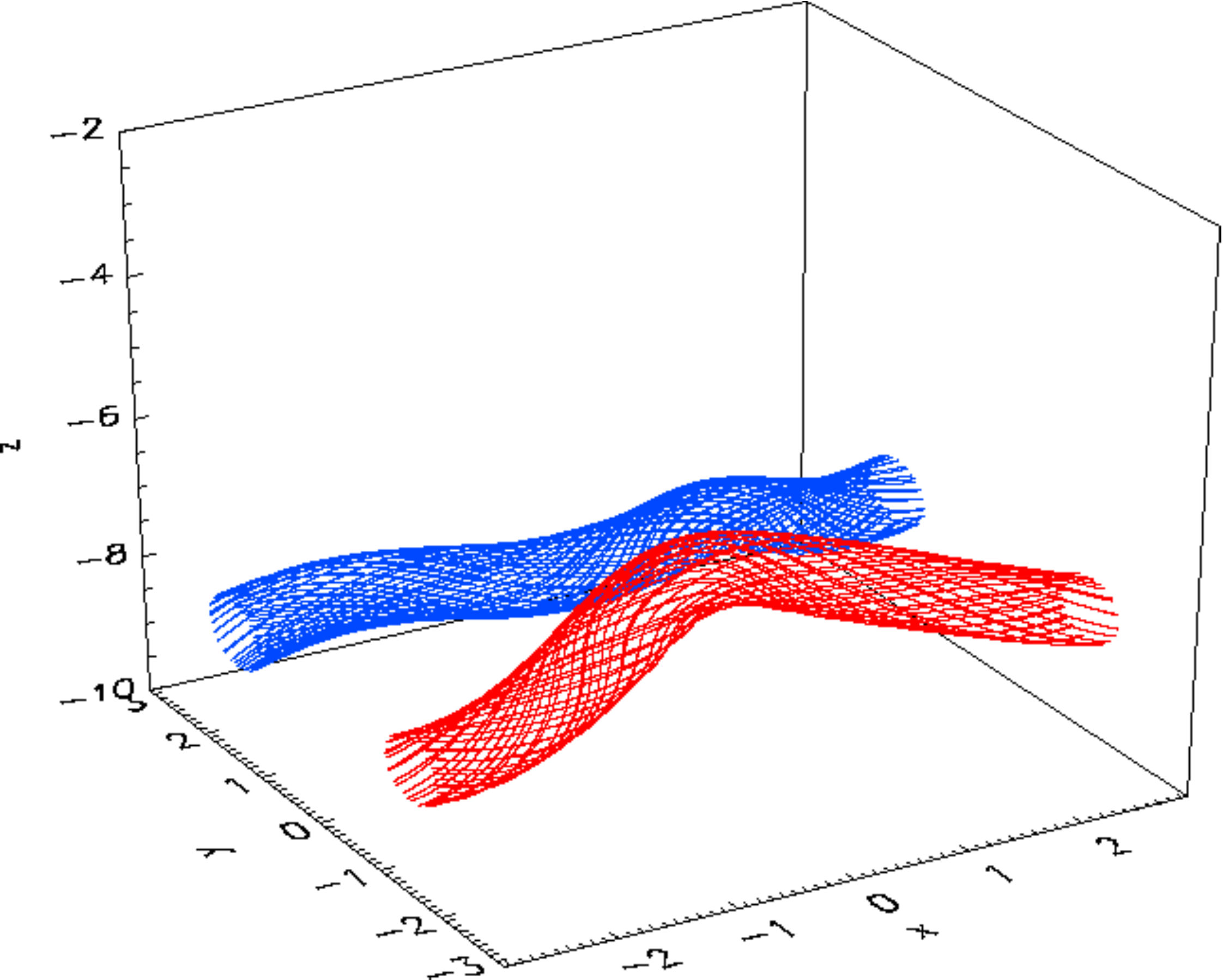}
(b)\includegraphics[width=0.4\textwidth]{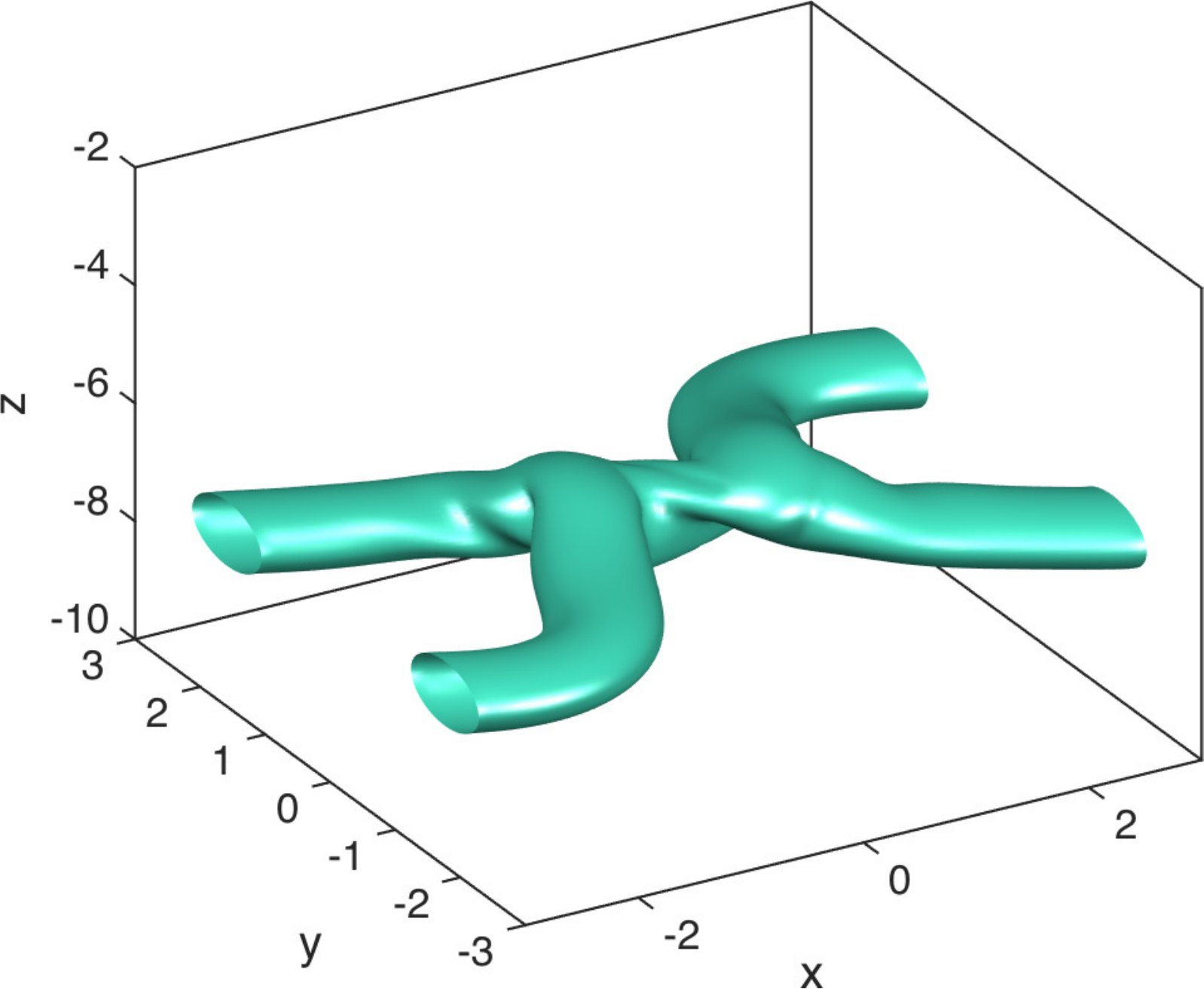}
\includegraphics[width=0.4\textwidth]{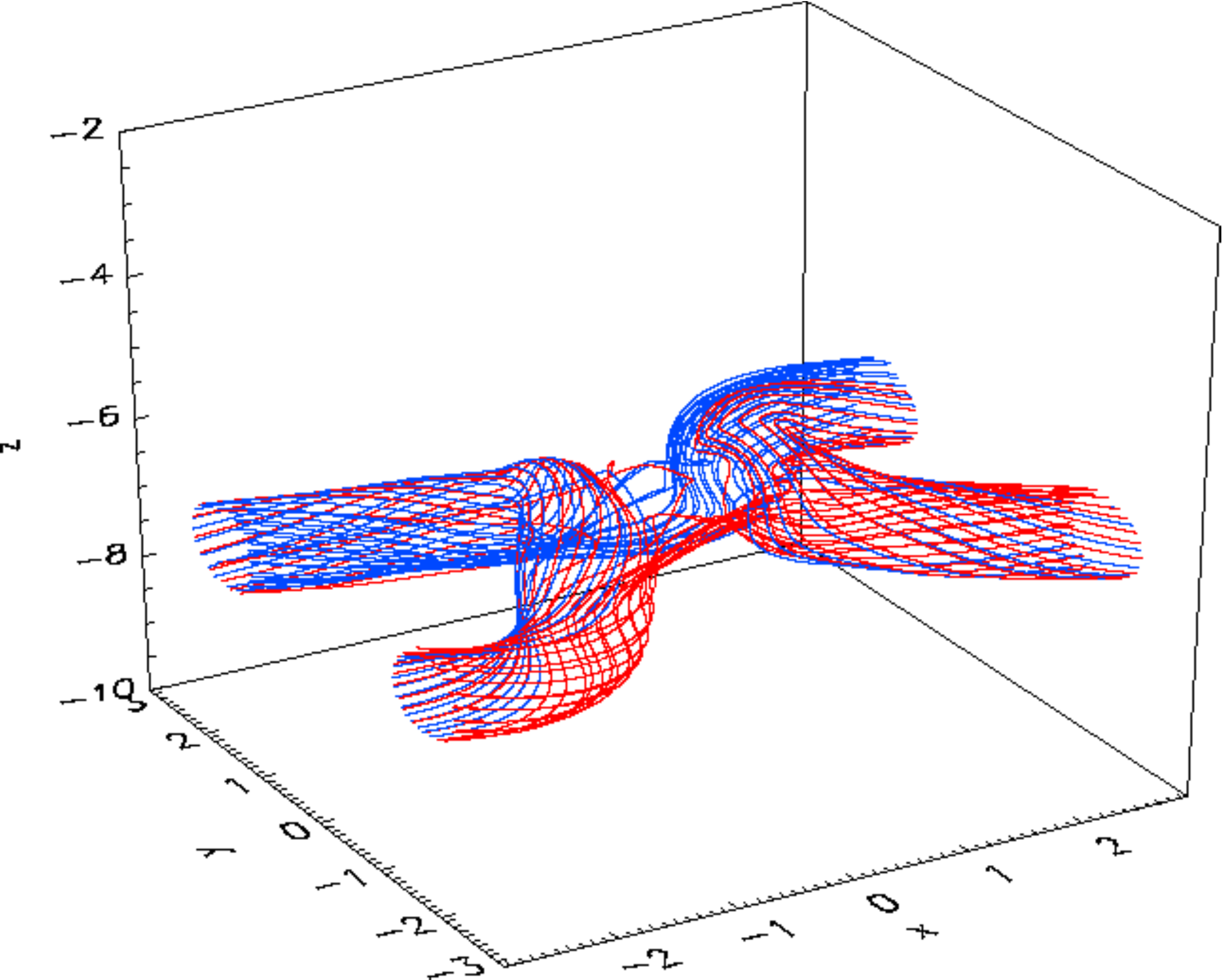}
(c)\includegraphics[width=0.4\textwidth]{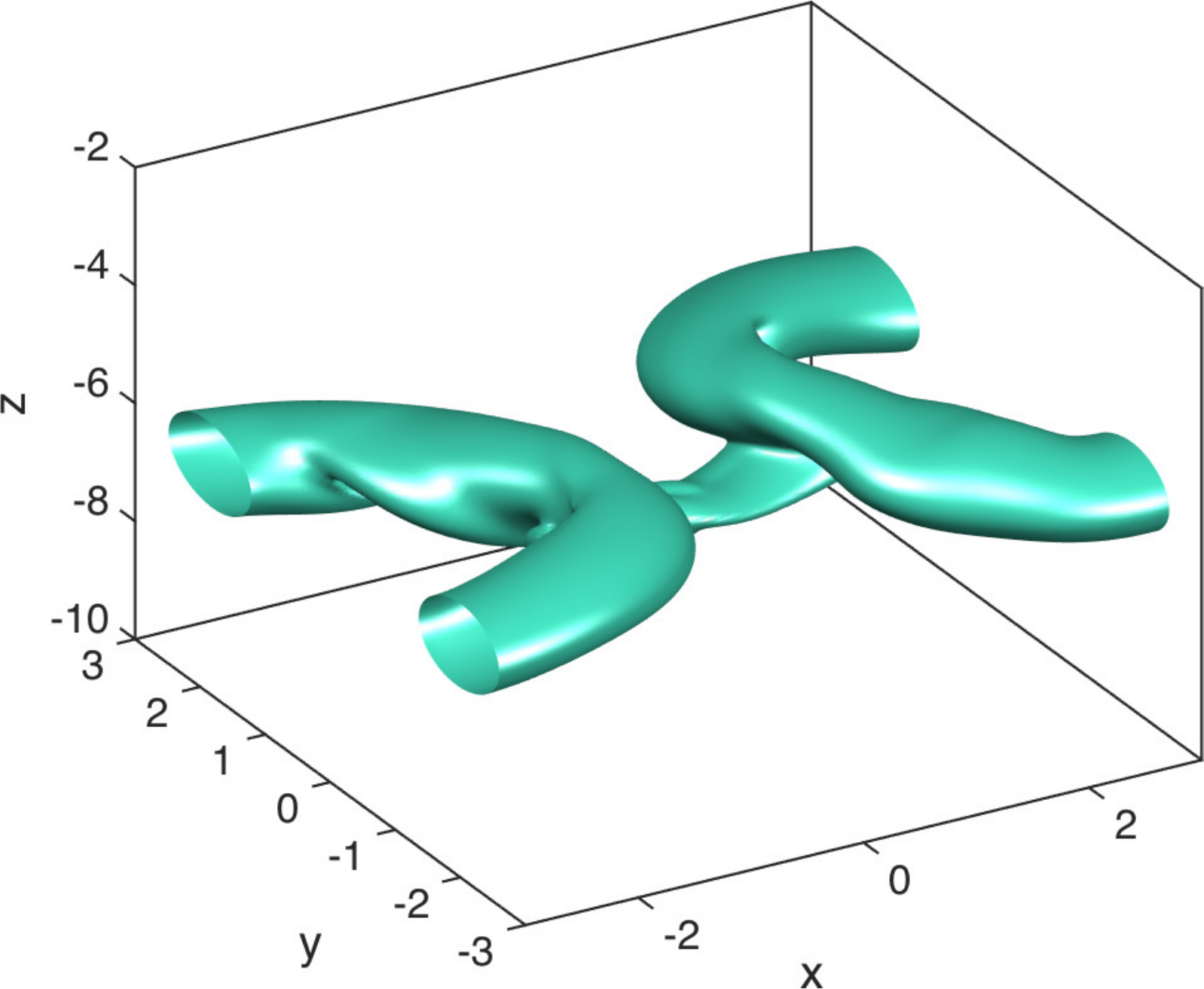}
\includegraphics[width=0.4\textwidth]{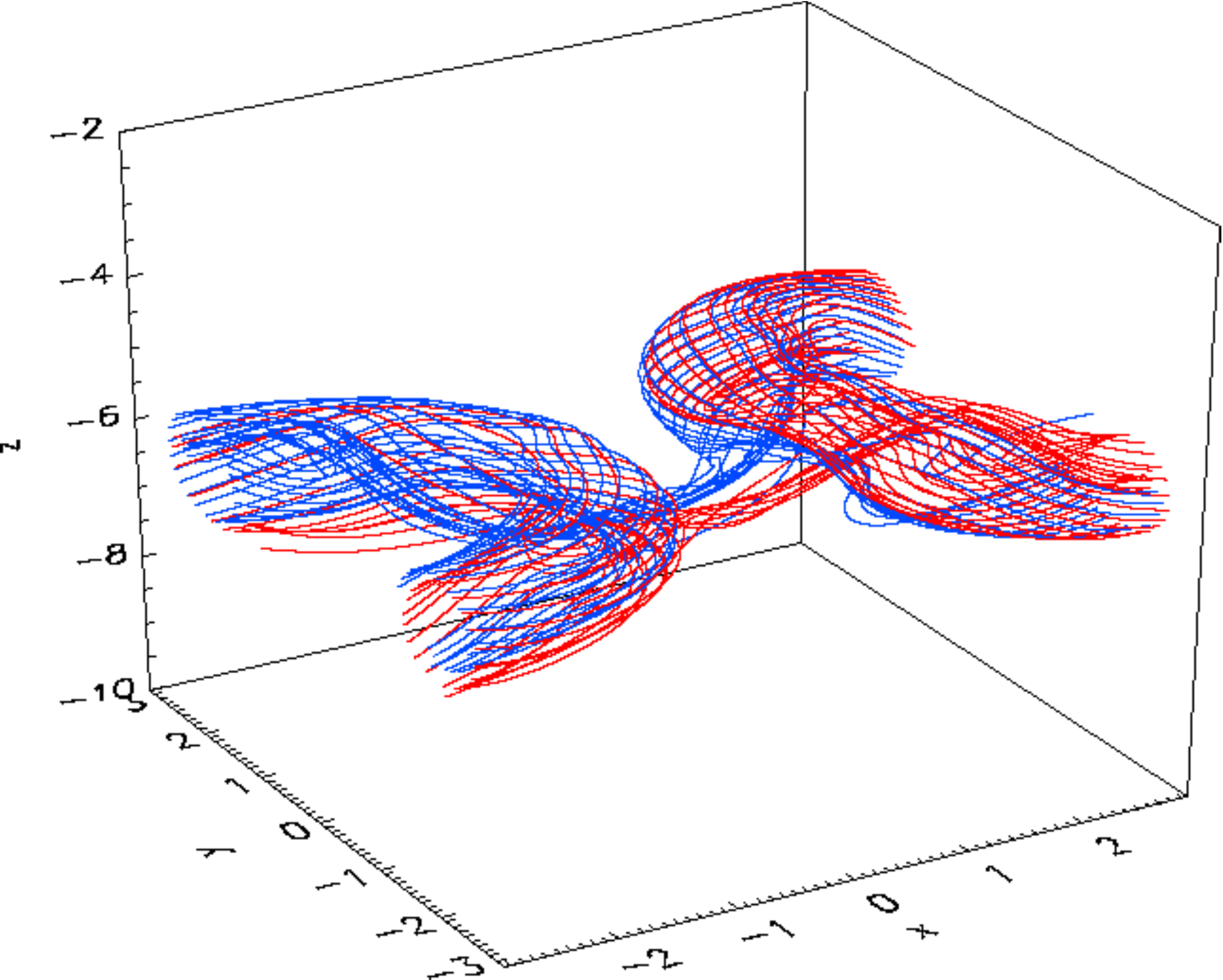}
(d)\includegraphics[width=0.4\textwidth]{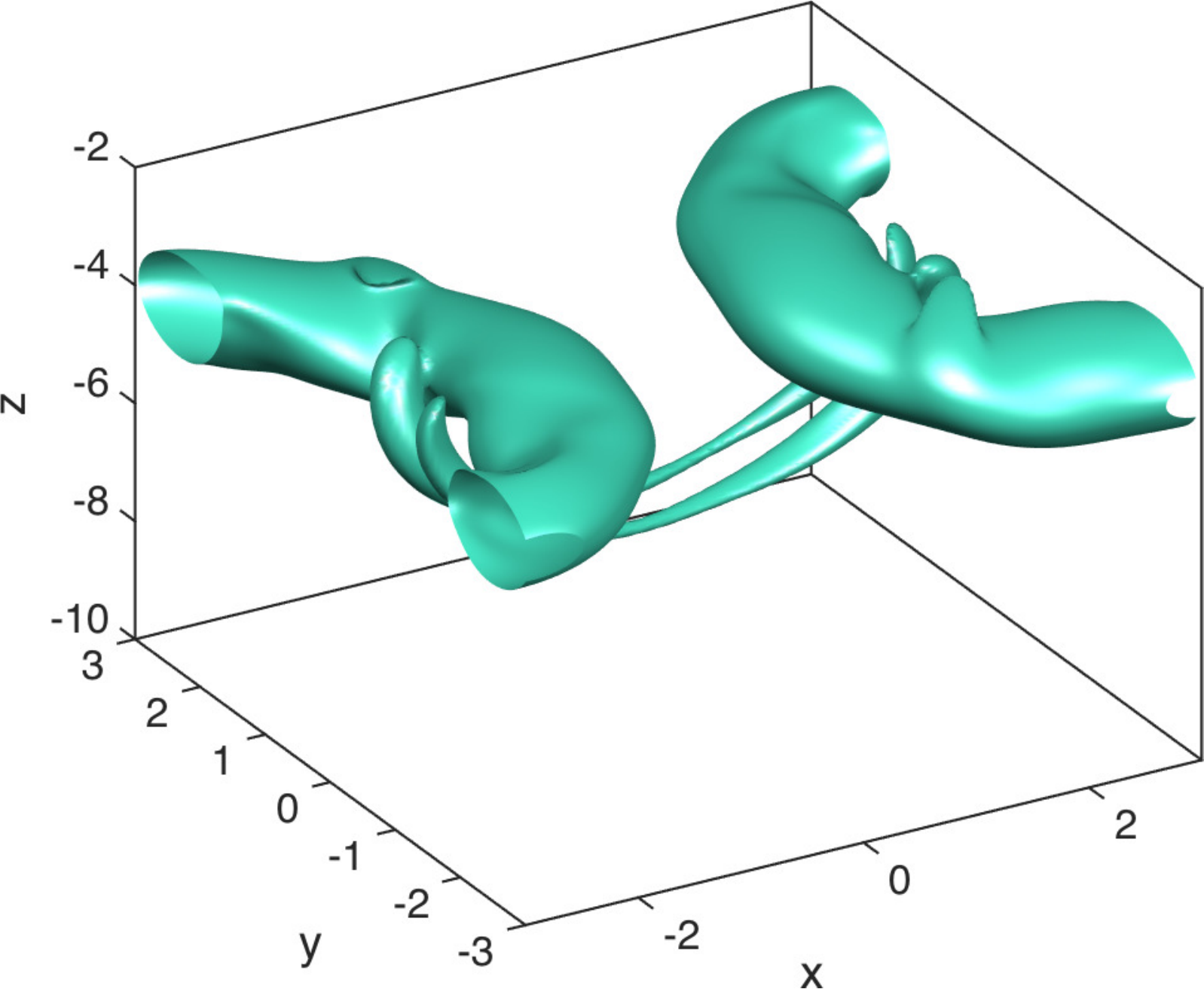}
\includegraphics[width=0.4\textwidth]{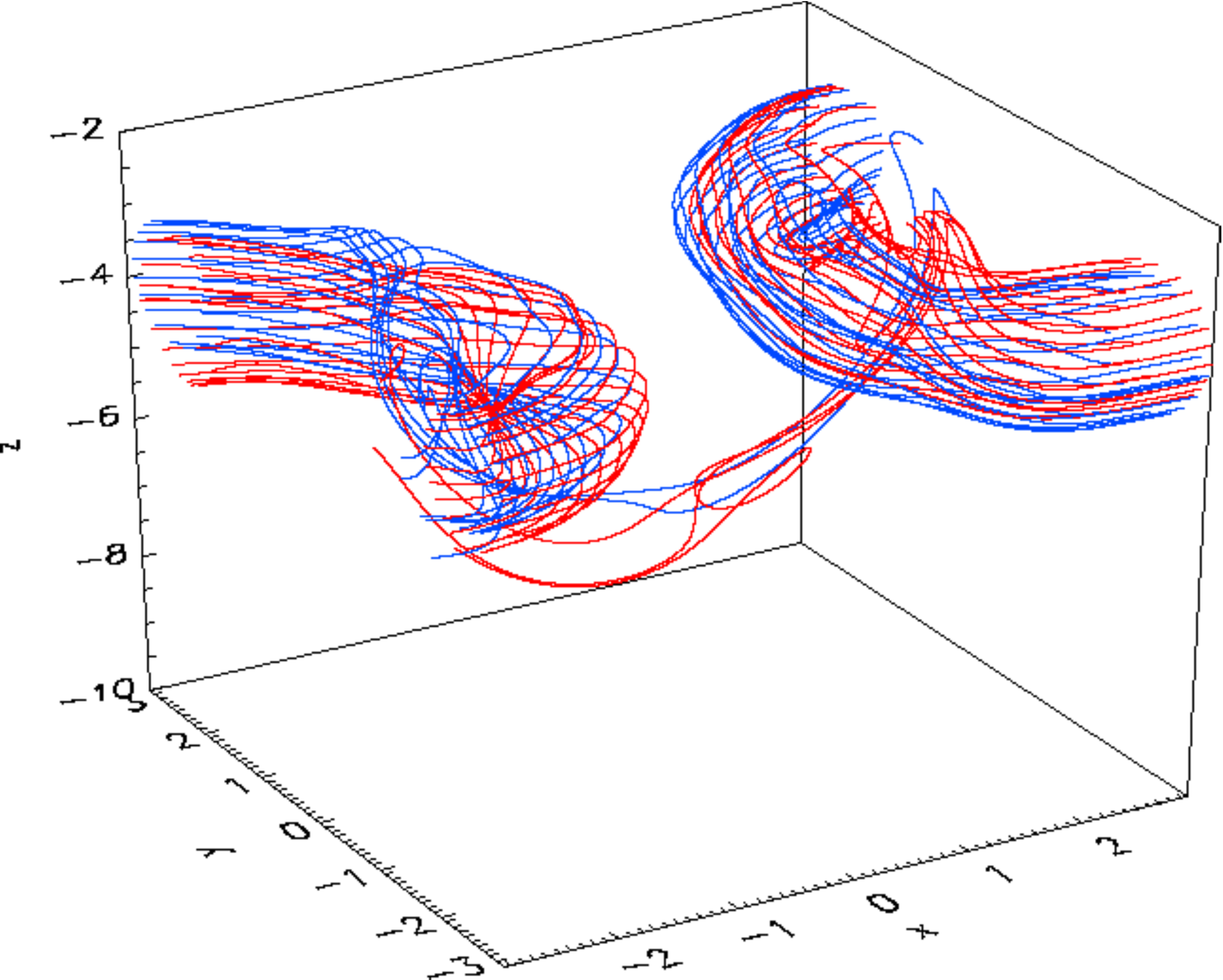}
\caption{$30\%$ vorticity isosurfaces (left) and vortex lines plotted from contours at 30\% of the maximum vorticity on the $x=\pm 3$ boundaries (right), at (a) \tb{$t^*=0.47$}, (b) \tb{$t^*=0.94$}, (c) \tb{$t^*=1.41$} and (d) \tb{$t^*=2.34$}. From the simulation with net helicity for the tube pair, and \tb{$q=0.970$ ($v_0=2$)}.}
\label{fig:afnht130isosurf}
\end{figure*}

At \tb{$t^*\approx 0.5$} the flux tubes press together to form a double vortex layer, and reconnection commences. The reconnection process can be observed qualitatively by examining the vorticity isosurfaces in Figure~\ref{fig:afnht130isosurf}. 
We note that for this and subsequent figures we typically depict the simulations with \tb{$q=0.970$} for clarity; however, \tb{these are representative of the overall evolution for other values of $q$ (for all runs with non-zero net helicity).}
\tb{Many aspects of the} reconnection process are similar to the reconnection of untwisted tubes -- in particular the bridges still form on top of the threads and split apart after reconnection. The main difference visible is the skew introduced to the configuration, due to the combination of the rotation of the `bends' in the tubes together with their propagation along the tubes. The perturbations propagate in opposite directions (for the case of finite net helicity described in this section), breaking the symmetry about $y=0$. Examining the vortex lines (Figure~\ref{fig:afnht130isosurf}), it is clear that a net twist remains within the bridge vortex rings.
 {Note that these plots are constructed by at each time integrating 50 fieldlines from seed points in the $x=3$ and $x=-3$ planes. Specifically, the vortex lines are initiated from starting points that are equally spaced along contour lines of $|\vort|$ in the plane in question, at $30\%$ of the maximum vorticity in the plane.}


\subsection{Reconnection Regions}\label{subsec:afnhrecregions}

\begin{figure}
\centering
(a)\includegraphics[width=0.42\textwidth]{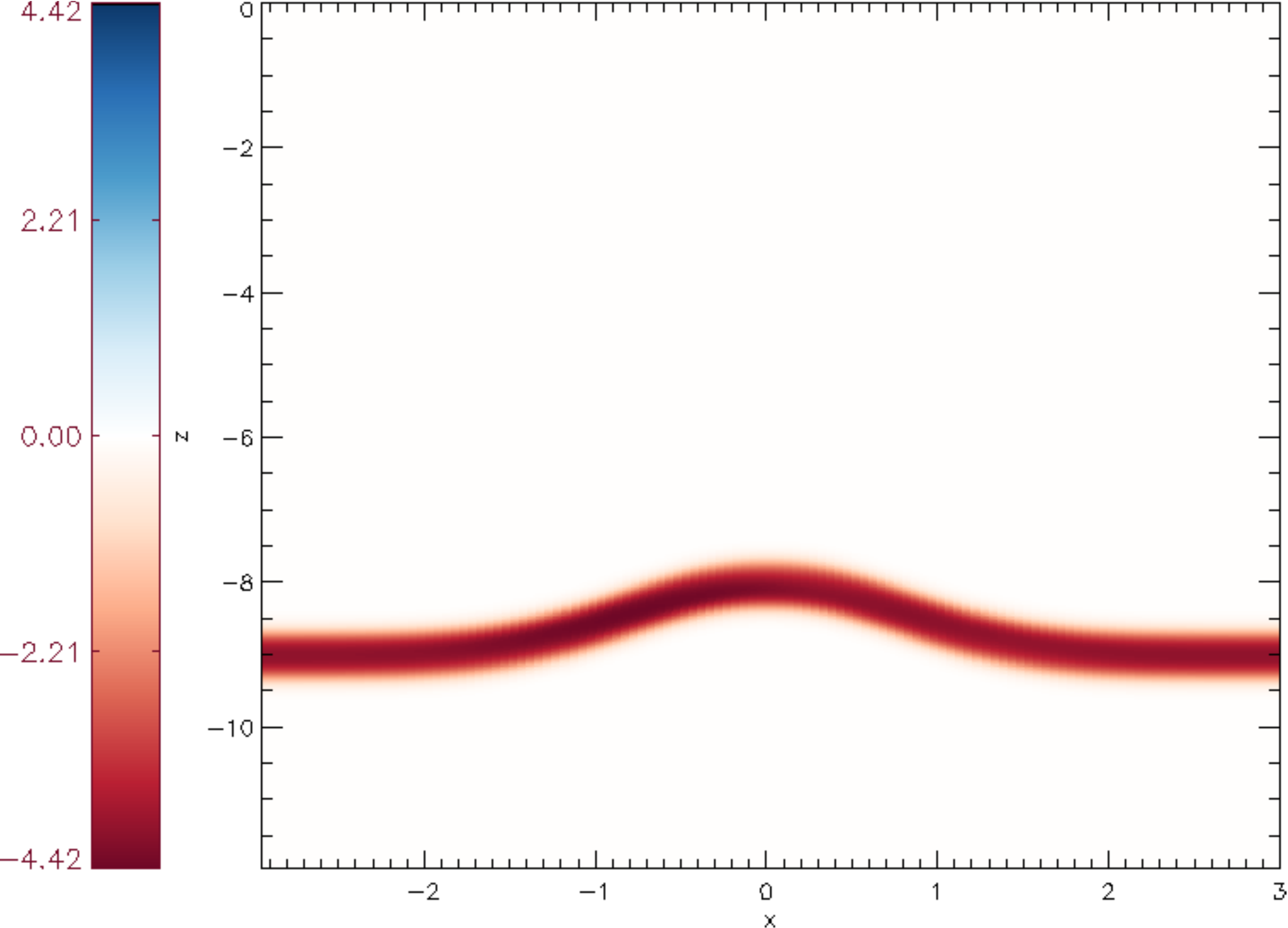}
(b)\includegraphics[width=0.42\textwidth]{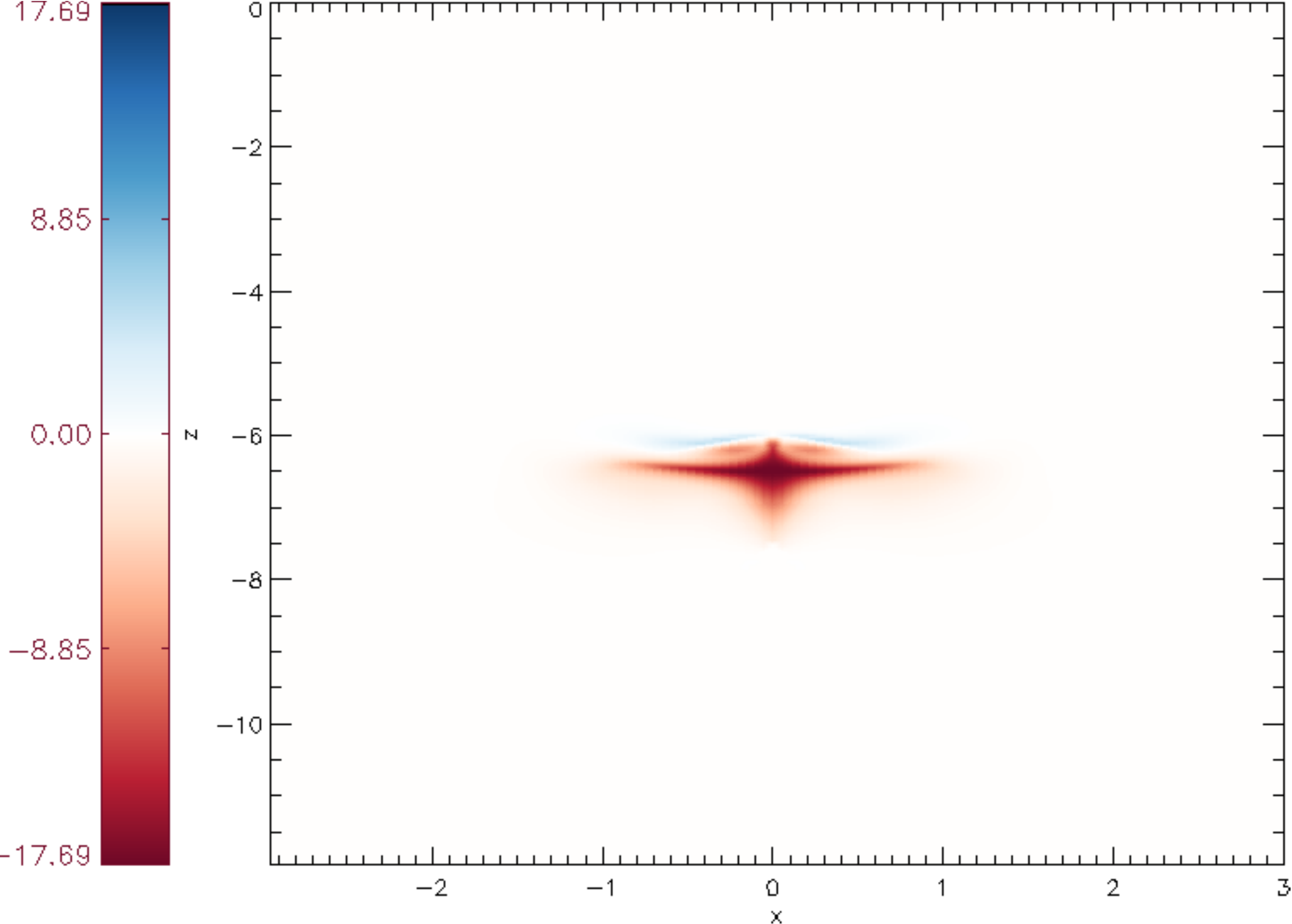}
(c)\includegraphics[width=0.42\textwidth]{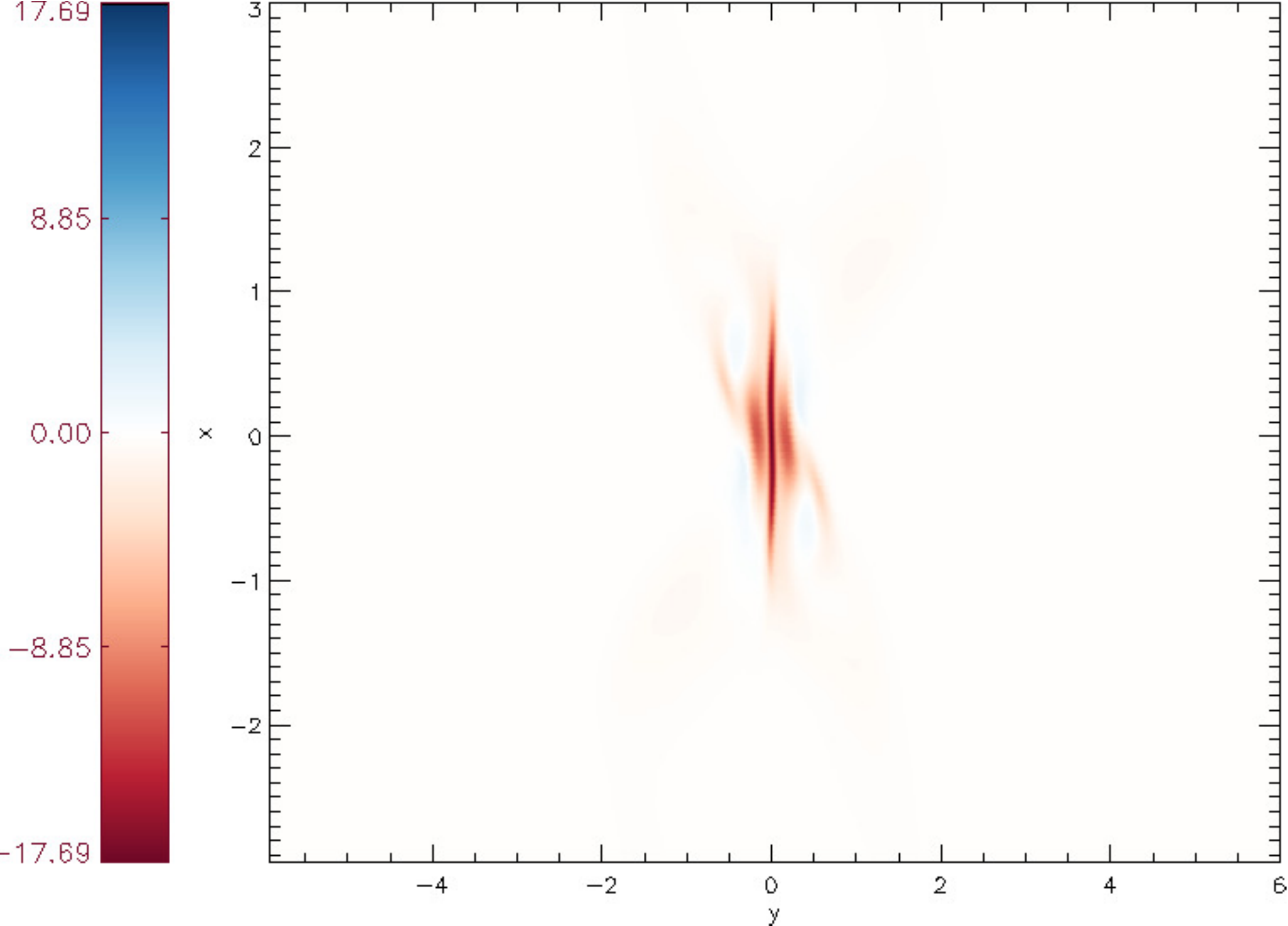}
(d)\includegraphics[width=0.42\textwidth]{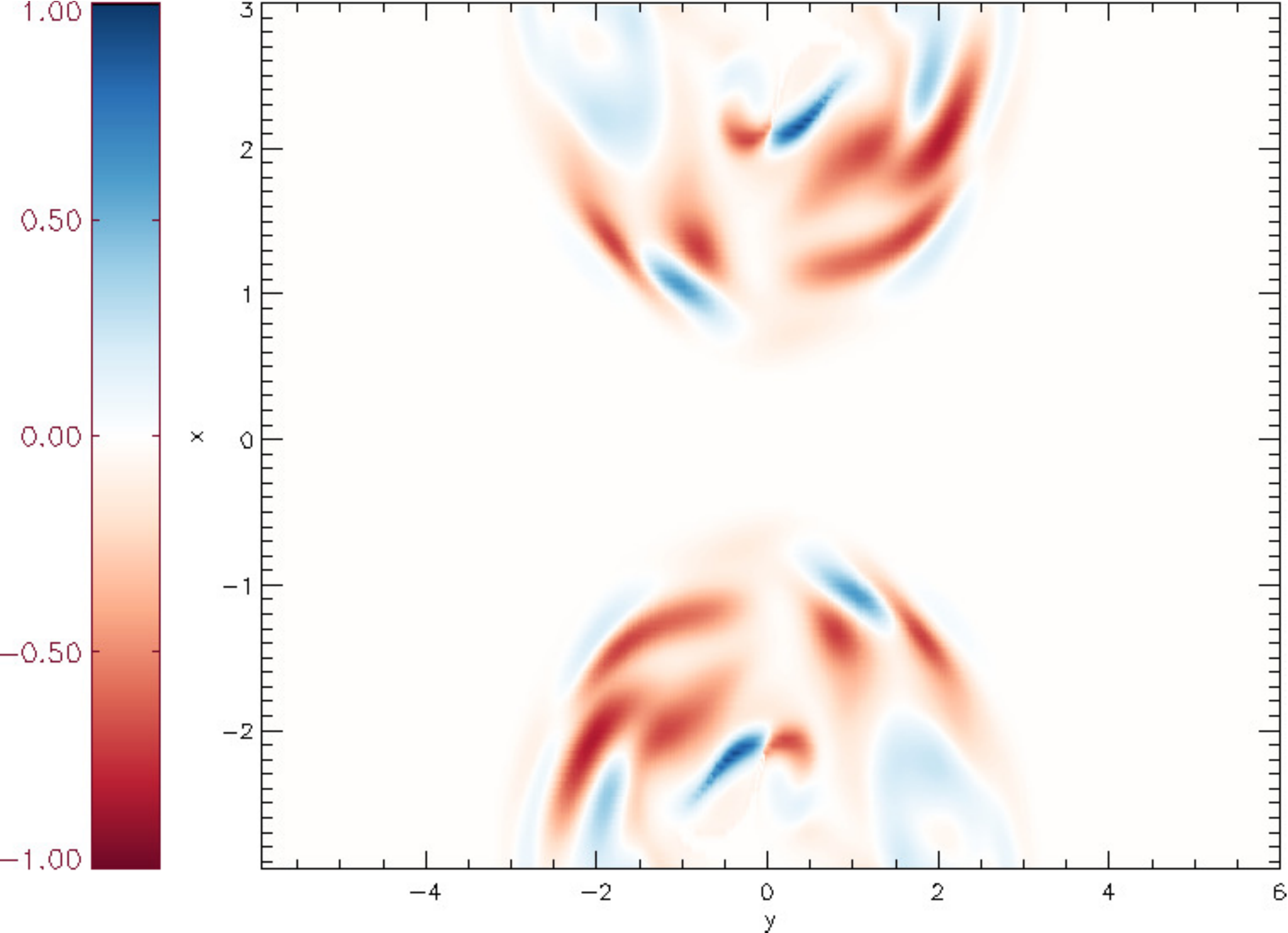}
\caption{Contour plots of $(\nabla\times\vort)_\parallel$, {for \tb{$q=0.970$}}. In all cases the contours are shown in planes that pass through the location of the spatial maximum over the domain  of $(\nabla\times\vort)_\parallel$ at the corresponding time. (a) \tb{$t^*=0$}, $y=-1$, (b) \tb{$t^*=0.89$}, $y=0$, (c) \tb{$t^*=0.89$}, $z=-6.65$ and (d) \tb{$t^*=2.39$}, $z=-3.8$.}
\label{fig:afnhcontcwdw}
\end{figure}

As indicated in the previous section, the presence of an axial flow leads to a skew in the vortex tubes as they approach (specifically, a local rotation around the $z$-axis). As such, the $y=0$ plane (``dividing plane") is not a symmetry plane, and therefore flux measurements in that plane cannot be used to quantify the change of flux.
Instead, to understand where reconnection occurs in our system we examine in Figure~\ref{fig:afnhcontcwdw} the distribution of $(\nabla\times\vort)_\parallel$ within the domain.

In Figure~\ref{fig:afnhcontcwdw}(a), $(\nabla\times\vort)_\|$ is shown to be uniformly distributed along the tubes at $t=0$, leading to the loss of twist at early times described above. By \tb{$t^*=0.89$} (Figures \ref{fig:afnhcontcwdw}(b) and (c)) we see that $(\nabla\times\vort)_\parallel$ has become concentrated into a thin, intense sheet structure between the two  vortex tubes. In Figure~\ref{fig:afnhcontcwdw}(c) the slight rotation of the tubes is seen to lead to a bending of the vortex sheet out of the dividing plane.  This well-localised region within which the reconnection occurs is centred on the $z$-axis by symmetry. As such, the rate of reconnection of flux may be measured by integrating $(\nabla\times\vort)_\parallel$ along the $z$-axis -- see the Appendix. 

\begin{figure}
\centering
\includegraphics[width=0.5\textwidth]{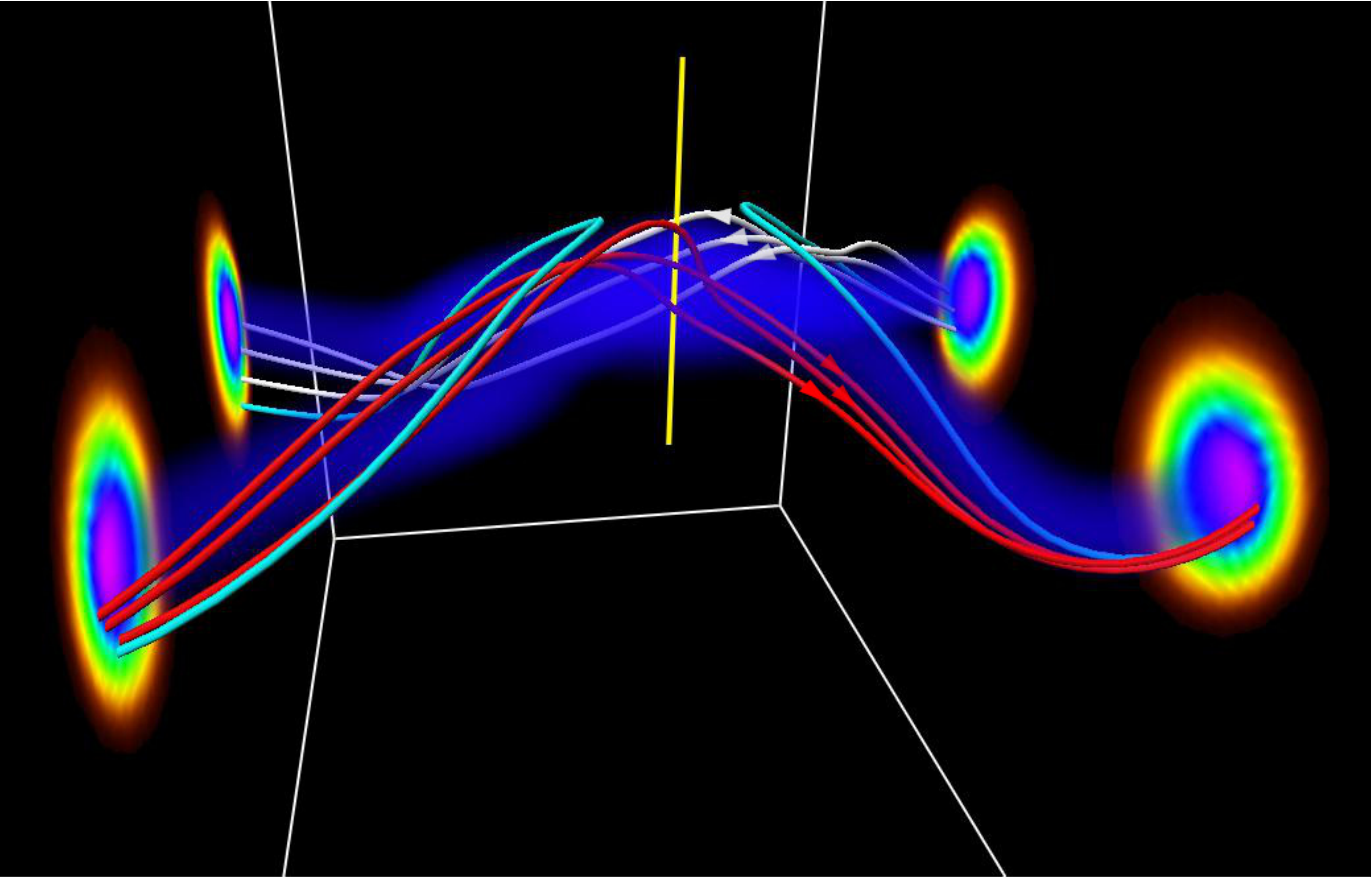}
\caption{Selected vortex lines during the interaction of vortex tubes with finite net helicity, with \tb{$q=0.970$}, at \tb{$t^*=0.70$}. Red and white: thread vortex lines in the vortex sheet that will soon reconnect. Cyan: reconnected bridges. Shading on the end planes and in the volume shows $|\vort|$.}
\label{fig:vapor_d1}
\end{figure}

Due to the twist of the vortex lines within the tubes, there is a non-zero vorticity component along the $z$-axis within the reconnection region. As such, the vortex lines do not reconnect anti-parallel to each other at a vorticity null line, but instead they reconnect at a finite angle, as shown in Figure \ref{fig:vapor_d1}.
In Figure~\ref{fig:afnhcontcwdw}(d) we see that after reconnection much smaller scales have developed in the contours of $(\nabla\times\vort)_\parallel$ within the bridge vortex rings as the axial flow begins to oscillate \citep[also observed for \tb{the untwisted case $q=0$}, see][]{mcgavin2018a}. This indicates that in contrast to the ordered, uni-directional slippage (loss of twist) prior to reconnection, field lines are now changing connectivity by slipping in both directions at various locations throughout the tubes. Helicity variations associated with such oscillations have been considered by \cite{scheeler2017}. Further related ideas appear in the work of \cite{takaoka1996}, who demonstrated the generation of helicity during internal reconnection within a helical vortex structure.

\subsection{Local topology change: creation of null points}

\begin{figure}
\centering
(a)\includegraphics[width=0.33\textwidth]{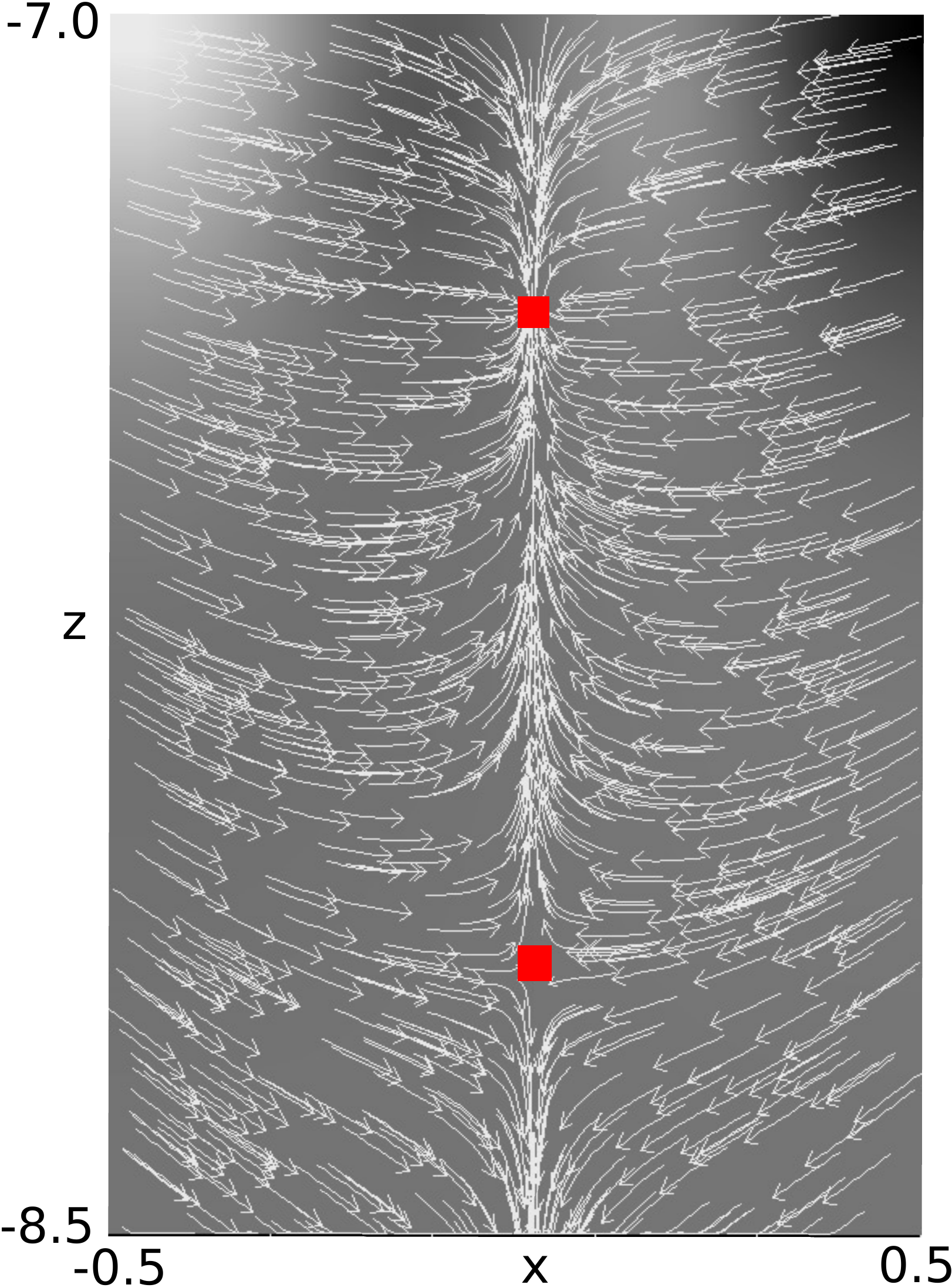}
(b)\includegraphics[width=0.33\textwidth]{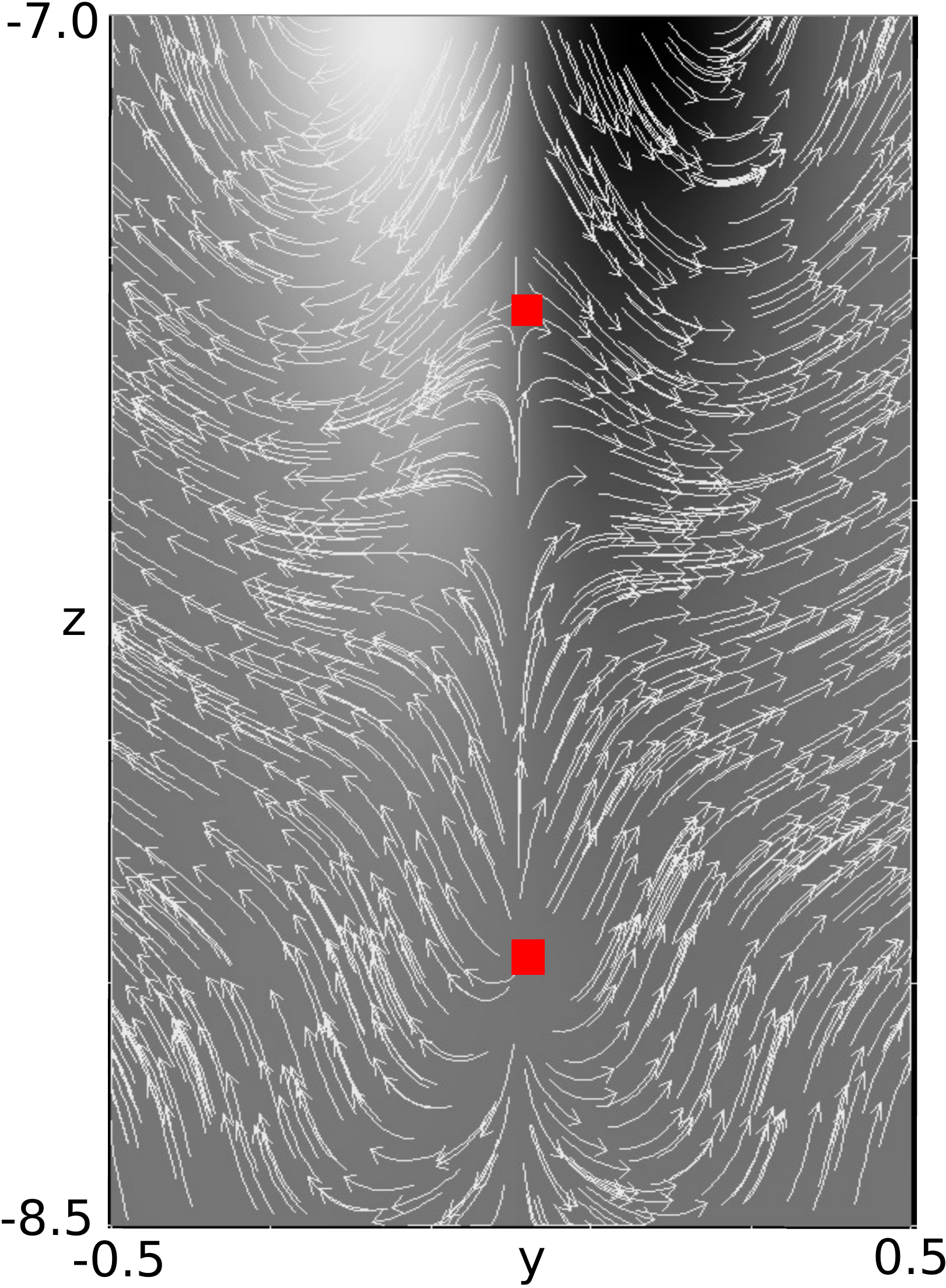}
\caption{3-D null points (red squares) and the surrounding unit-vector vorticity field $\hat{\vort}$ in (a) the $y=0$ plane and (b) the $x=0$ plane at \tb{$t^*=0.70$}. Background shading shows the out-of-plane component of $\vort$. From the simulation with net helicity for the tube pair, and \tb{$q=0.485$}.}
\label{fig:afnhfanspine}
\end{figure}

As the vortex tubes begin to press against each other and reconnect, the vorticity along the $z$-axis becomes non-zero in the region where the tubes meet. This vorticity is always parallel to the $z$-axis by symmetry and initially $\omega_z<0$ for all twists. However, as the reconnection proceeds, we observe an increase in the topological complexity of the vorticity field at the reconnection site, as pairs of 3D vortex nulls are formed. These vorticity nulls are located by applying a numerical implementation of the method described by \cite{haynes2007}. The null pair creation occurs when $\omega_z$ becomes positive along some portions of the $z$-axis. The structures of these nulls are shown in Figure~\ref{fig:afnhfanspine} where for clarity we plot the vorticity field as a unit vector. The main vorticity distribution in the core of the vortex tubes at the top of the images locates the nulls with respect to the main  reconnection region. 
The reason for this local reversal of the vorticity and coincident null pair generation is not clear, and warrants further study in the future. It is worth noting that null points also occur for other values of \tb{$q$} and of $R_m$, and their locations relative to the main vortex tubes were found to vary between simulations.

The structure of the vorticity field in the vicinity of a vortex null point is characterised by a one-dimensional \emph{spine line} along which field lines approach (or recede from) the null, and a two-dimensional \emph{fan surface} within which field lines recede from (or approach) the null -- see \cite{1993PhFlB...5.2355G,parnell1996}. 
Examining Figure~\ref{fig:afnhfanspine} we see that the null point located at $z\approx-7.3$ has field lines that approach the null in the fan surface (that is approximately coincident with the $y=0$ plane), and leave the null along the spine. This is reversed for the  null at $z\approx-8.2$. This is expected since when two nulls are created, they must be of opposite topological degree \citep{1993PhFlB...5.2355G}. The orientation of the null point spine and fan structures is such that their fan surfaces intersect to form a \emph{separator} field line along the portion of the $z$-axis between them.

\subsection{Quantitative change of flux}\label{subsec:afnhflux}

\begin{figure}
\centering
(a)\includegraphics[width=0.45\textwidth]{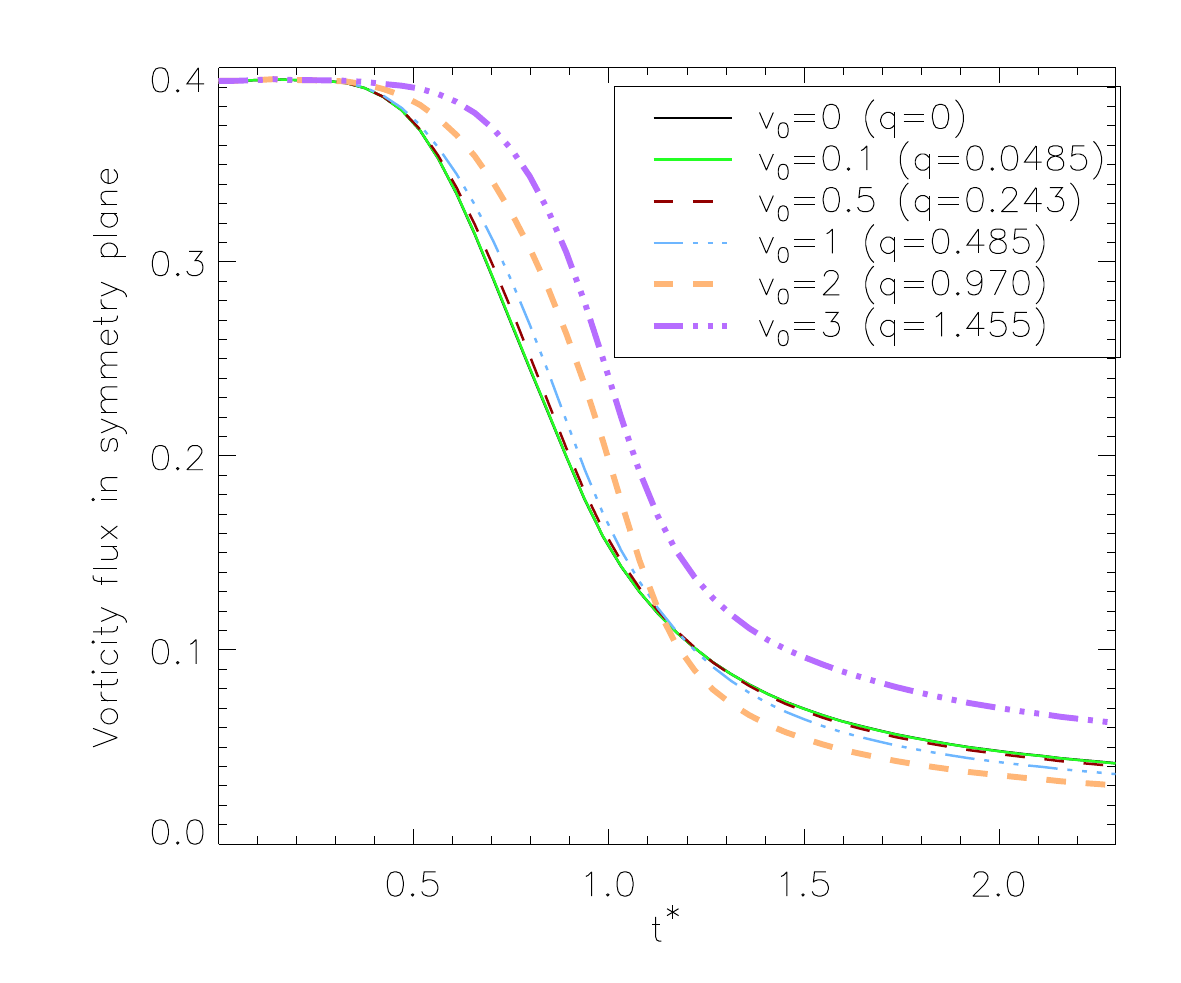}
(b)\includegraphics[width=0.45\textwidth]{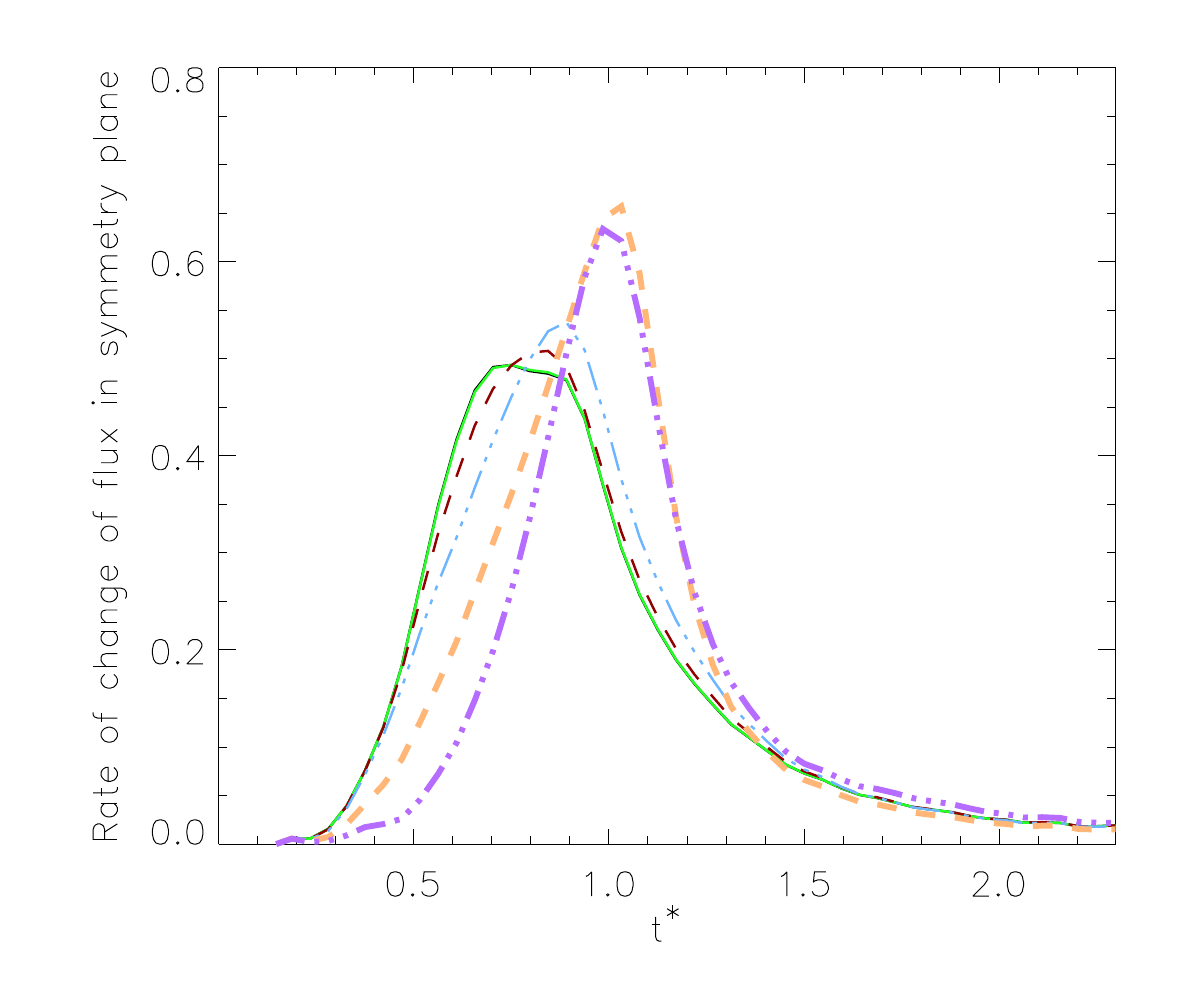}
(c)\includegraphics[width=0.45\textwidth]{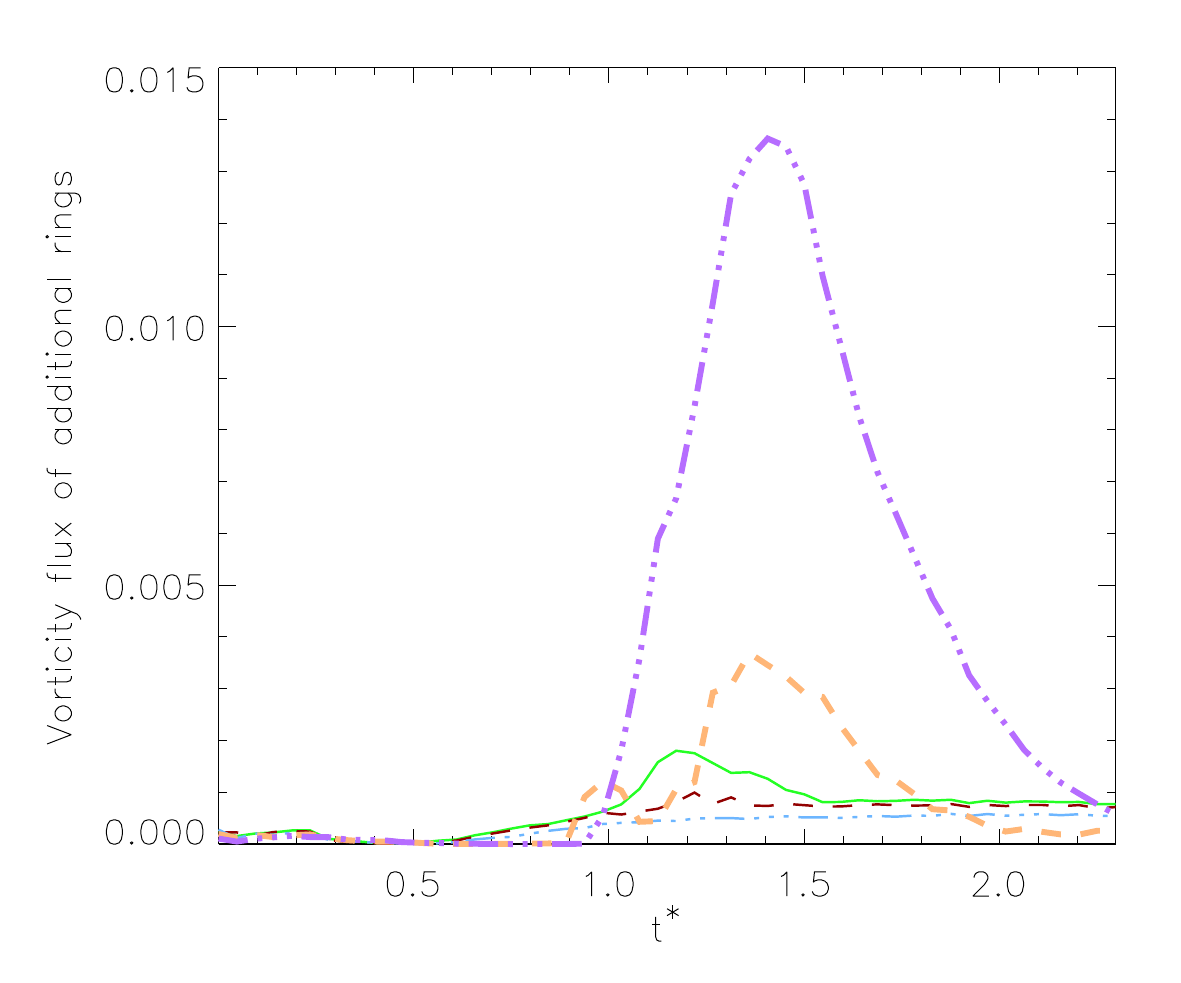}
%
\caption{(a) Vorticity flux in the symmetry plane. (b) Rate of change of {vorticity} flux in the symmetry plane \tb{(note that here and in the following figures this is a rate of change with respect to $t^*$)}. (c) Vorticity flux of the \tb{flux rings around the $z$-axis}.}
\label{fig:afnhvortflux}
\end{figure}

We now discuss the change in flux connectivity during the reconnection, that can be obtained either by analysis of the symmetry plane or by integrating $(\nabla\times\vort)_\parallel$ along the $z$-axis as described above.
These measurements are plotted in Figure~\ref{fig:afnhvortflux}(a). 
\tb{For increasing $q$ we observe a general trend in which the reconnection occurs later, with more flux being reconnected by the end of the simulation  (consistent with the results of \citep[][]{2012PhFl...24g5105V}). (This is first noticeable from $q=0.243$ -- the plots for $q=0$ and $q=0.0485$ are almost indistinguishable on this scale due to the very weak twist, compounded by twist dissipation prior to reconnection, for $q=0.0485$.) }
However, \tb{when the axial flow is increased to $q=1.455$}, we find that this pattern does not continue. Instead, the reconnection is later, the maximum reconnection rate is marginally lower than for \tb{$q=0.970$}, and the overall flux reconnected is lower than in all the other simulations. This trend was also observed by \cite{2012PhFl...24g5105V}, and we hypothesise that it comes about because the perturbations travel sufficiently fast in opposite directions along the tubes that by the time they have rotated towards one another at \tb{$q=1.455$}, they catch one another in a `glancing blow' rather than meeting head-on. 

For reconnection between untwisted vortex tubes at $Re=800$, additional vortex rings form. For the low twist simulations (\tb{$q\leq 0.485$}) the perturbations still collide close to head-on as in the $\tb{q=0}$ case, suggesting that vortex rings are likely to form also in these simulations. 
Indeed, it could be expected that significantly larger additional flux rings appear at higher twists, as follows. As noted already the perturbations travel in opposite directions along the tubes, and for $\tb{q \geq 0.485}$ are significantly displaced from one another along $x$ by the time the tubes meet. As such, when the tubes initially come into contact, they may do so at two points displaced symmetrically from the symmetry plane in $x$. This is of course dependent on the curvature of the tubes at this point, but if it occurs, it would mean that reconnection takes place in two locations along the tubes. This naturally leads to the formation of the standard large flux rings, together with an additional flux ring between the two reconnection points, centred on the $z$-axis.
In Figure~\ref{fig:afnhvortflux}(c) we plot the flux of the vortex ring around the central axis (defined as all field lines that intersect both the $x=0$ and $y=0$ planes twice). There is a small flux measured in this ring for all values of the twist, which is expected since this was also seen in the $\tb{q=0}$ simulations. We see as hypothesised a significantly larger flux ring for the $\tb{q=1.455}$ simulation. After forming, this vortex ring undergoes a rapid self-annihilation for all values of \tb{$q$}.

\subsection{Global Topology}\label{subsec:afnhtopology}

\begin{figure}
\centering
(a)\includegraphics[width=0.38\textwidth]{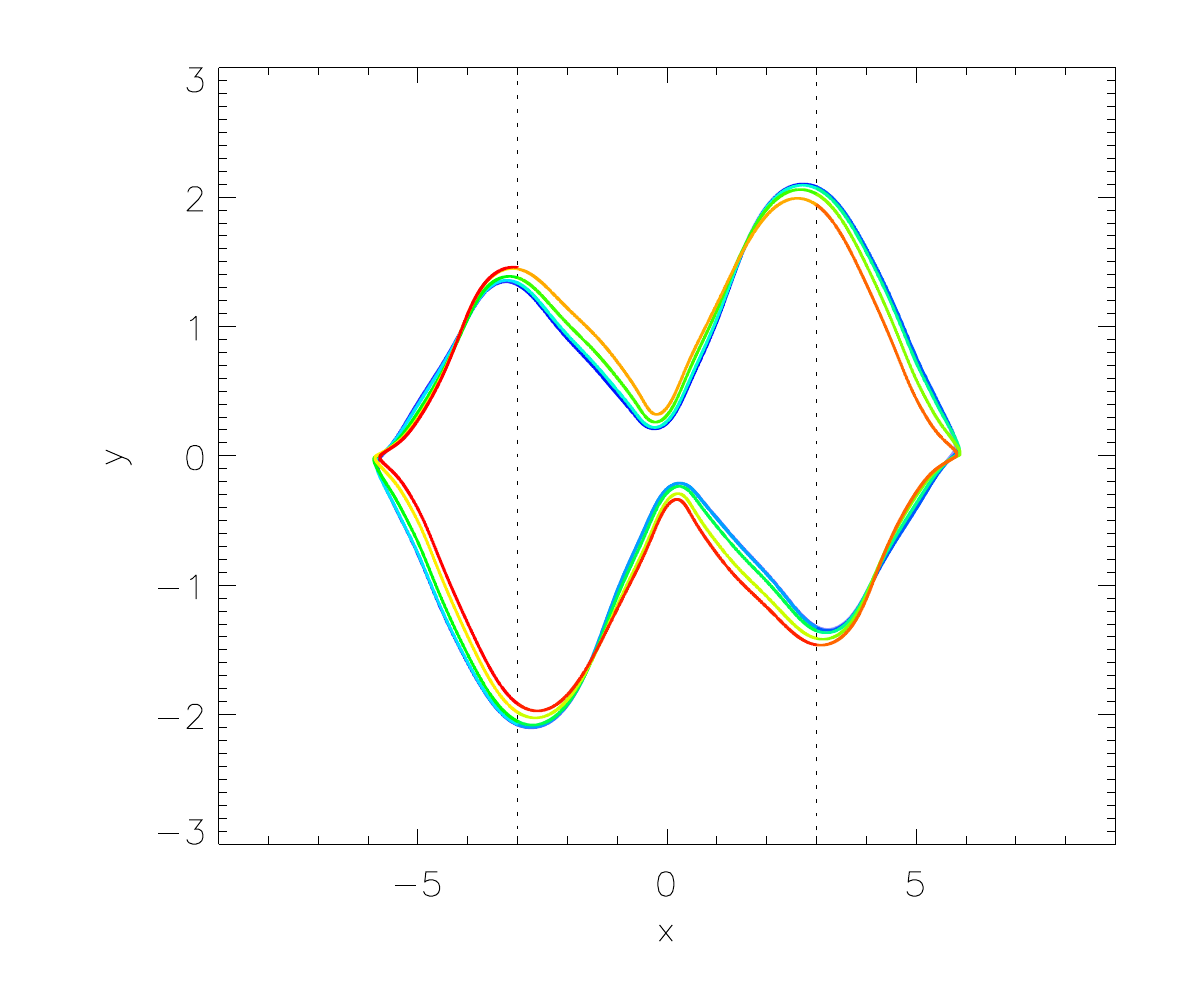}
(b)\includegraphics[width=0.38\textwidth]{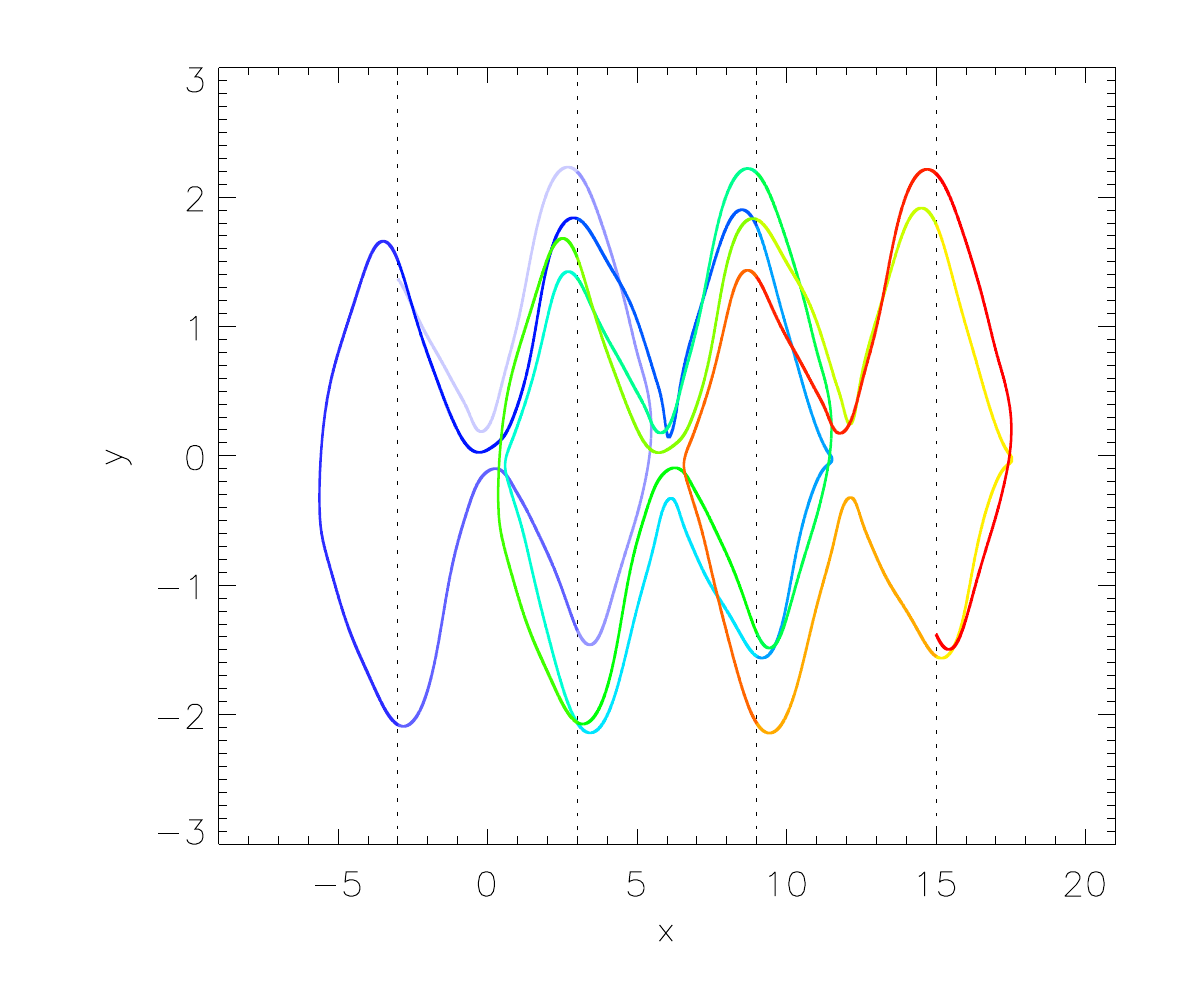}
(c)\includegraphics[width=0.38\textwidth]{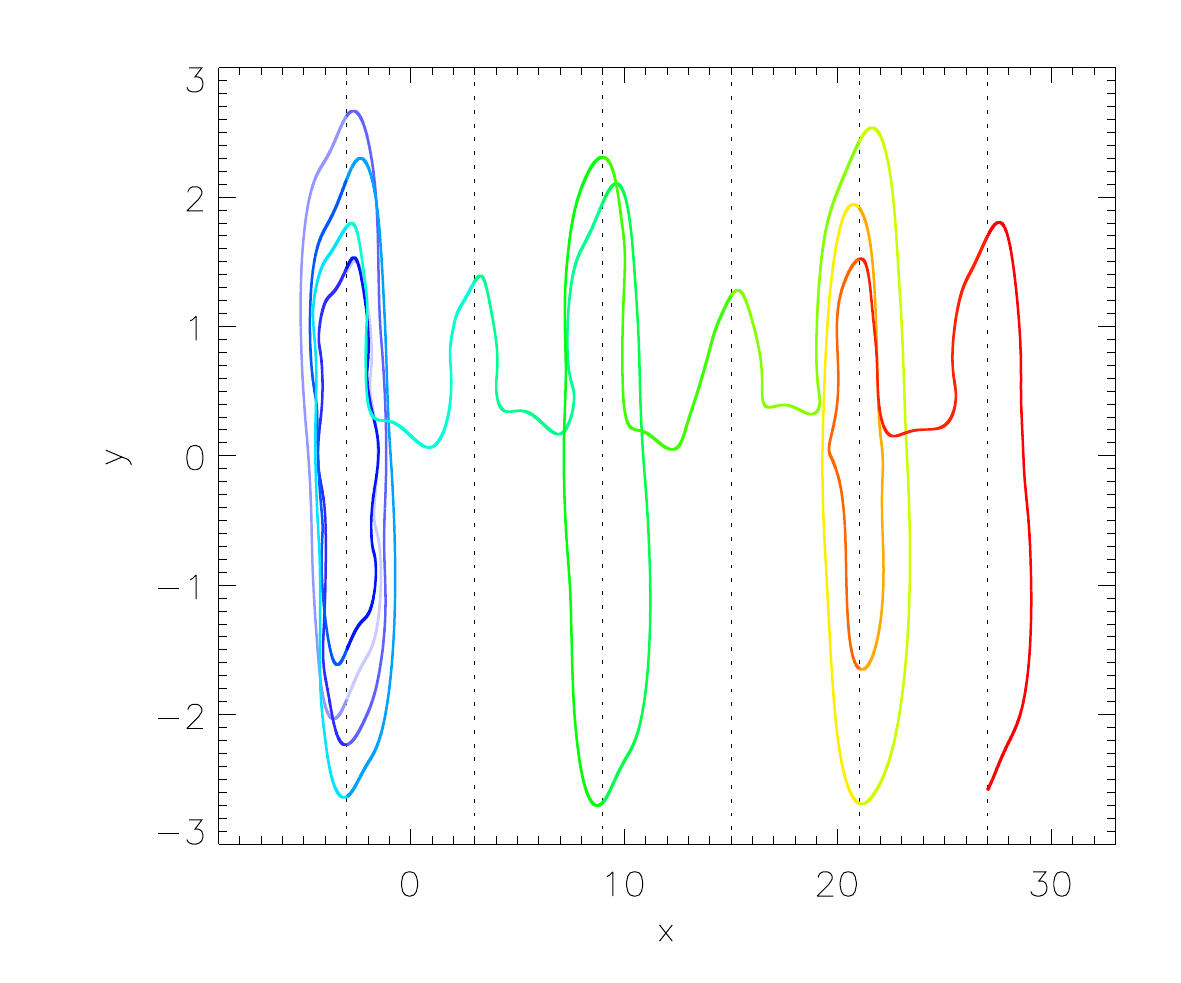}
(d)\includegraphics[width=0.38\textwidth]{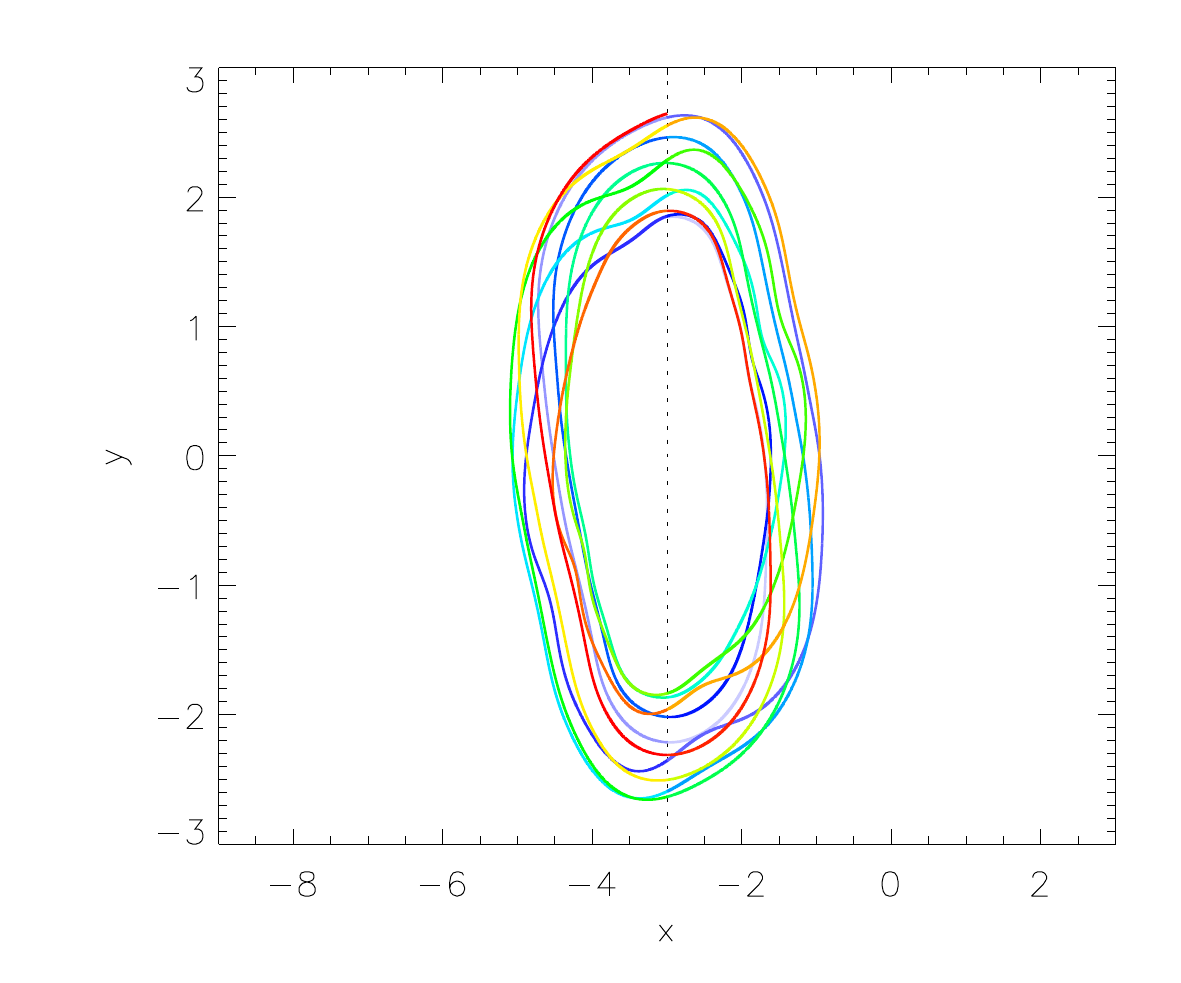}
\caption{Selected vortex lines plotted from the $30\%$ maximum vorticity contour at $x=-3$ at (a) \tb{$t^*=0.66$}, (b) \tb{$t^*=0.75$}, (c) \tb{$t^*=1.64$} and (d) \tb{$t^*=2.11$} with \tb{$q=0.485$}. Change in colour indicates crossing a plane $x=3n, n\in \mathbb{Z}$.}
\label{fig:afnhtopology}
\end{figure}

We examine now the global topology that results from the reconnection process in our simulations (i.e. the topology when we consider multiple periods in the $x$-direction). With zero initial axial flow (\tb{$q=0$}) the vortex lines that are not reconnected (threads) stretch to infinity and the reconnected vortex lines (bridges) form closed rings intersecting the dividing plane and centred at $x=3n, n\in \mathbb{Z}$, due to the symmetries of the system. With the introduction of axial flow, it is clear that a vortex line with non-integer twist within the domain will not map opposite points on $x=\pm 3$ (for a thread), or  mirror points with respect to the dividing plane for a bridge vortex line. This leads to a more complicated topology in which vortex rings extend over several periods of the domain, and vortex lines may wind in and out of many flux rings (these flux rings no longer being bounded by closed flux surfaces of $\vort$). We plot some example vortex lines in Figure~\ref{fig:afnhtopology} taken from various times for the simulation with \tb{$q=0.485$}. These show vortex lines that are period-1 (i.e.~periodic over one period of the domain, Figure~\ref{fig:afnhtopology}d), period-2 (Figure~\ref{fig:afnhtopology}a), and period-4 (Figure~\ref{fig:afnhtopology}b). Also shown are vortex lines that appear to wind in and out of many ring structures in Figure~\ref{fig:afnhtopology}(c).

\section{Results: tube pair with zero net helicity}\label{sec:zh}

\subsection{Qualitative description of the reconnection process}

\begin{figure*}
\centering
(a)\includegraphics[width=0.4\textwidth]{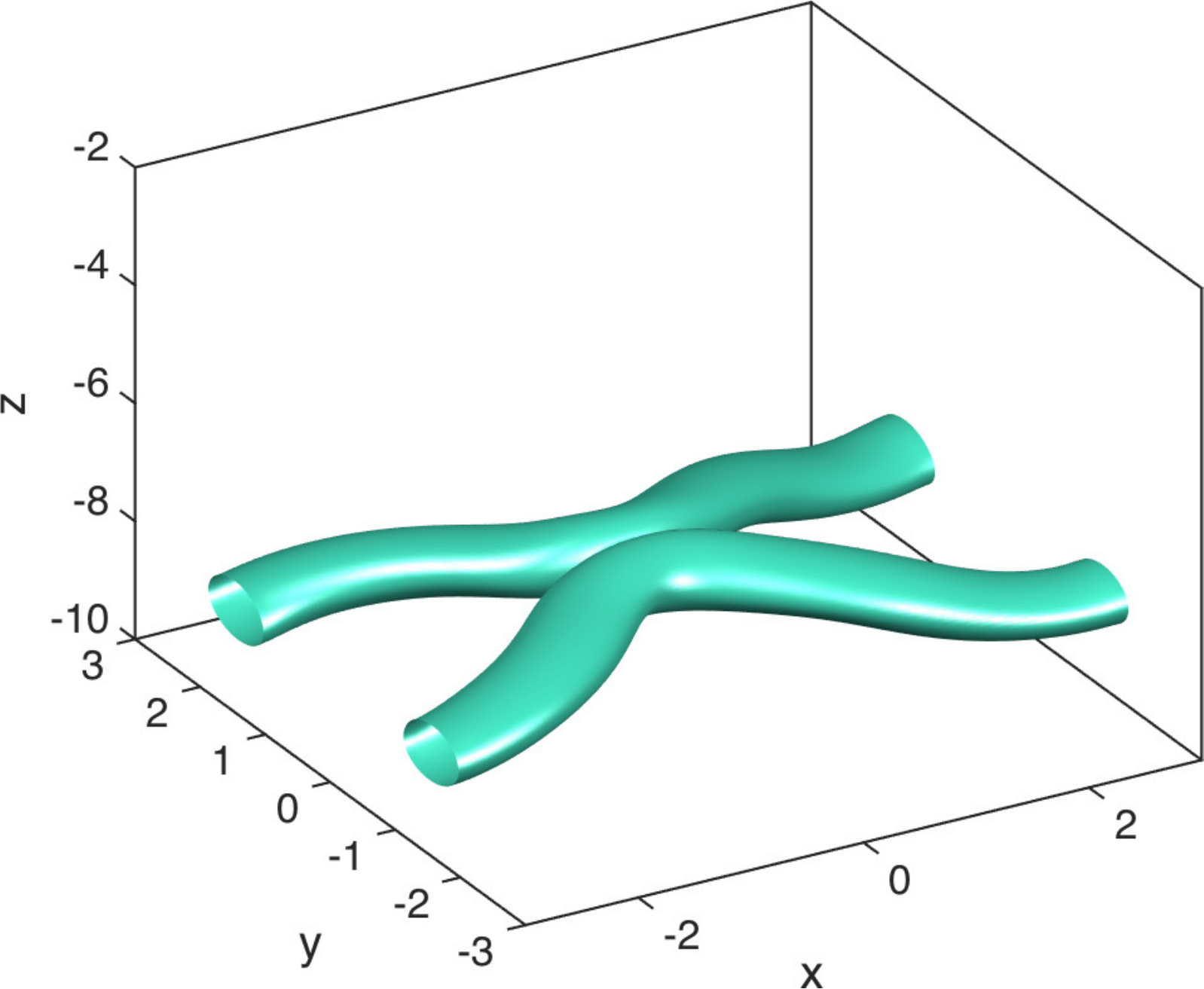}
\includegraphics[width=0.4\textwidth]{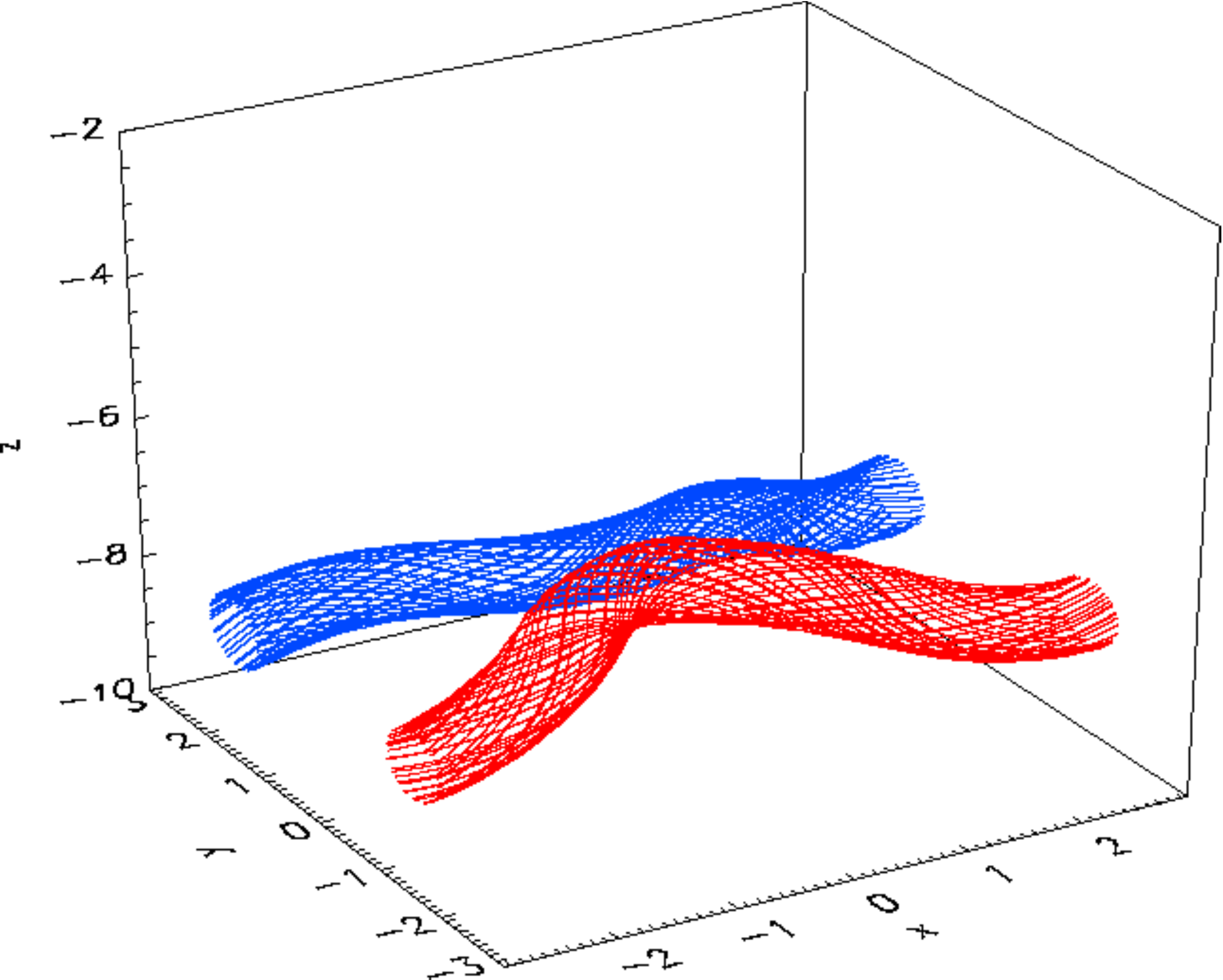}
(b)\includegraphics[width=0.4\textwidth]{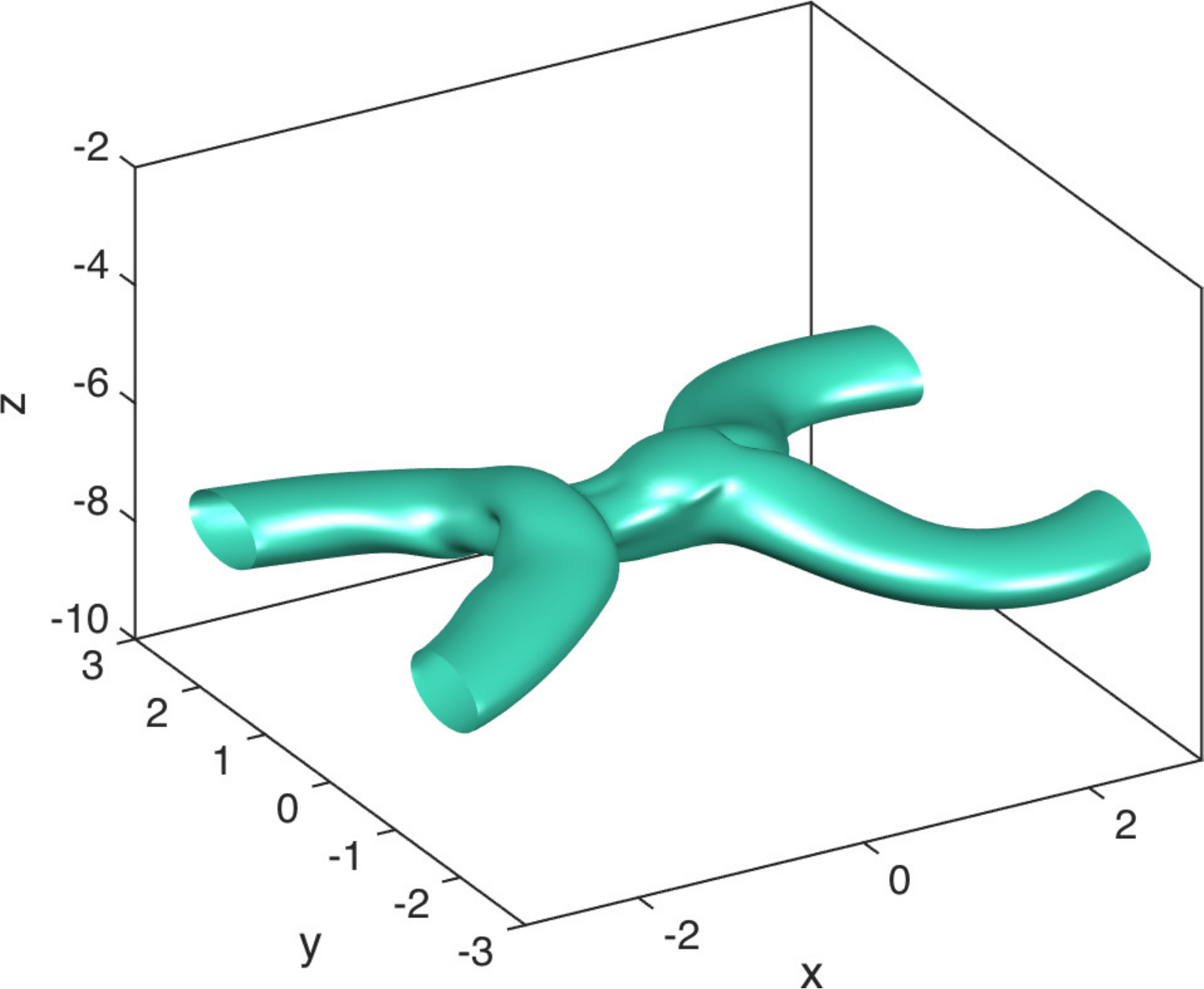}
\includegraphics[width=0.4\textwidth]{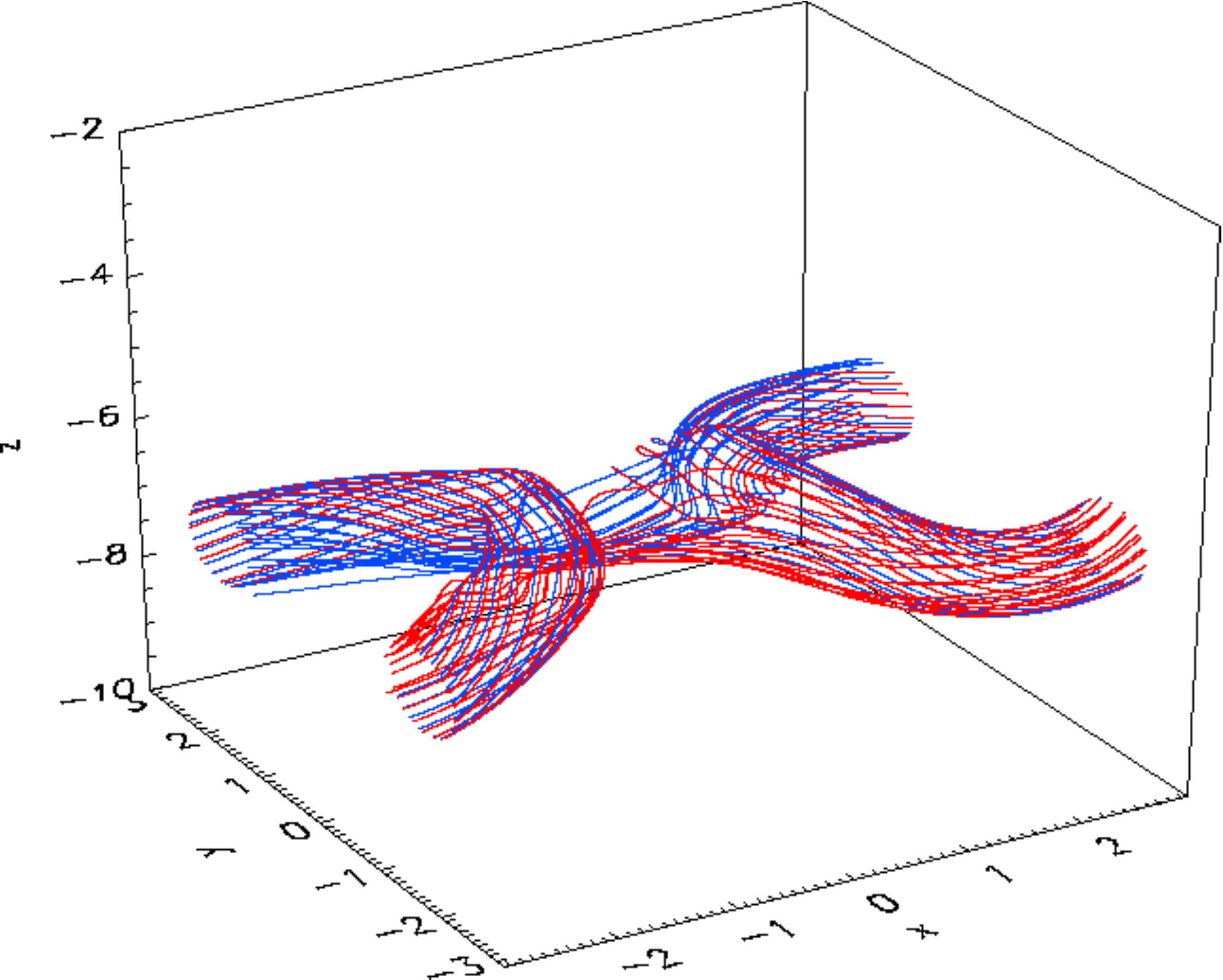}
(c)\includegraphics[width=0.4\textwidth]{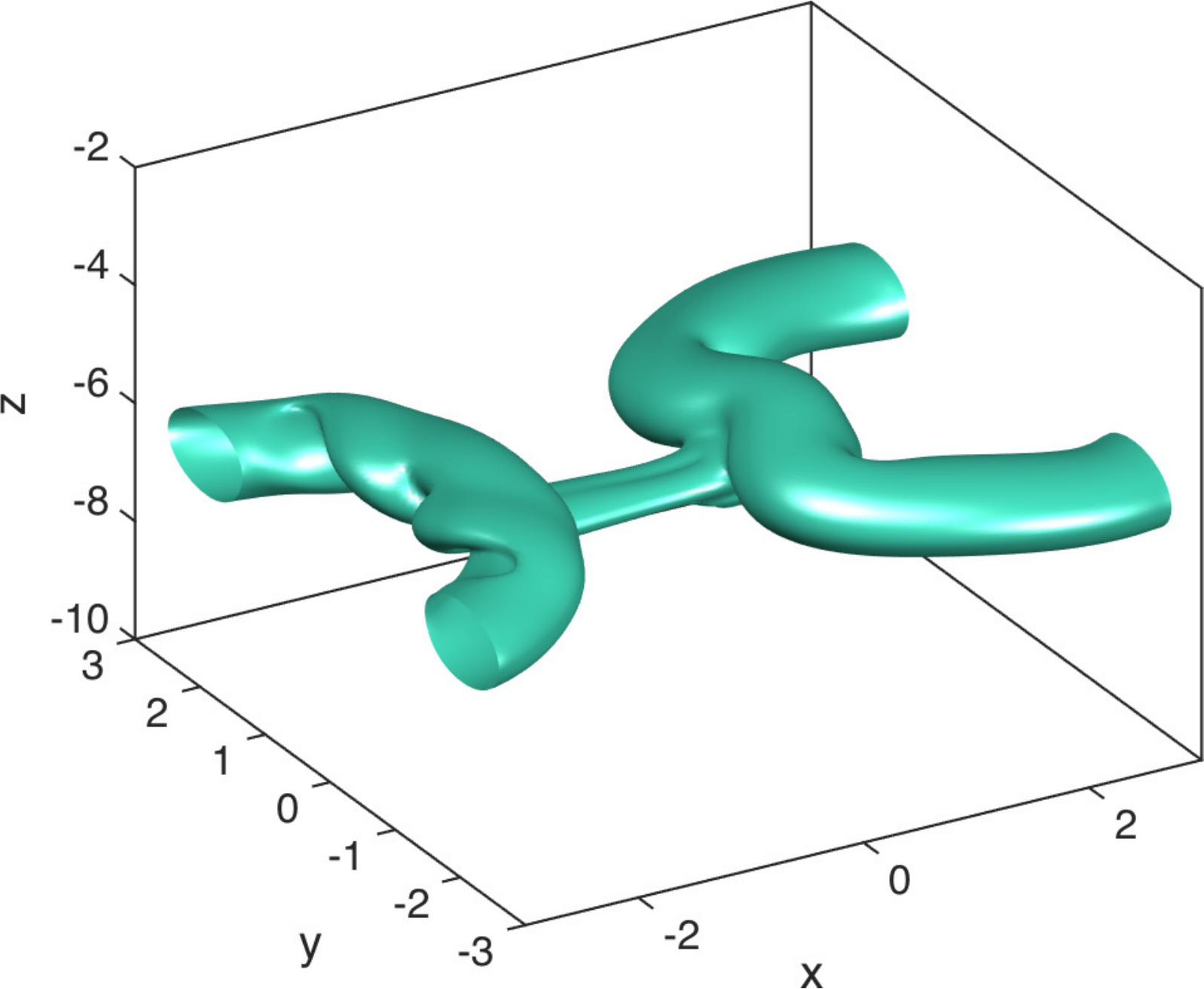}
\includegraphics[width=0.4\textwidth]{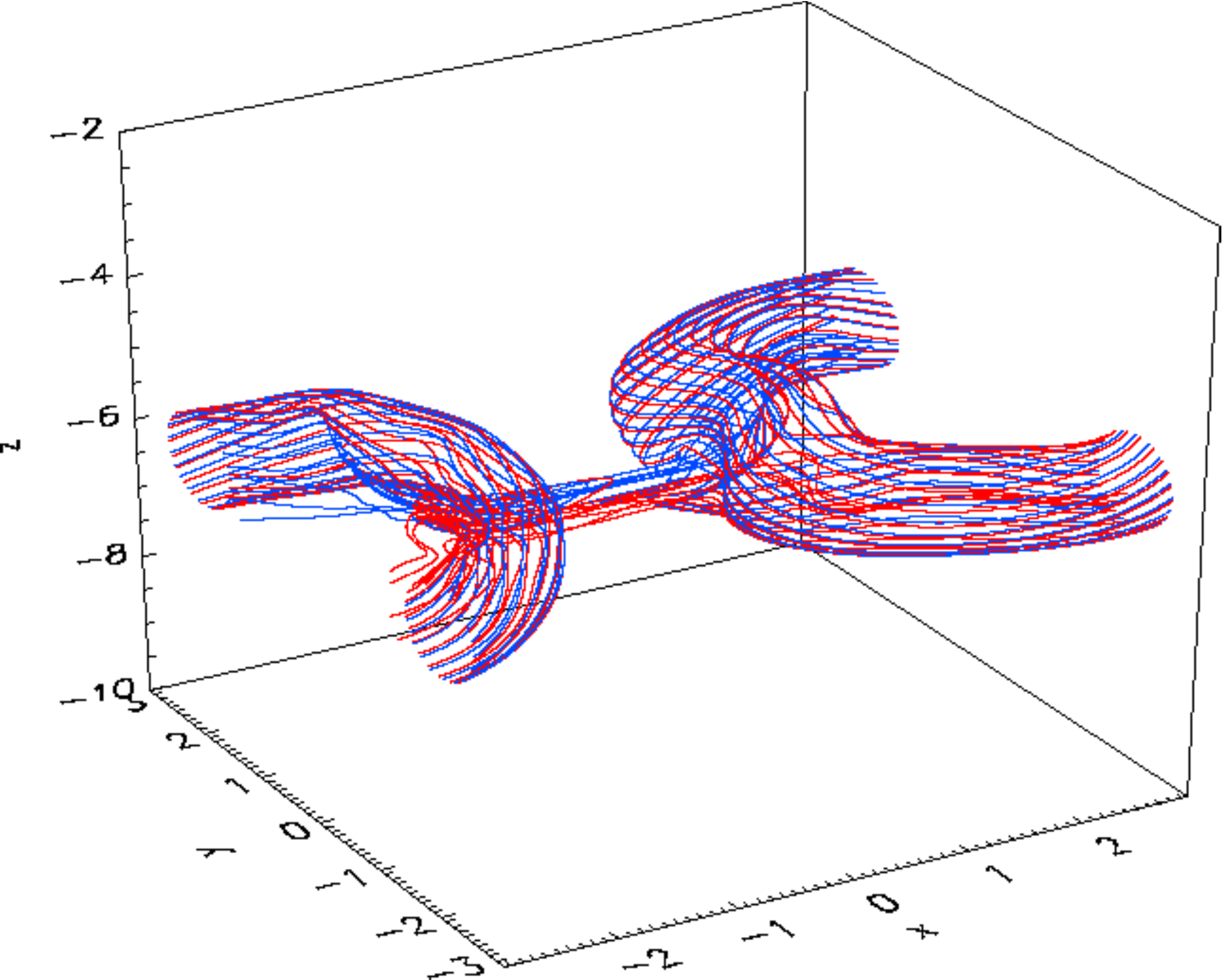}
(d)\includegraphics[width=0.4\textwidth]{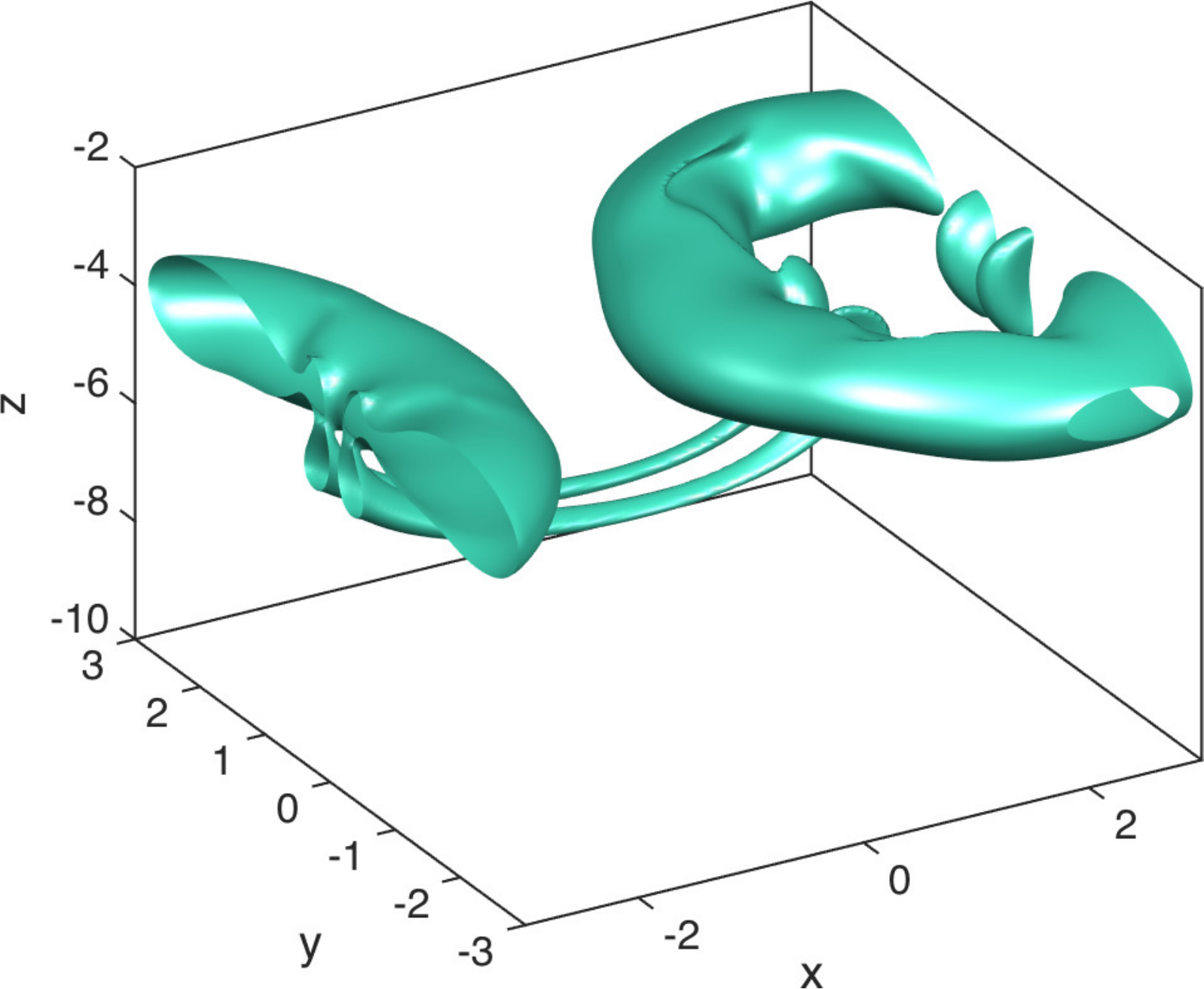}
\includegraphics[width=0.4\textwidth]{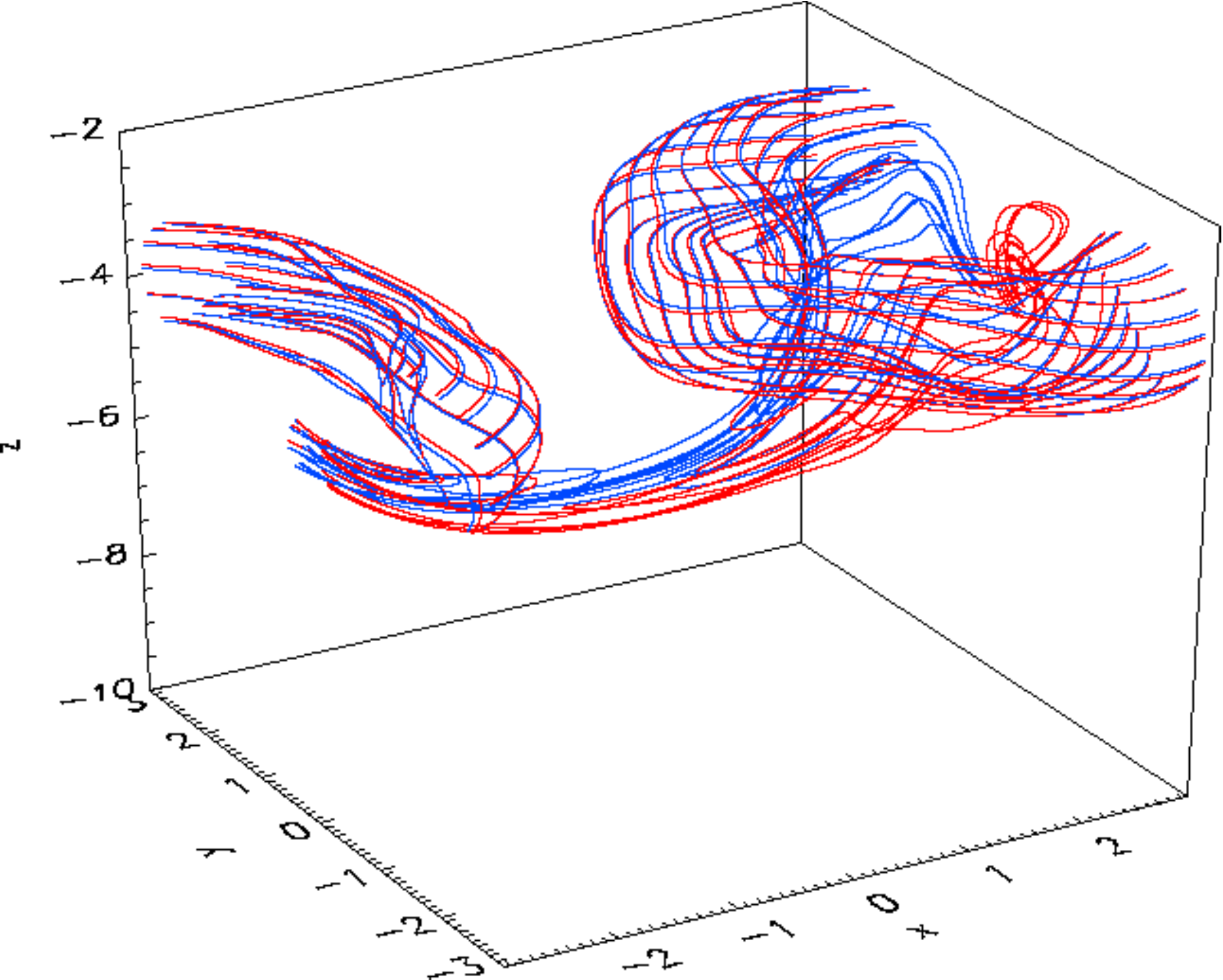}
\caption{$30\%$ vorticity isosurfaces (left) and vortex lines plotted from contours at 30\% of the maximum vorticity on the $x=\pm 3$ boundaries (right), at (a) \tb{$t^*=0.47$}, (b) \tb{$t^*=0.94$}, (c) \tb{$t^*=1.41$} and (d) \tb{$t^*=2.34$}. From the simulation with zero net helicity for the tube pair, and \tb{$q=0.970$}.}
\label{fig:afzht130isosurf}
\end{figure*}

We now consider simulations in which the direction of the axial flow in one of the tubes is reversed, such that the net helicity in the domain is zero. Each vortex tube loses twist at the same rate as the non-zero net helicity simulations (see Section~\ref{subsec:afngtwist}), which we do not discuss further here. 

As before, the perturbations are observed to travel along the tubes in Figure~\ref{fig:afzht130isosurf}, but in this case both perturbations move in the same direction. This perturbation movement means that the $x=0$ plane is no longer a plane of symmetry. However, the dividing plane $y=0$ is a symmetry plane in this case, and will be used later for {vorticity} flux measurements. 
\begin{figure}
\centering
\includegraphics[width=0.5\textwidth]{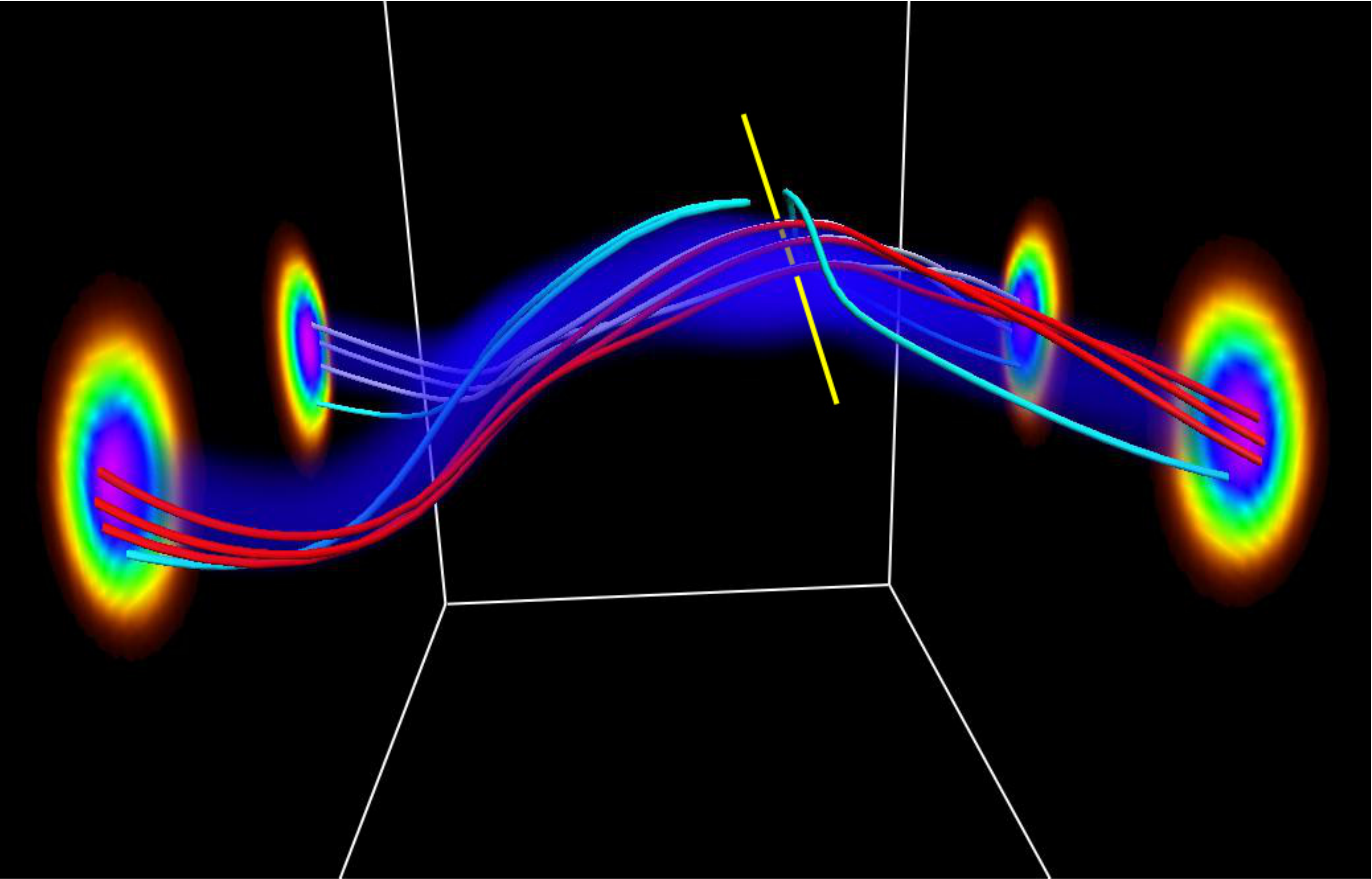}
\caption{Selected vortex lines during the interaction of vortex tubes with zero net helicity, with \tb{$q=0.970$}, at \tb{$t^*=0.70$}. Red and white: thread vortex lines in the vortex sheet. Cyan: reconnected bridges. Shading on the end planes and in the volume shows $|\vort|$. The yellow line indicates the approximate path of the null line.}
\label{fig:vapor_dm1}
\end{figure}
The qualitative properties of the reconnection process are shown by the isosurfaces and vortex lines in Figure~\ref{fig:afzht130isosurf}. 
The plots are based on the simulation run with \tb{$q=0.970$}.
In Figure~\ref{fig:afzht130isosurf}(b) we see the bridges forming, with an apparent imbalance between the larger left bridge and the smaller right bridge. This asymmetry occurs because the vortex lines  reconnect in an anti-parallel fashion at a null line that in this case lies at an angle to the vertical due to the twist of the vortex lines within the tubes. In Figure~\ref{fig:vapor_dm1} the anti-parallel orientation of the reconnecting field lines is highlighted -- the null line runs locally perpendicular to the plane in which the vortex lines lie (approximate position marked in yellow). In Figure~\ref{fig:afzht130isosurf}(c) the asymmetry between the bridges is maintained; a hairpin structure appears in the right bridge whereas the left bridge remains quite smooth (see also Figure \ref{fig:vapor_dm1}). Following reconnection each vortex ring has no net twist {-- the twist on average cancels between the tubes, as does the axial flow. However, neither the twist nor the axial flow is zero everywhere, but rather each fluctuates in both time and space, at the same time as the reconnected tubes (rings) undergo
helical `Kelvin wave' oscillations.}

\subsection{Dividing Plane Vorticity Contour Plots}

\begin{figure}
\centering
(a)\includegraphics[width=0.42\textwidth]{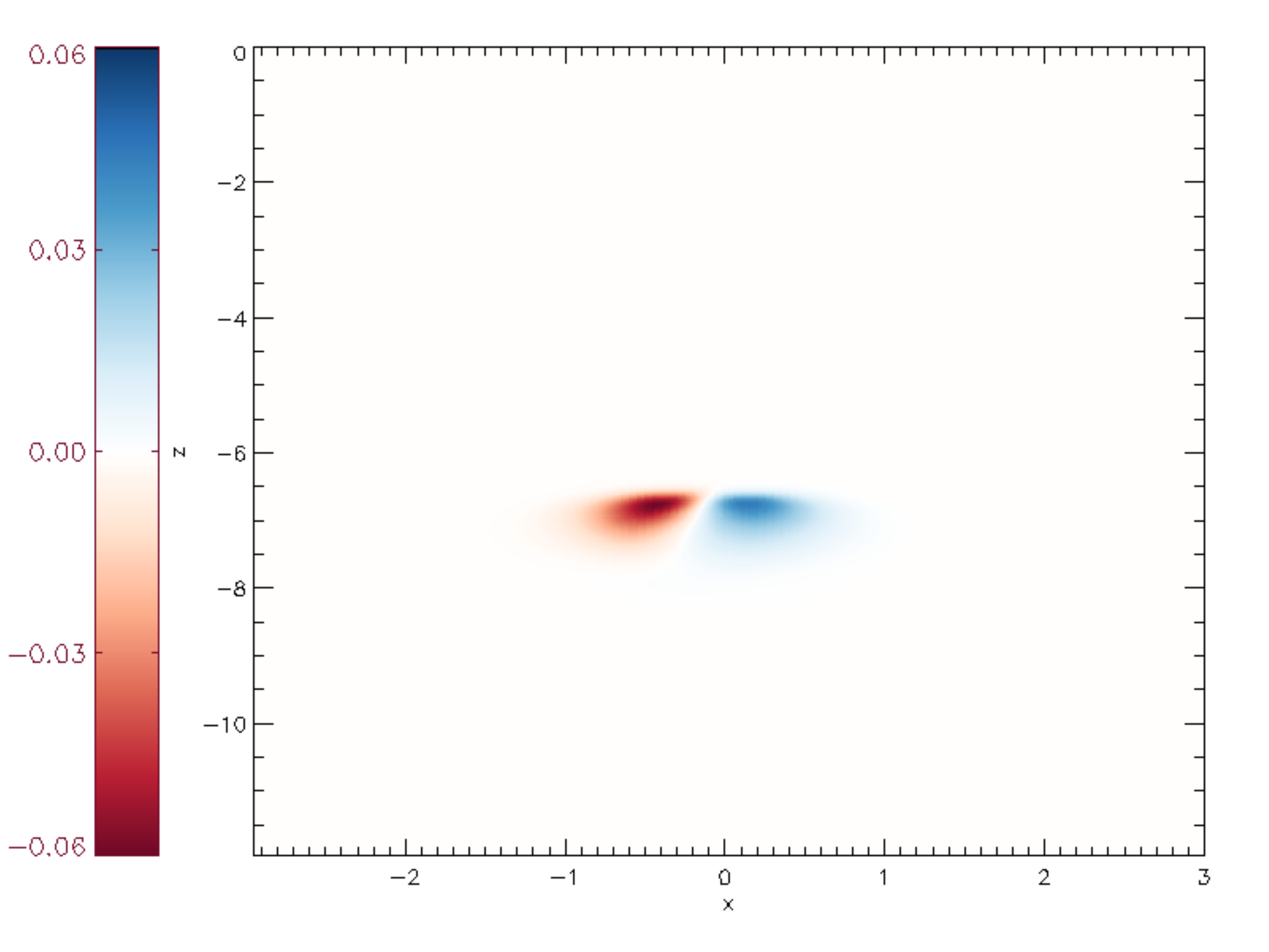}
(b)\includegraphics[width=0.42\textwidth]{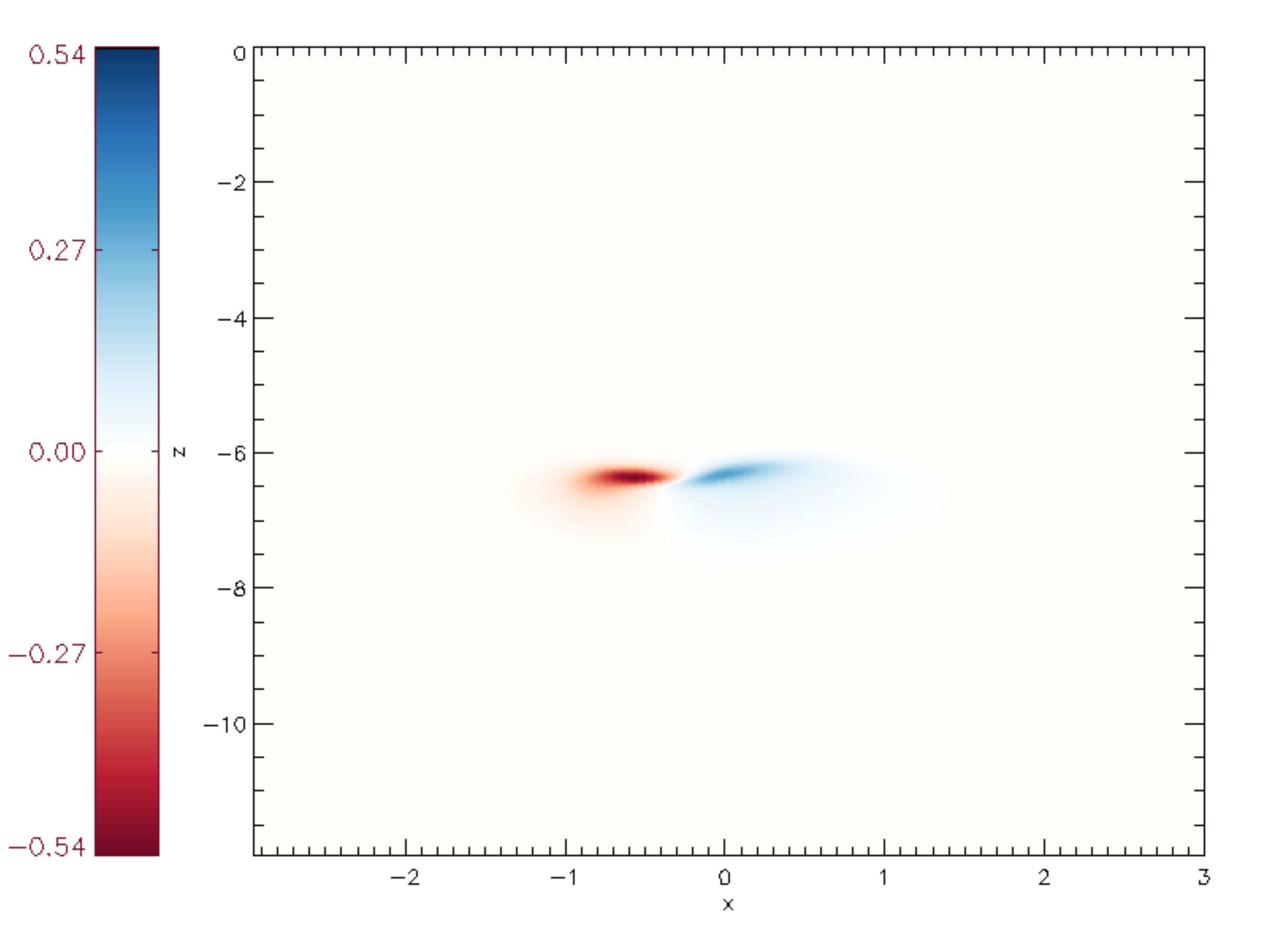}
(c)\includegraphics[width=0.42\textwidth]{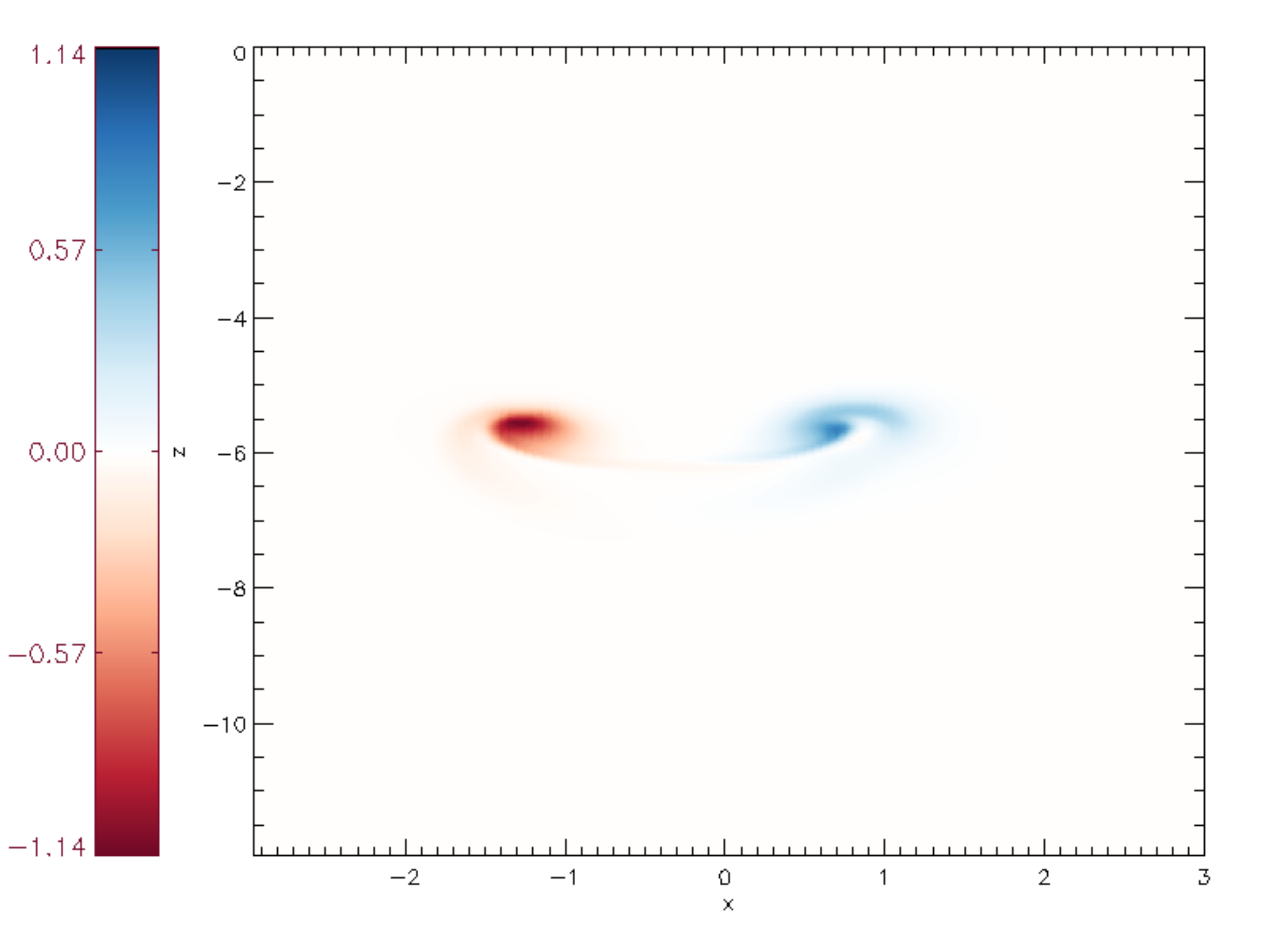}
(d)\includegraphics[width=0.42\textwidth]{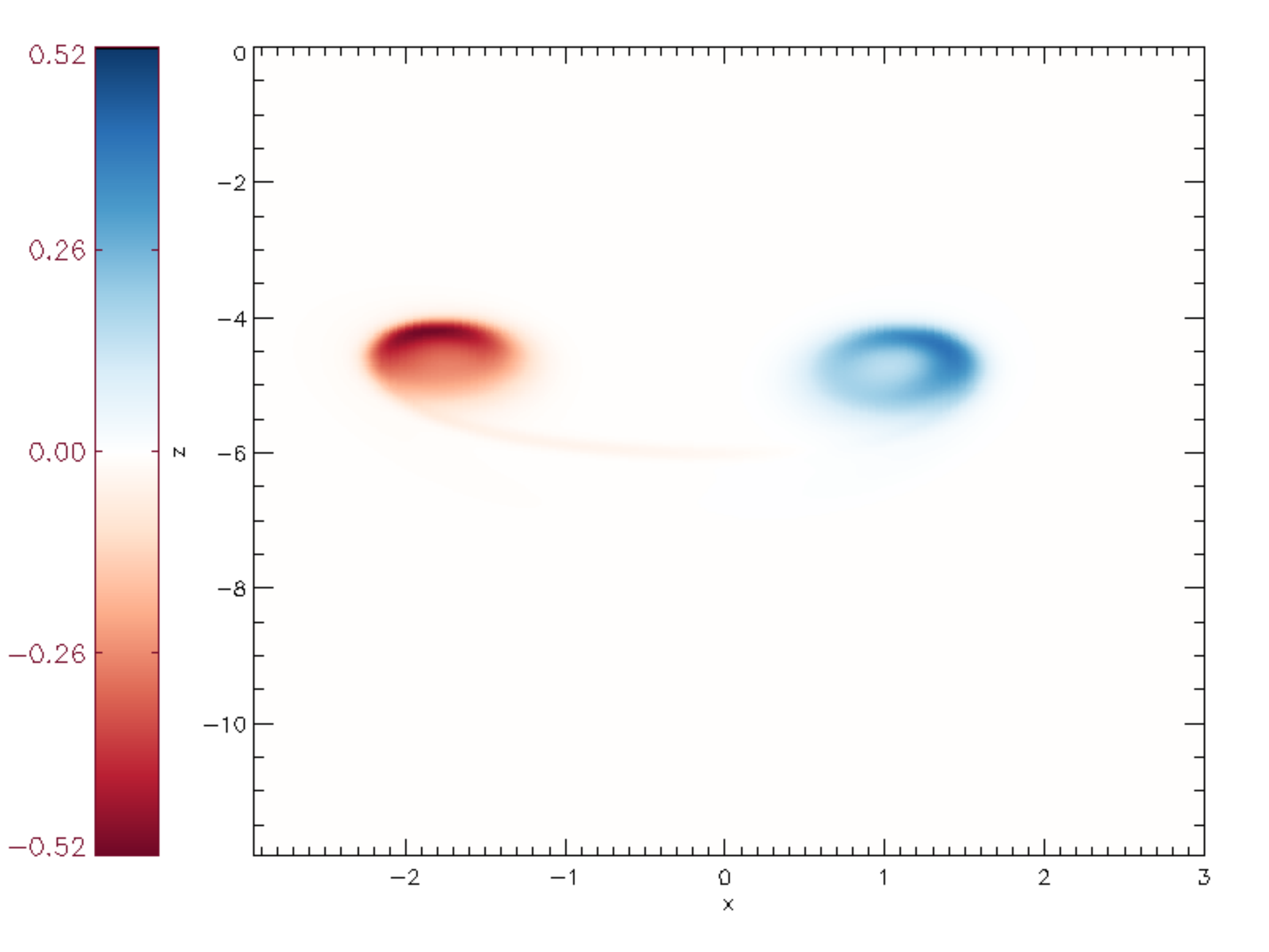}
\caption{$\omega_y$ contours in the dividing plane ($y=0$) for the simulation with \tb{$q=0.485$} at (a) \tb{$t^*=0.47$}, (b) \tb{$t^*=0.70$}, (c) \tb{$t^*=1.17$} and (d) \tb{$t^*=1.88$}.}
\label{fig:afzhdivplanecont}
\end{figure}

To observe the asymmetry in the bridges that has been introduced by the addition of axial flow we plot the vorticity contours in the dividing plane in Figure~\ref{fig:afzhdivplanecont}. At \tb{$t^*=0.47$} we can see that the null line between the two bridges  lies at an angle to the vertical, as previously discussed. It is also clear that the maximum vorticity is no longer the same in each bridge (as it is for \tb{$q=0$}). As the vorticity flux in the bridges must be equal, we see that the weaker vortex tube (the one that intersects the $y=0$ plane at $x>0$) has a larger diameter than the stronger one. The differences between the two bridges continue throughout the simulation as the angle in which the null line formed disappears. At later time the long hairpin vortex lines that have recently reconnected are much longer for the negative bridge, seen best in Figure~\ref{fig:afzhdivplanecont}(d). 

\begin{figure}
\centering
(a)\includegraphics[width=0.45\textwidth]{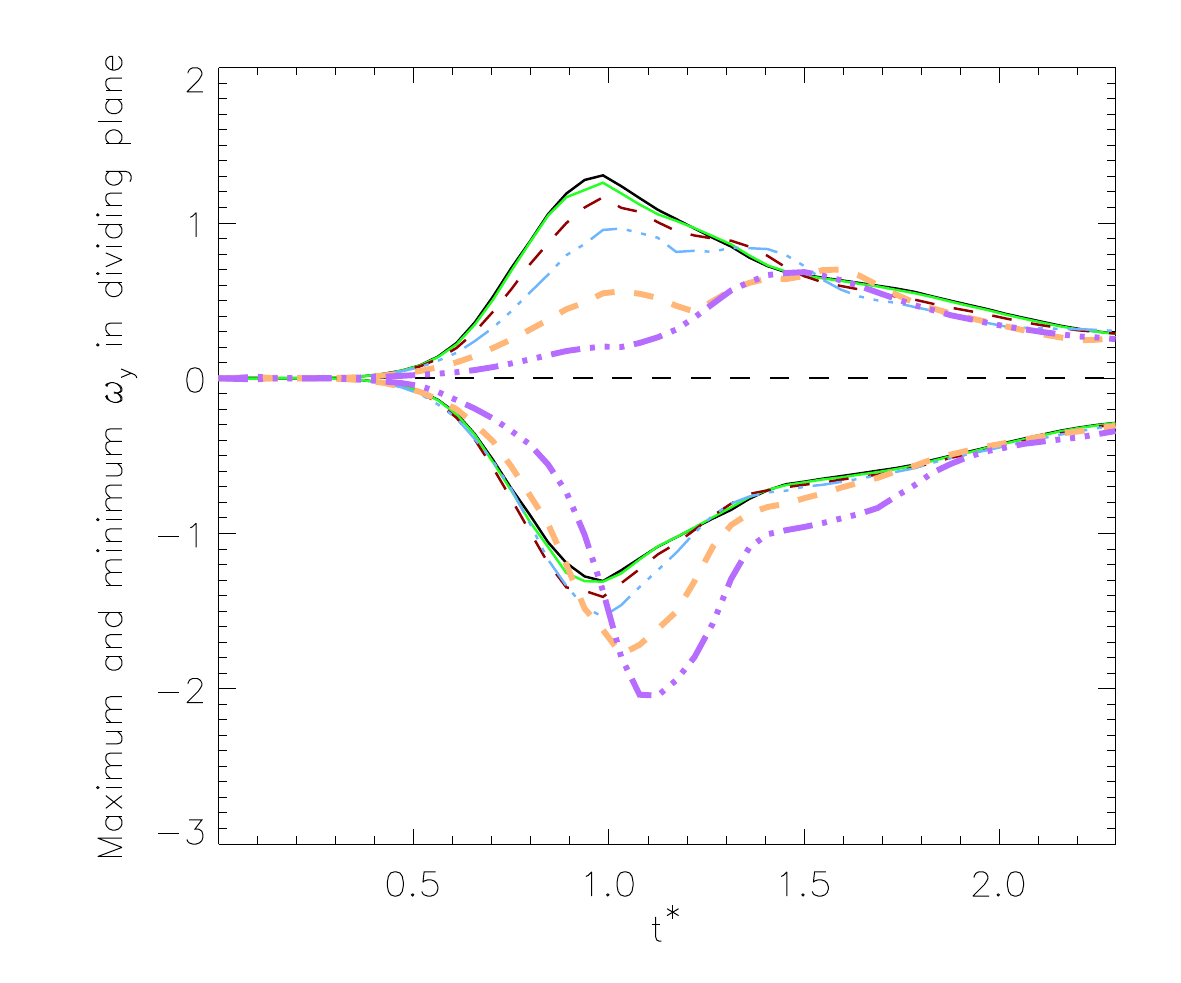}
(b)\includegraphics[width=0.45\textwidth]{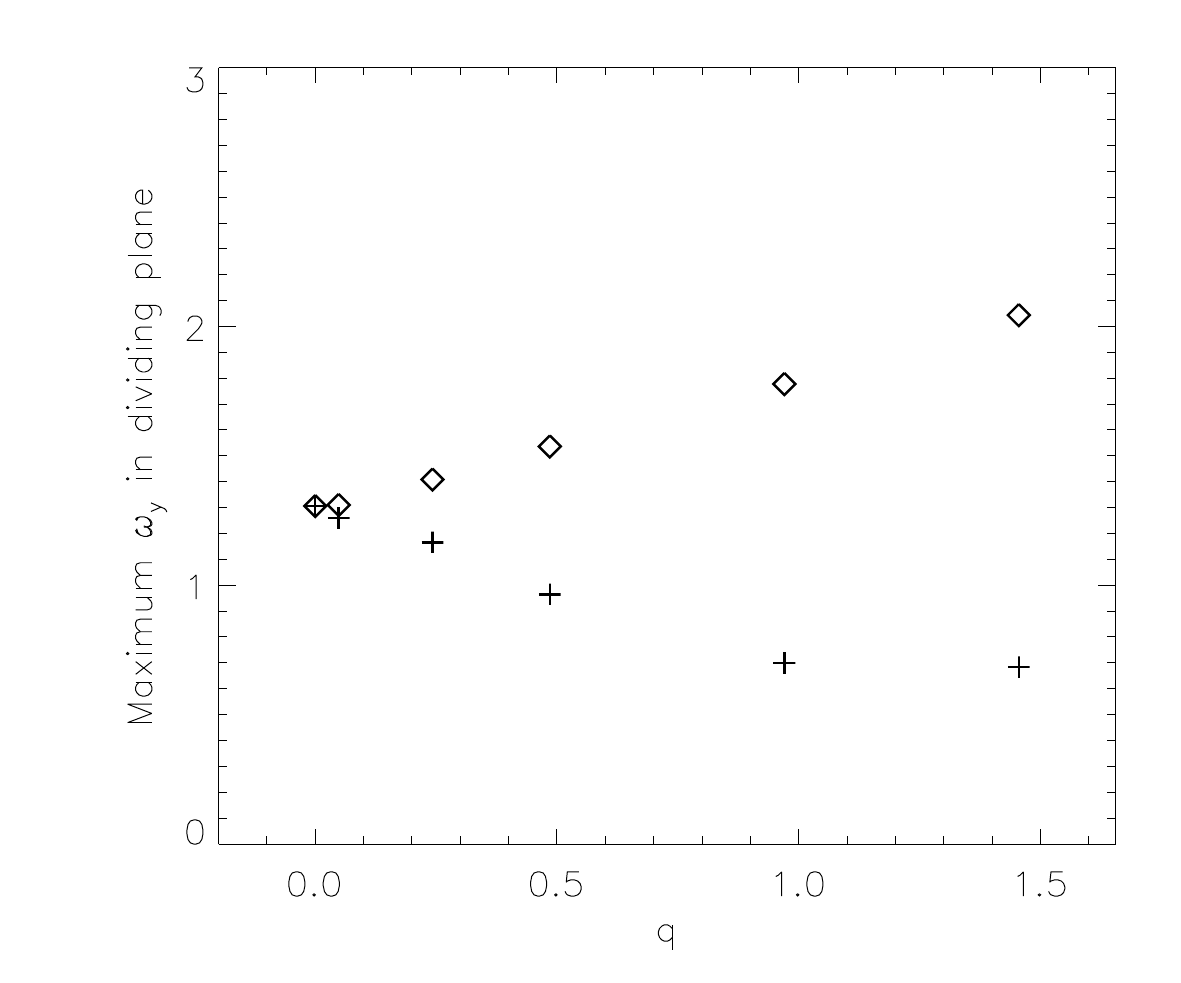}
\caption{(a) Maximum  values of $\pm\omega_y$ in the dividing plane as a function of time. \tb{Line styles as in Figure \ref{fig:afnhvortflux}}. (b) Max($\omega_y$), diamonds, and $-{\rm max}(-\omega_y)$, crosses, over time in the dividing plane as a function of \tb{$q$}. }
\label{fig:afzhmaxvort}
\end{figure}

To quantify the asymmetry between the bridges we plot in Figure~\ref{fig:afzhmaxvort} the maximum positive and negative values of $\omega_y$ in the dividing plane. With an increase in twist ($v_0$, \tb{equivalently $q$}) the positive bridge gets weaker, the negative bridge stronger and the spatio-temporal extremes occur later in time. This could be due to the increased angle of the null line during reconnection leading to different shapes of the bridges. Plotted as a function of twist in (b) we see that for the range of twists considered these extreme values scale approximately linearly with \tb{$q$}. However, this trend cannot continue to higher twist values (at least for the maximum of $-\omega_y$), and indeed a break in the trend is seen for the \tb{$q=1.455$} simulation.



\subsection{Nature and rate of reconnection}

\begin{figure}
\centering
(a)\includegraphics[width=0.45\textwidth]{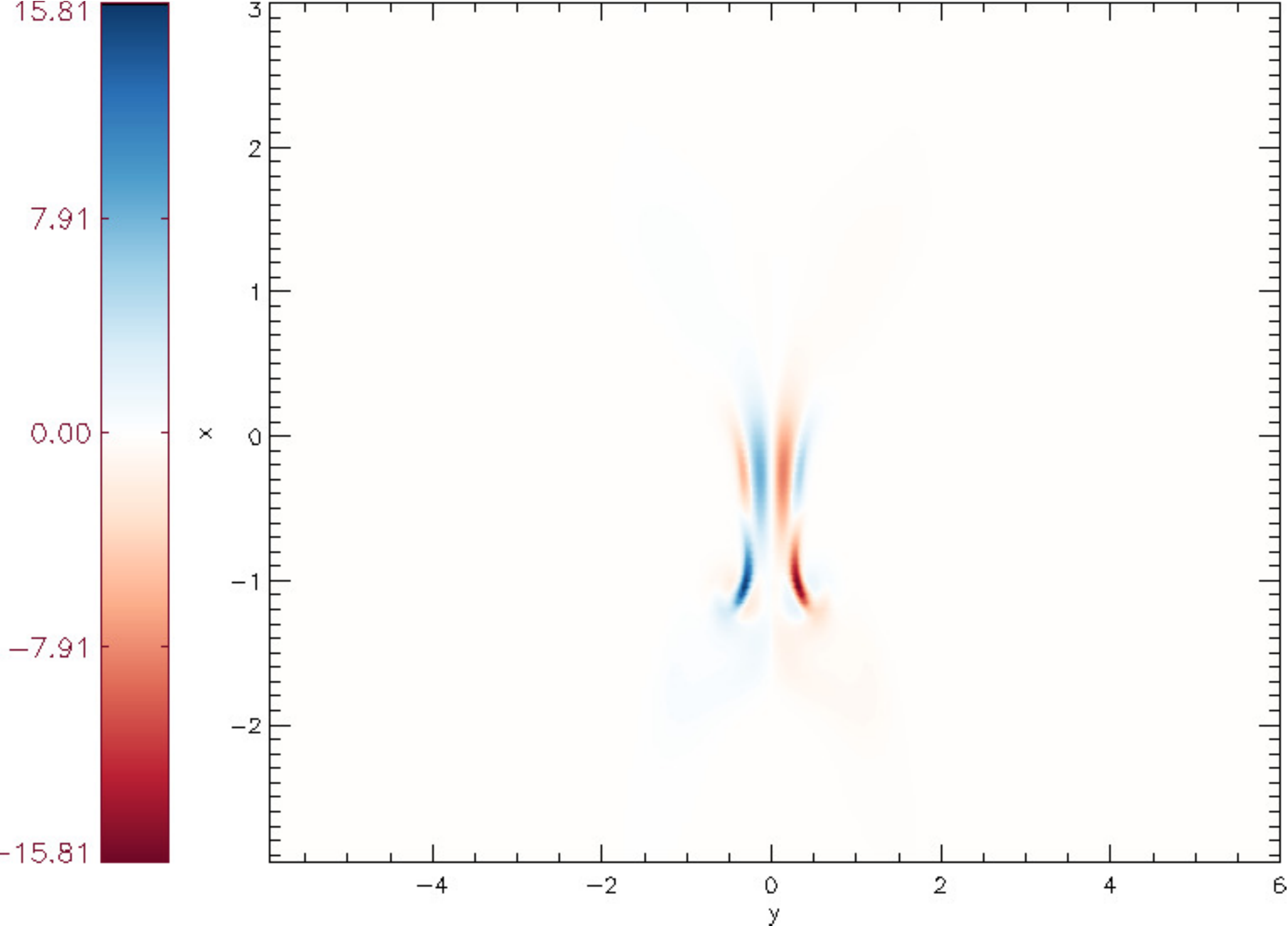}
(b)\includegraphics[width=0.45\textwidth]{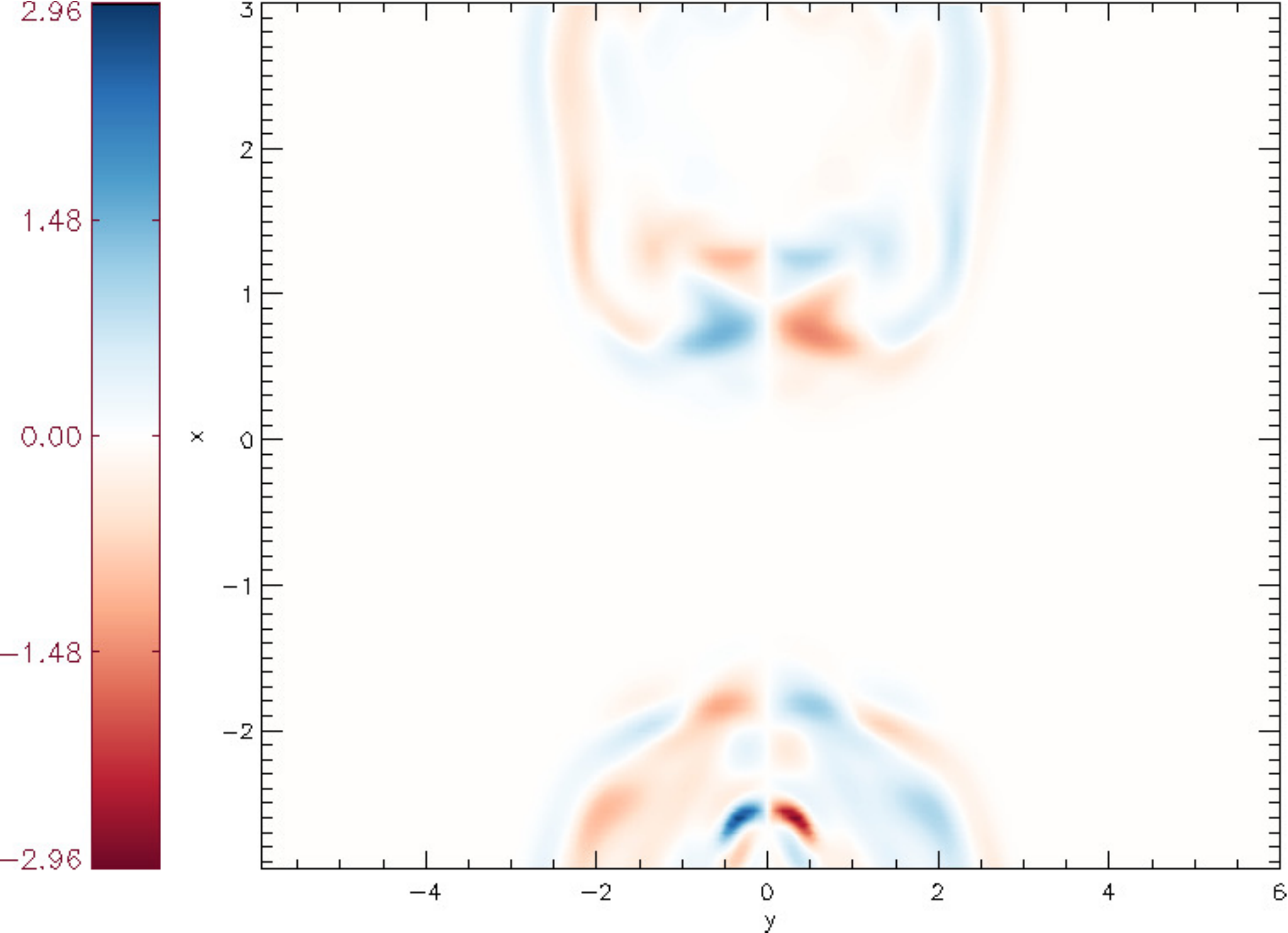}
\caption{Contours of $(\nabla\times\vort)_\parallel$ for the simulation with zero net helicity  and \tb{$q=0.970$}, at (a) \tb{$t^*=0.94$} and $z=-6.35$ and (b) \tb{$t^*=1.88$} and $z=-4.75$. In each case the contours are shown in planes that pass through the location of the spatial maximum over the domain  of $(\nabla\times\vort)_\parallel$ at the corresponding time.}
\label{fig:afzhcontcwdw}
\end{figure}

While the perturbations in these simulations travel along the tubes such that $x=0$ is no longer a symmetry plane, the dividing plane is still a symmetry plane, such that any vortex line passing through it must have been reconnected (there is no significant Kelvin-Helmholtz instability at this value of $R_m$, that may lead to a breaking of this symmetry).
As demonstrated in Figure~\ref{fig:afzhcontcwdw}, $(\nabla\times\vort)_\parallel=0$ in the dividing plane by symmetry, indicating that the reconnection process between the tubes is locally two-dimensional. This is consistent with our earlier discussion of the presence of an X-line in this plane, albeit one that is tilted with respect to the $z$-axis. We still see reconnection regions (local concentrations of $(\nabla\times\vort)_\parallel$) within the tubes that are responsible for ``internal reconnection" that changes the twist, as described for the \tb{$q=0$} simulations by \cite{mcgavin2018a}. In (b) we observe a number of localised regions of $(\nabla\times\vort)_\parallel$, that are associated with the twist oscillations (Kelvin waves). Note that the intense concentrations around $x=-2.5$ being stronger than their ``mirror images" around $x=+2$ is a projection effect, since the rings do not lie in planes of constant $z$.

\begin{figure}
\centering
(a)\includegraphics[width=0.45\textwidth]{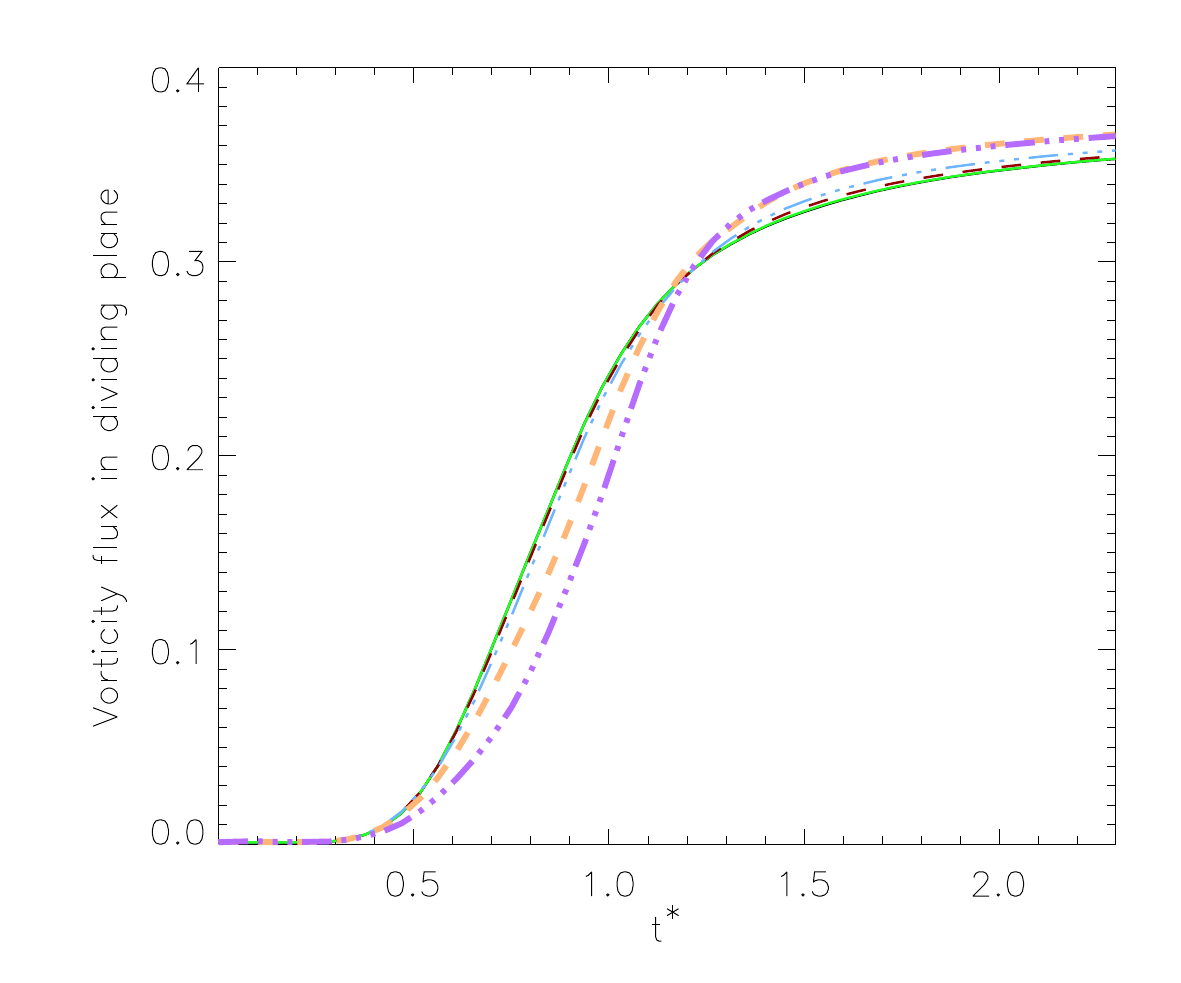}
(b)\includegraphics[width=0.45\textwidth]{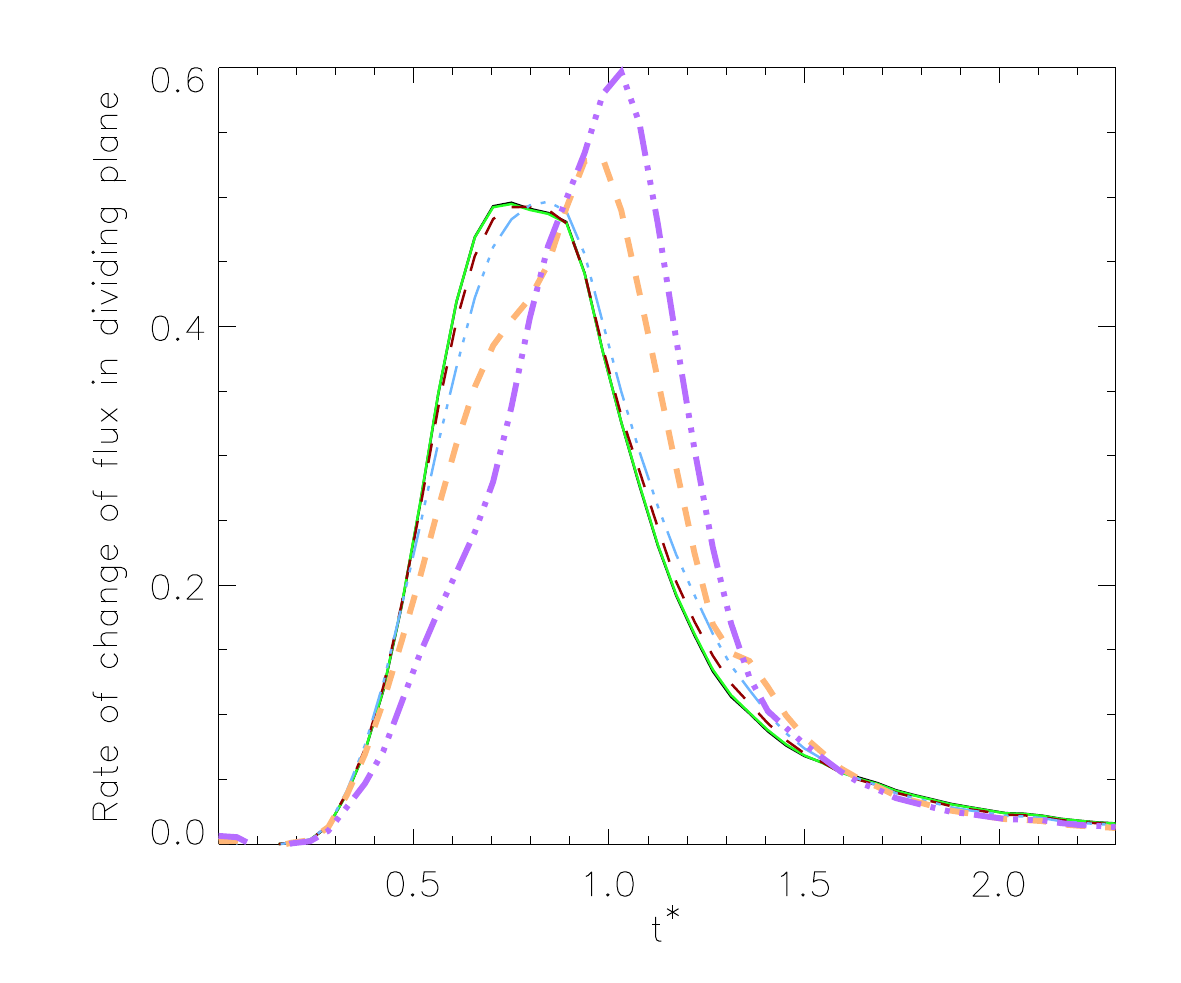}
(c)\includegraphics[width=0.45\textwidth]{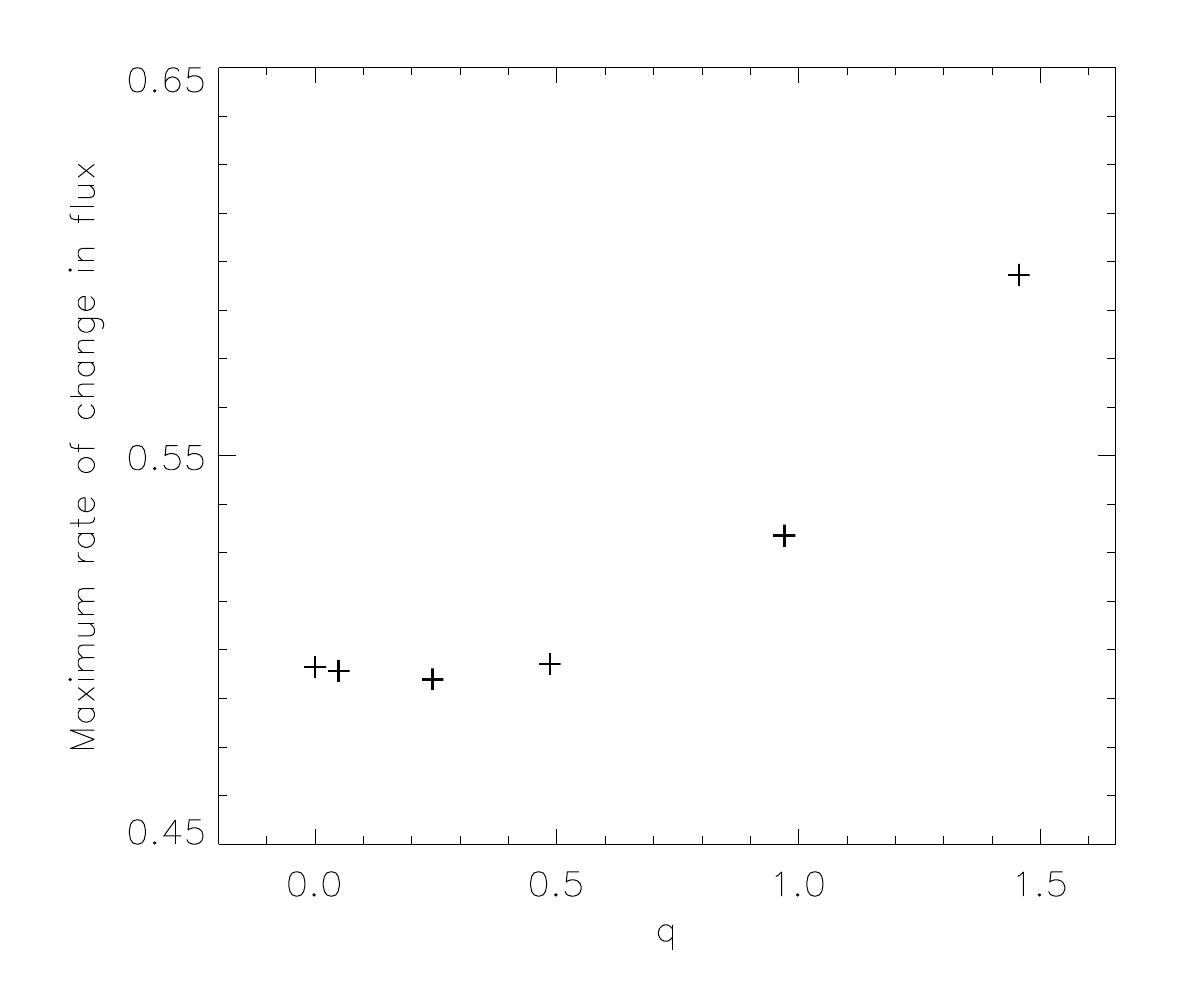}
\caption{(a) {Vorticity} flux through the dividing plane over time, and (b) rate of change of this flux (\tb{derivative with respect to $t^*$}), \tb{with line styles as in Figure \ref{fig:afnhvortflux}}. (c) Maximum rate of change of the {vorticity} flux as a function of \tb{$q$}.}
\label{fig:afzhvortflux}
\end{figure}

In Figure~\ref{fig:afzhvortflux} we plot the {vorticity flux through the dividing plane and its rate of change} as a measure of reconnection.
\tb{As for the simulations with non-zero net helicity we observe a general trend that for increasing $q$ the reconnection occurs later, and the total flux reconnected by the end of the simulation increases (at least up to $q=0.970$). }
We hypothesise that this later, `more complete' reconnection for larger \tb{$q$} could be due to the different shape of the vortex sheet formed and/or the change in angle of the null line at which the reconnection takes place.
\tb{For $q>0.485$ the peak reconnection rate is found to increase with $q$, however it appears to be almost independent of $q$ for $q<0.485$.}

\subsection{Global Topology}

\begin{figure}
\centering
(a)\includegraphics[width=0.45\textwidth]{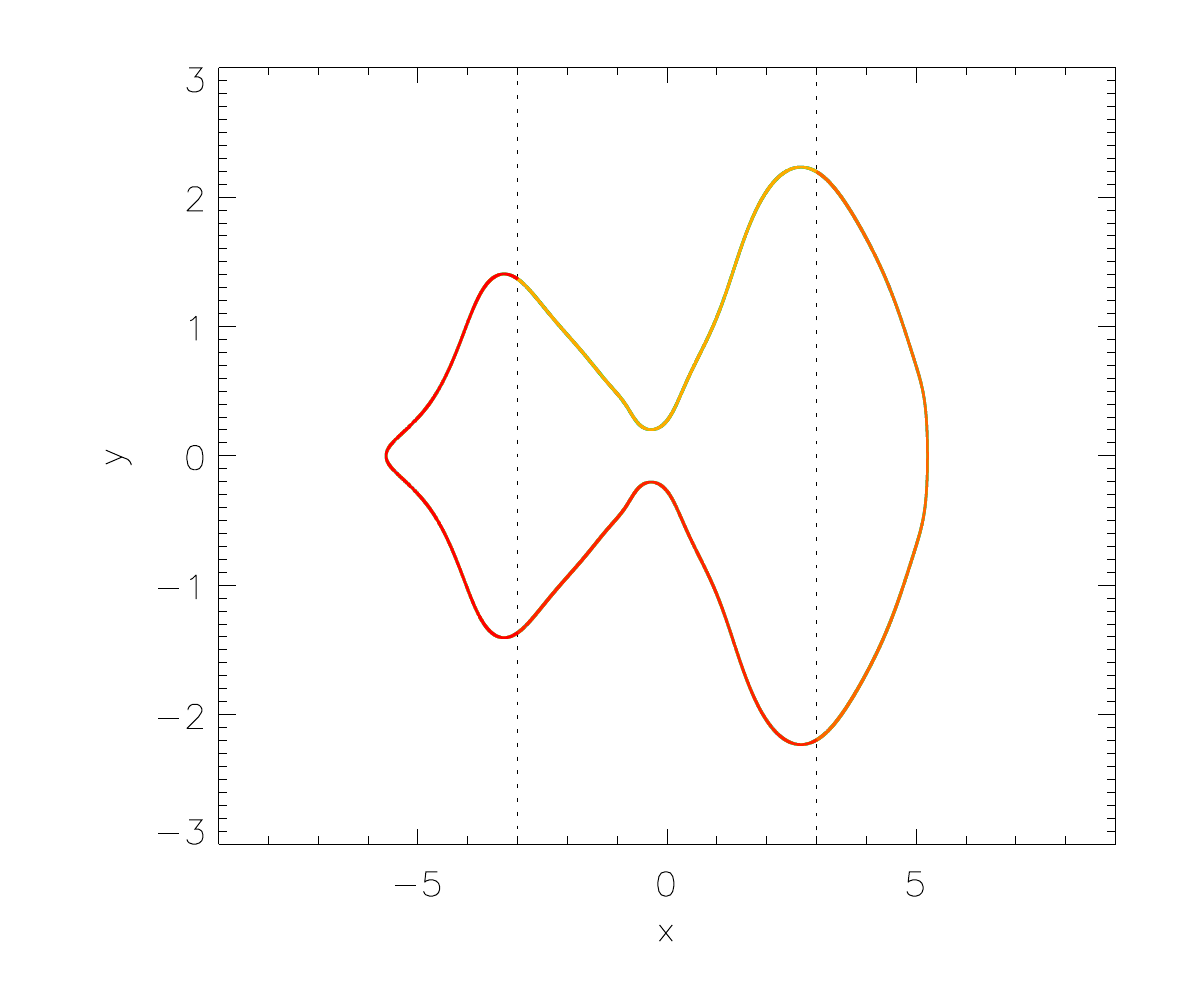}
(b)\includegraphics[width=0.45\textwidth]{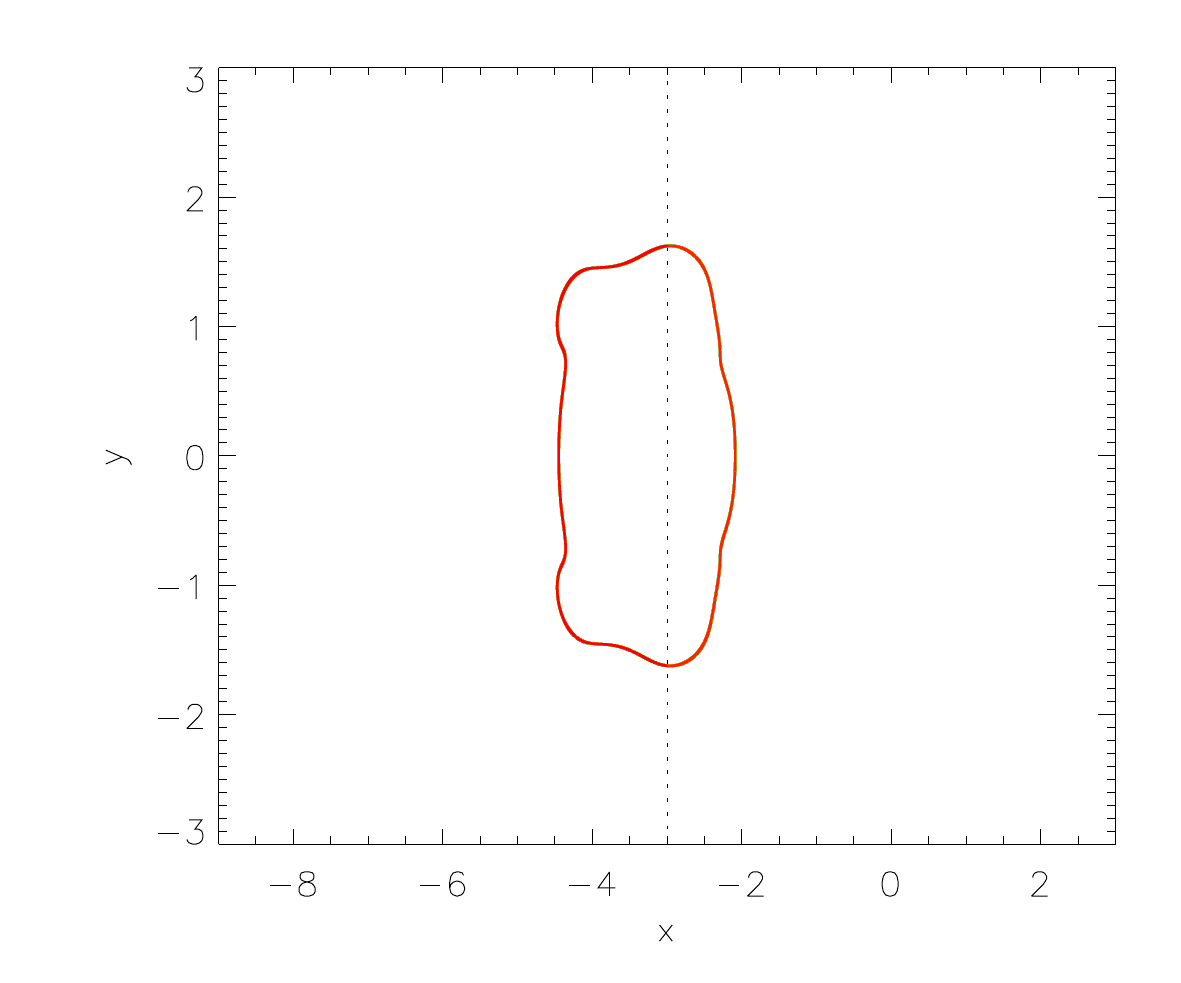}
\caption{Selected vortex lines plotted from the $30\%$ maximum vorticity contour at $x=-3$, at (a) \tb{$t^*=0.75$} and (b) \tb{$t^*=1.88$}. The change in colour indicates crossing a plane $x=3n, n\in \mathbb{Z}$.}
\label{fig:afzhtopology}
\end{figure}

As discussed in Section~\ref{subsec:afnhtopology} the topology of the system with axial flow after reconnection is not as simple as the  untwisted (\tb{$q=0$}) case. However, for the present simulations with zero net helicity, the symmetry with respect to the dividing plane means that any vortex ring formed after reconnection will always be symmetric, and thus have no net twist. Therefore the global topology is simpler than for the finite-net-helicity simulations. We plot some sample vortex lines  in Figure~\ref{fig:afzhtopology}: in (a) is an example of a thread that becomes a bridge and vice versa between adjacent periodic domains along $x$, leading to a ``period-2" vortex ring. In (b) is an example of a period-1 vortex ring, with the Kelvin waves clearly visible.

\section{Comparison of zero and non-zero net helicity simulations}\label{sec:comp}
\subsection{Flux Evolution}

\begin{figure}
\centering
\includegraphics[width=0.5\textwidth]{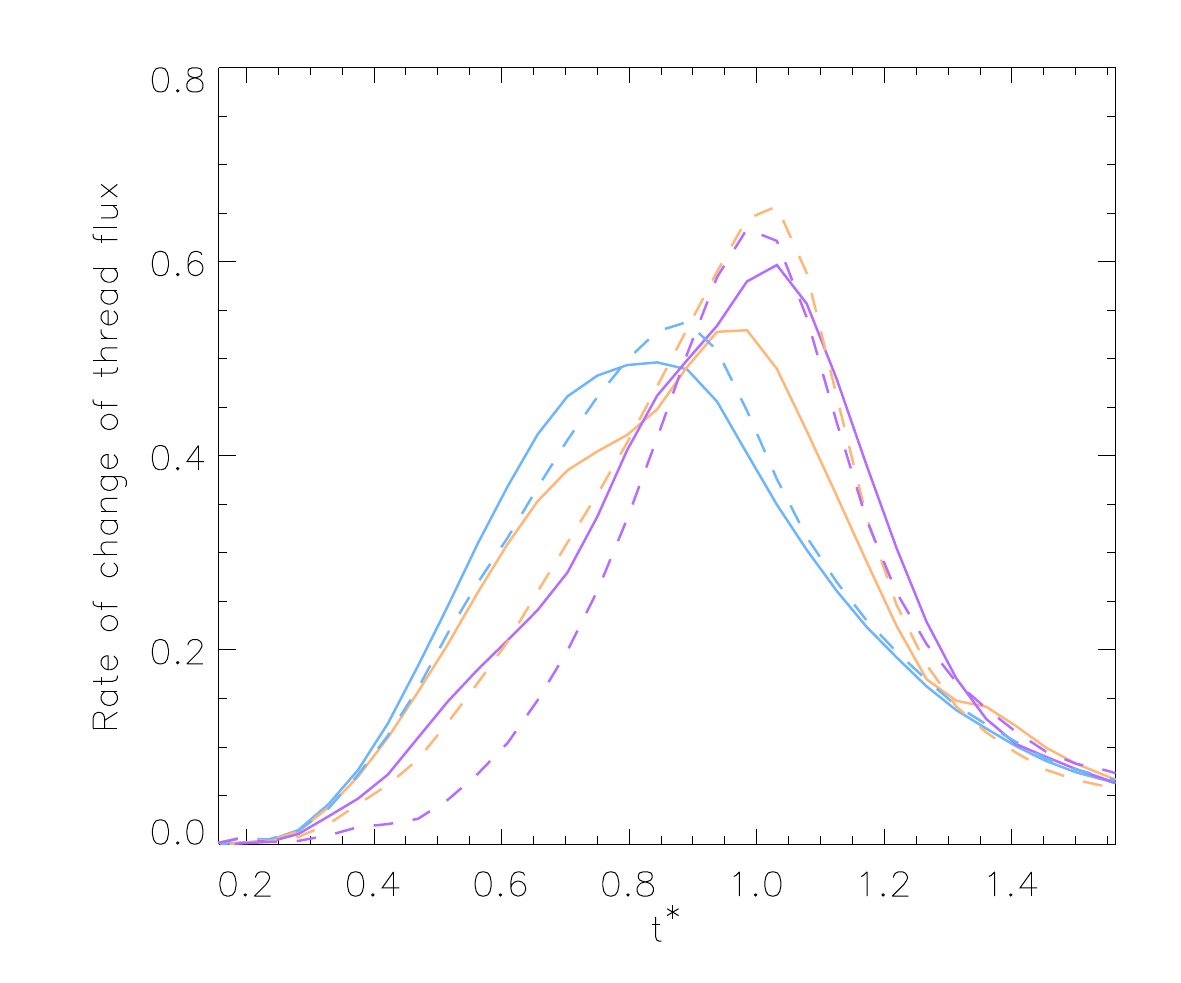}
\caption{Comparison of the reconnection rate over time for the zero helicity runs (solid) and non-zero helicity runs (dashed). Colours as in Figure \ref{fig:aftwistloss}.}
\label{fig:afbothflux}
\end{figure}
We now briefly make a direct comparison between the two different sets of simulations; those with non-zero net helicity and zero net helicity. We have shown in the previous sections that this change in axial flow has led to a topologically different reconnection process due to the change in symmetry. We consider first the rate of reconnection, plotted in Figure~\ref{fig:afbothflux} \tb{for $q\geq 0.485$}. The non-zero helicity runs are seen to exhibit higher maximum reconnection rates than the corresponding runs with zero helicity. Due to the movement of the perturbations along the tube this becomes difficult to compare at even higher twists with the \tb{$q=1.455$} non-zero helicity run being a clear outlier since the perturbations do not collide head-on, as described above.

\subsection{Volume Integrals}

\begin{figure}
\centering
(a)\includegraphics[width=0.46\textwidth]{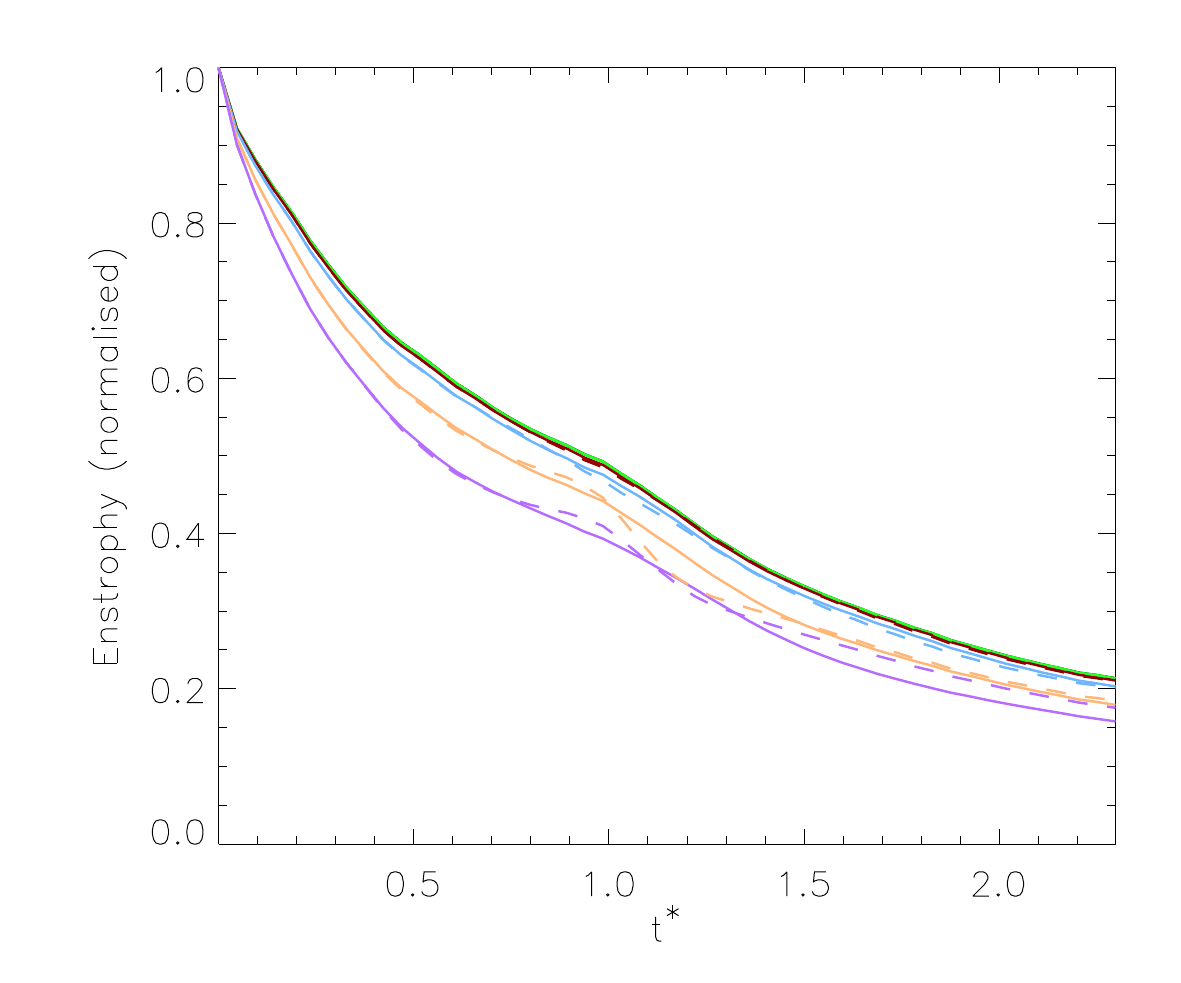}
(b)\includegraphics[width=0.46\textwidth]{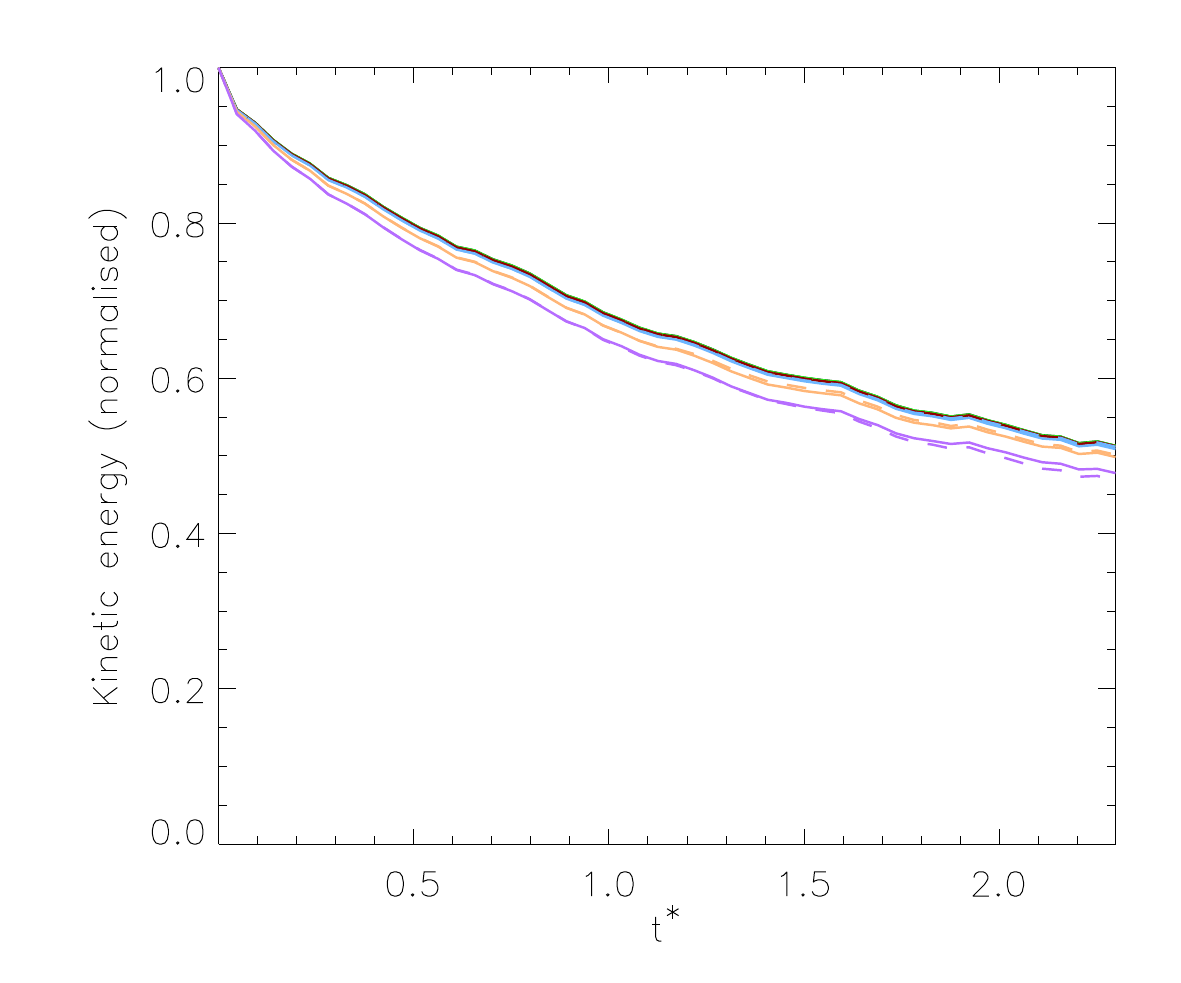}
(c)\includegraphics[width=0.46\textwidth]{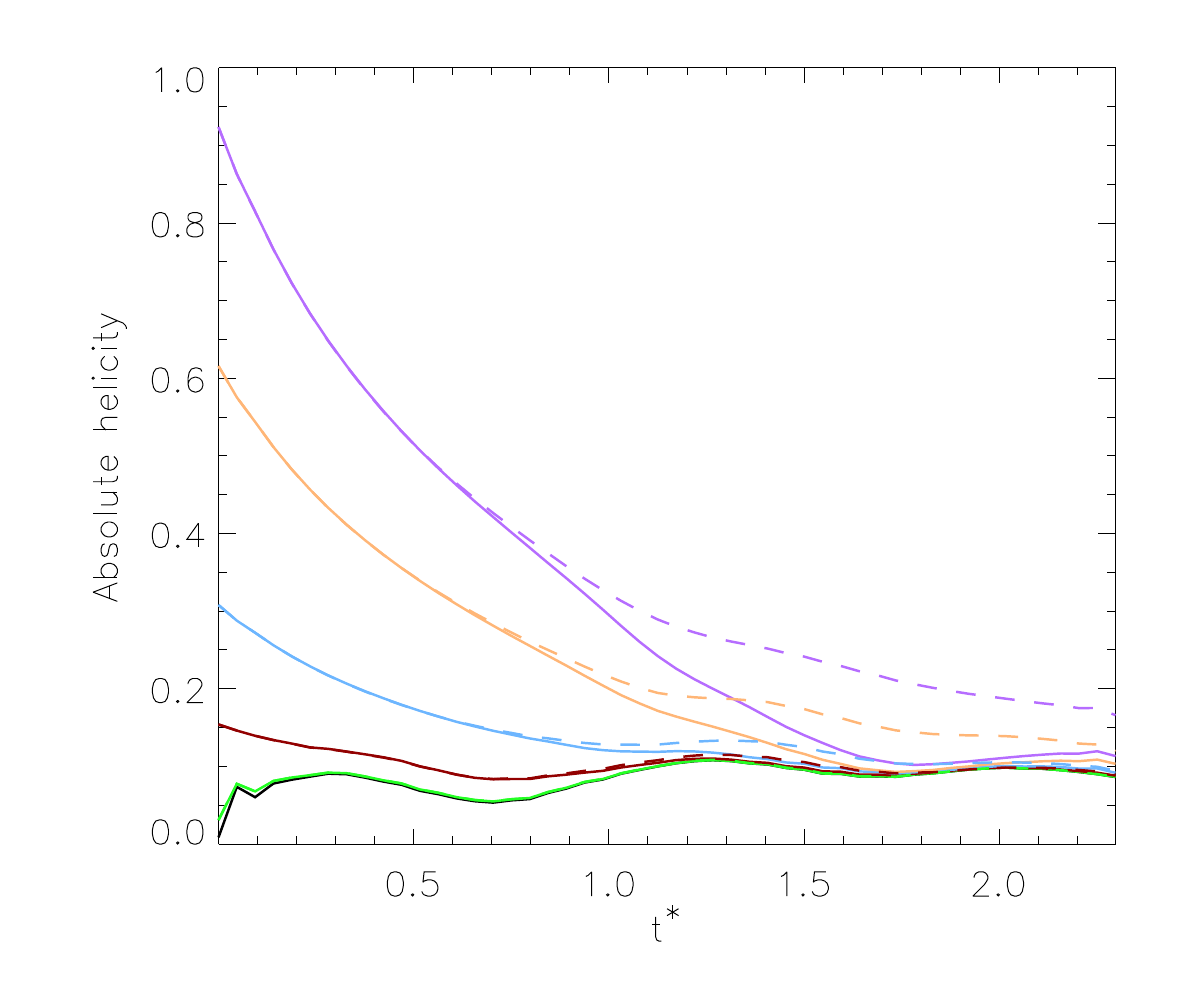}
\caption{Comparison of volume integrated quantities for the zero helicity runs (solid) and non-zero helicity runs (dashed). (a) Normalised enstrophy, (b) normalised kinetic energy, (c) net absolute helicity $|\vort\cdot\vel|$. Colours as in Figure \ref{fig:aftwistloss}.}
\label{fig:volint}
\end{figure}

We next consider the evolution of various volume-integrated quantities for all simulations. Plotting the enstrophy for all the simulations in Figure~\ref{fig:volint}(a) \tb{the differences} between the zero and non-zero helicity cases \tb{for $q<0.485$ are not visible on the scale of the overall time variation}, suggesting that there are only minor differences of the vortex sheet morphology between the two cases. However, for \tb{$q \geq 0.485$}, especially noticeable at \tb{$q=0.970$ and $q=1.455$}, we see a larger enstrophy `bump' in the non-zero helicity simulations at \tb{$t^*\approx1$}. We suggest that this may be due to the rotation of the vortex tubes leading to a longer vortex sheet in the non-zero net helicity case. From Figure~\ref{fig:volint}(b) we see that by contrast the change in twist has little effect on the evolution of the kinetic energy.



We plot the net absolute (or unsigned) helicity ($\int_V|{\bf v}\cdot\vort|dV$) during the simulations in Figure~\ref{fig:volint}(c) to help visualise the evolution of the twist once the vortex tubes reconnect.  We observe that the net absolute helicity diverges for the two different sets of simulations from $\tb{t^*\approx0.7}$ when the reconnection begins. All simulations retain significant helicity density in the thread field lines at late time. In the simulations with zero net helicity, the helicity (twist) tends to cancel since reconnecting field lines have opposite twist. However, as the bridges in the non-zero helicity runs have net twist, they preserve their helicity. The discrepancy between the two cases is more pronounced for higher values of \tb{$q$}.

\section{Conclusions}\label{sec:conc}

While the majority of studies of reconnecting vortex tubes have considered vortex tubes without twist, reconnection in the presence of an axial flow within the tube may be more relevant for understanding various physical phenomena. Interaction of helical vortex tubes is relevant for understanding processes in helical turbulent flows, as well as secondary reconnections in interactions of initially untwisted tubes (since the primary reconnection process acts to insert twist onto the bridges and threads).
We have shown here that the introduction of axial flow along the tubes (and thus twist of the vortex lines around the tube axes) significantly affects the reconnection process by breaking the symmetry, in different ways depending on whether the tubes are oppositely twisted (leading to a zero net helicity for the tube pair) or twisted in the same sense (finite net helicity).

In both cases the initial perturbations propagate along the tubes. Considering first the case with finite net helicity, the two perturbations travel in opposite directions along the tubes, meaning that they do not meet exactly `head-on'. When the tubes press against one another there is a non-zero component of $\vort$ along the reconnection line -- field lines reconnect at a finite angle rather than anti-parallel, this angle depending on the twist in the tubes. In the heart of the reconnection region pairs of vorticity null points may be formed. The generation of these null points is of interest given their relevance for the problem of finite-time blow-up \citep[e.g.][]{bhattacharjee1992,pelz2002}. We anticipate a much more complex topology, containing for example many more nulls, at higher Reynolds number where rapid shear-flow instabilities are expected \citep{2012PhFl...24g5105V}.

When the reconnecting tubes are oppositely twisted (giving zero net helicity) the applied perturbations move in the same direction as one another, and thus meet `head-on', but away from the original plane of symmetry. The two bridges that are formed by the reconnection process differ from one another -- for example in that one bridge exhibits much better-defined hairpin vortex lines \citep{1989PhFl....1..633M}. This stems from the fact that the reconnection is two-dimensional -- occurring along an X-type null line of $\vort$ -- and that this null line is not oriented perpendicular to the initial global direction of the tube axes (the relative angle depending on the twist). Unlike the untwisted case -- in which the reconnection line is perpendicular to the tube axes -- as the twist is increased the reconnection line tilts progressively further away from this perpendicular direction (this angle may also vary during the process if there is a variation of the local field line twisting angle across the tube radius).

For both helicity configurations we find a more complex post-reconnection topology than for the untwisted case, since field lines with non-integer winding numbers within the periodic domain do not connect symmetric points on the boundaries. The simpler case is that with zero net helicity, in which the twists of the vortex tubes cancel during reconnection such that the reconnected tubes are untwisted. The remaining twist of the thread vortex lines does however lead to the possibility of vortex rings closed over multiple periods. For reconnecting vortex tubes with net helicity, both the threads and bridges are twisted,  and multiple-period closed vortex lines are found, together with lines that stretch over as many periods as analysed while wrapping through multiple ring structures.
{It should be noted that substantial additional topological complexity is expected to be generated during the reconnection process at higher Reynolds numbers than those considered here, when for example numerous additional vortex rings may form, and the Kelvin-Helmholtz instability leads to filamentation of the vortex tubes during reconnection \citep[e.g.][]{2011PhFl...23b1701H,2012PhFl...24g5105V,mcgavin2018a}.}

\tb{For both zero and non-zero net helicity, the presence of axial flow/vortex line twist leads the reconnection process to occur later, and reconnect more flux overall.} Comparing the quantitative properties between the two configurations, we found 
\tb{that the} reconnection occurs later for the non-zero net helicity case, but reaches a higher maximum reconnection rate \citep[see also][]{2012PhFl...24g5105V}. Moreover, the \tb{peak} enstrophy is greater (perhaps due to a longer vortex sheet geometry), and the net absolute helicity retained after reconnection is greater (being zero in the bridges in the zero net helicity case due to cancellation).

Future work should include studying the interactions at higher Reynolds numbers, expected to lead to instabilities in the vortex sheet and the formation of additional vortex rings \citep{2012PhFl...24g5105V}. In Section \ref{subsec:afngtwist} we have analysed the loss of twist in the tubes prior to reconnection. However, in the future work it would be interesting to observe the difference in twist loss for the bridges and threads during  and after reconnection, though this is more challenging due to the absence of a well-defined axis. Furthermore, the post-reconnection topology could be analysed in further detail, to determine the relative locations of vortex rings of period one, and period greater than one (i.e.~vortex lines that close after traversing multiple periods of the domain). In order to simulate reconnection at higher twists, it will be important in future to offset the perturbations of the tubes initially (in $x$) such that they collide head-on.

\appendix
\section{}\label{appA}

In an inviscid barotropic fluid vortex lines are material lines, meaning that all fluid elements that initially lie on the same vortex line will remain connected by a vortex line at later time, since
\begin{equation}\label{eq:ideal}
\frac{D}{Dt}\bigg(\frac{\vort}{\rho}\bigg)=\bigg(\frac{\vort}{\rho}\cdot\nabla\bigg)\vel,
\end{equation}
which can be compared to the evolution equation of a material line element \citep{1978magnetic,1994AnRFM..26..169K}.
In a perfectly conducting plasma the magnetic field obeys an identical equation, where $\vort$ is replaced by the magnetic induction $\Bvec$. In each case a finite dissipation may lead to a breakdown in the `frozen-in' condition for the relevant vector field. The parallels between magnetic reconnection in plasmas and vortex reconnection are described in detail by \cite{2001LNP...571..373H}. 
Briefly, they can be understood by comparing the Navier-Stokes equation for a barotropic fluid in the form
\begin{equation} \label{eq:dvdt}
-\frac{\partial\vel}{\partial t}-\nabla\cubr{\tilde{p}+\frac{\vel^2}{2}-\frac{4}{3}\nu\nabla\cdot\vel}+\vel\times\vort=\nu\nabla\times\vort,
\end{equation}
where $\nabla\tilde{p}=(1/\rho)\nabla p$, with the following plasma equation obtained from Maxwell's equations and Ohm's law:
\begin{equation}
-\frac{\partial \Avec}{\partial t} - \nabla\phi +\vel\times \Bvec = \frac{1}{\mu_0\sigma}\nabla\times \Bvec,
\end{equation}
where $\Bvec=\nabla\times\Avec$, $\phi$ is a gauge potential for the electric field, and $\vel$ is the plasma velocity. The magnetic permeability $\mu_0$ and electrical conductivity $\sigma$ can be combined into the magnetic diffusivity, $\eta=1/(\mu_0\sigma)$.
Comparing these two equations leads to the following associations:
\begin{eqnarray}\label{eq:analog}
\Avec~\leftrightarrow~\vel&\qquad \Bvec~\leftrightarrow~\vort,\nonumber
\\ \phi~\leftrightarrow~\tilde{p}+\frac{\vel^2}{2}-\frac{4}{3}\nu(\nabla\cdot\vel)&\qquad \eta=\frac{1}{\mu\sigma}~\leftrightarrow~\nu.
\end{eqnarray} 

{We note that Equation (\ref{eq:dvdt}) is not identical to those solved in our simulations.  If a barotropic fluid is assumed then Equation (\ref{momentum}) is equivalent to Equations (\ref{eq:dvdt}), (\ref{mass}), up to vector identities. We have performed the simulations both using the full energy equation (\ref{energy}) and an adiabatic (barotropic) equation of state, and find both the qualitative and quantitative differences to be negligible. Here we chose to present the results of the simulations using the full system (\ref{momentum}--\ref{energy}).}



\begin{figure}
\begin{center}
(a)\includegraphics[width=0.46\textwidth]{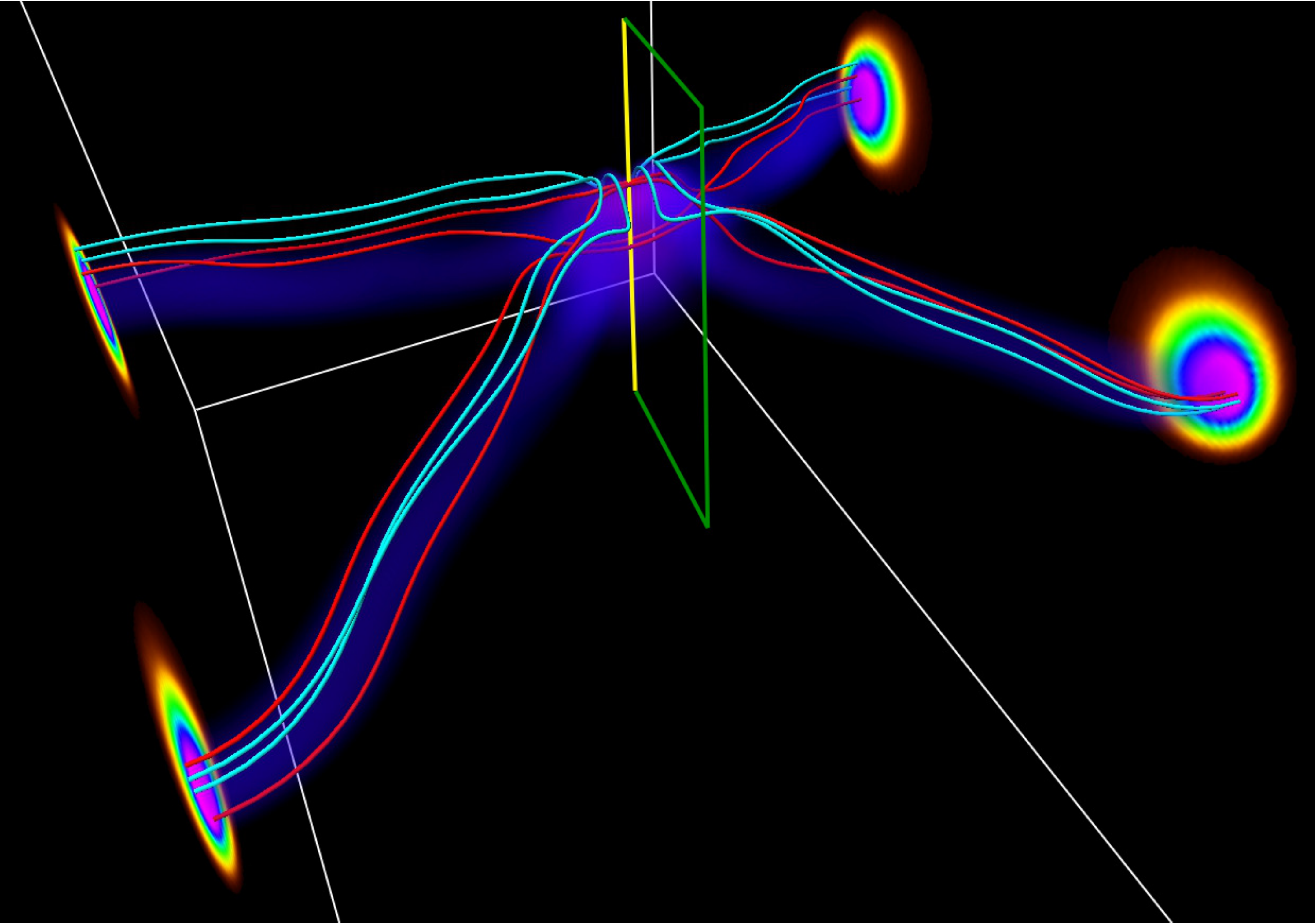}\\
(b)\includegraphics[width=0.45\textwidth]{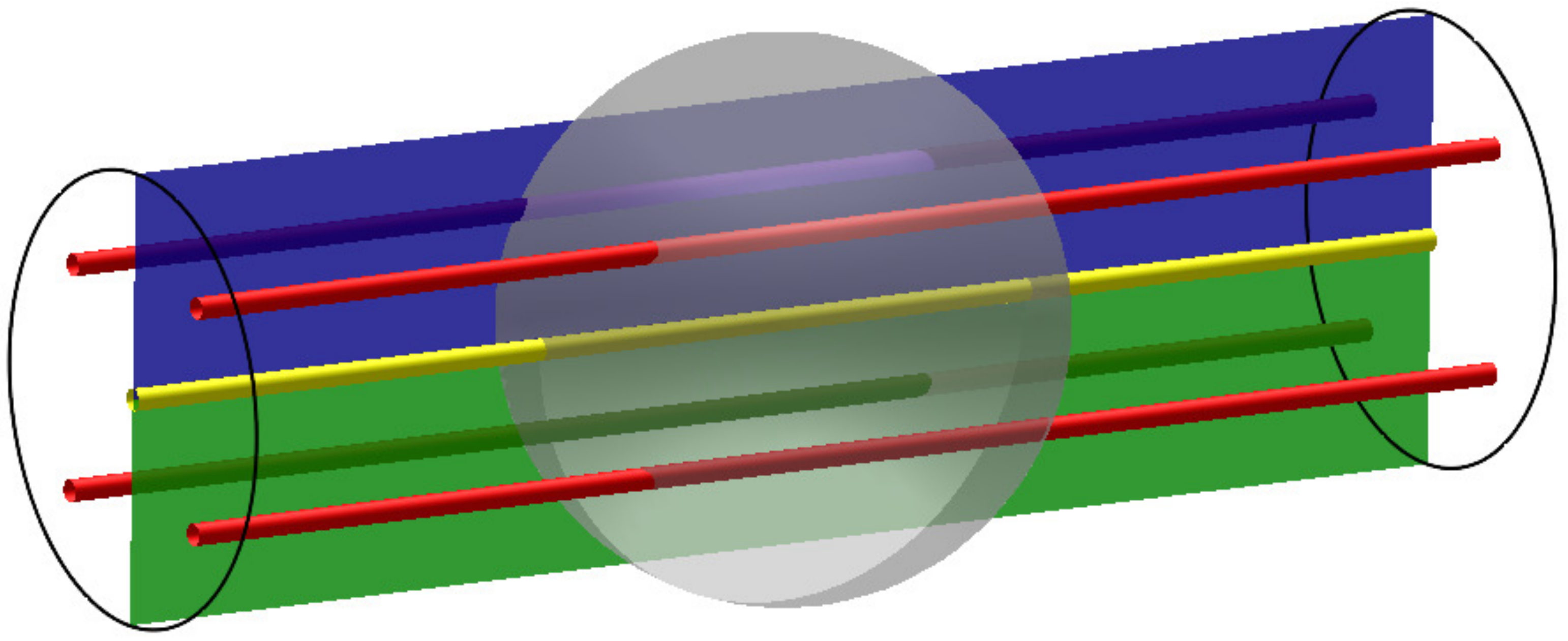}
(c)\includegraphics[width=0.45\textwidth]{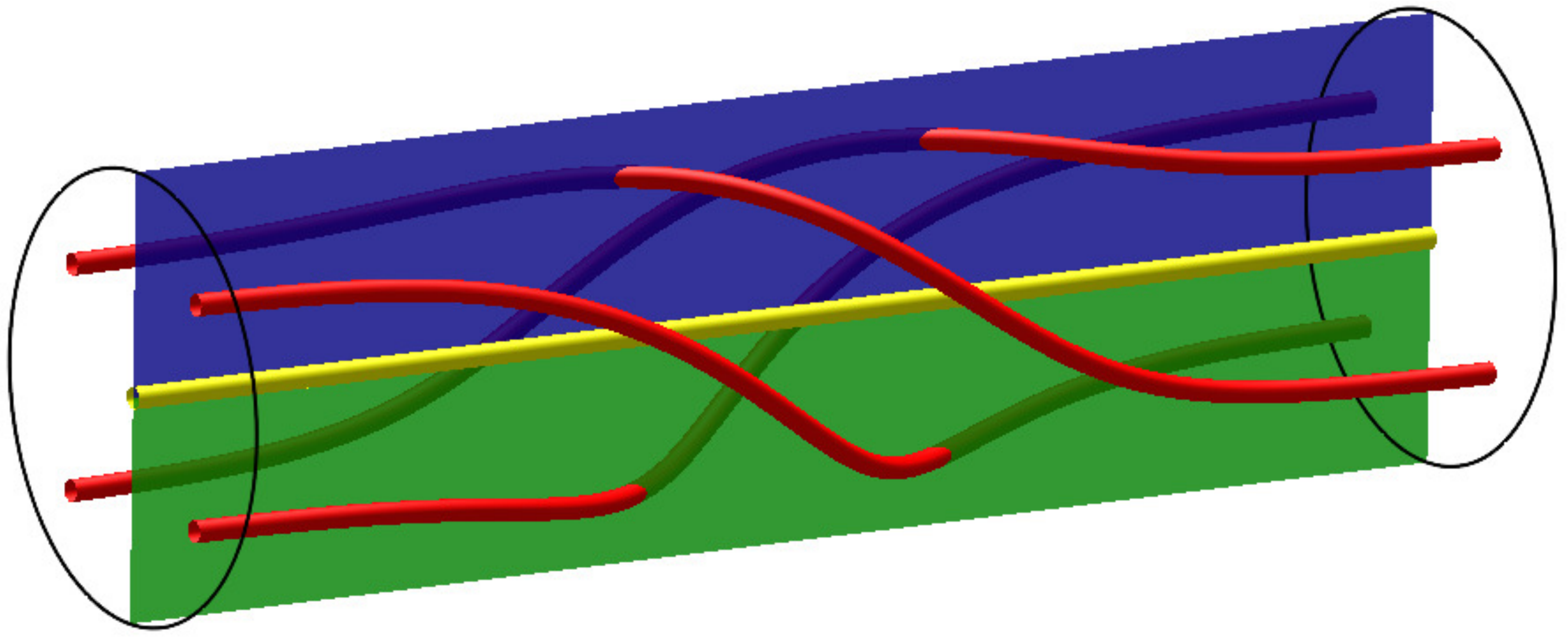}
\caption{(a) Selected vortex lines during the interaction of anti-parallel vortex tubes (red, threads; cyan, reconnected bridges). Shading on the end planes and in the volume shows $|\vort|$. Taken from the simulation described in \cite{mcgavin2018a}. Vorticity fieldlines (b) before and (c) after internal 3D reconnection. The shaded sphere in (b) represents the {\it non-ideal} region, within which $(\nabla\times\vort)\cdot\vort\neq 0$.}
\label{fig:intrectype3drec}
\end{center}
\end{figure}

Reconnection processes may be distinguished as either {\it 2D reconnection} -- in which the vortex lines lie locally in a plane -- and {\it 3D reconnection}. 
To understand 3D reconnection processes, the framework of {\it general magnetic reconnection} in a magnetised plasma was developed by \cite{1988JGR....93.5559H}. They demonstrated that this process requires the existence of a localised region within which the electric field ${\bf E}$ has a non-zero component parallel to the magnetic field, denoted $E_\|$. The rate of change of connectivity between plasma elements is then measured by finding the supremum of this quantity over all magnetic field lines passing through the region in which $\Evec\cdot\Bvec\neq0$:
\begin{equation}
\bigg(\int \Evec_\parallel dl\bigg)_{max}. 
\end{equation}
Noting that ${\bf E}=-\partial \Avec/\partial t-\nabla\phi$ and using the analogies drawn in Equation (\ref{eq:analog}), the 3D vortex reconnection rate is therefore given by
\begin{equation}\label{eqn:vortrecrate}
\bigg(\nu\int\nabla\times\vort \cdot {\bf dl}\bigg)_{max},
\end{equation}
the integral being performed with respect to arc length along vortex lines, and the supremum being taken over all vortex lines threading a localised reconnection region, i.e.~a region in which $(\nabla\times\vort)\cdot\vort\neq 0$.

The rate of change of flux during vortex tube reconnection in both the 2D and 3D cases can be understood within the above framework as follows. Consider the  rate of change of vorticity flux through the surface, $S$, whose boundary is the yellow-green loop {in Figure \ref{fig:intrectype3drec}(a)}. In the 2D case the yellow curve is taken to be the null line of the vorticity field ($\vort={\bf 0}$ along its length), while in the 3D case the yellow curve is a vortex line (i.e.~is locally tangent to $\vort$). In both cases $\vort={\bf 0}$ along the green curves. This flux change is given by
\begin{eqnarray}
\frac{\partial}{\partial t}\int_S\vort\cdot \nrm\diff S 
&=& \int_S[\nabla\times(\vel \times \vort) - \nu \nabla\times(\nabla\times\vort)]\cdot \nrm \diff  S \nonumber\\
&=& -\nu\oint_{\partial S}(\nabla\times\vort)\cdot\diff {\bf l}
\end{eqnarray}
where the final equality follows by applying Stokes' Theorem and noting that $\vort\times {\bf dl}={\bf 0}$ along the integration path. In the 3D case, the rate at which vorticity flux is converted from threads (red vortex lines in Figure \ref{fig:intrectype3drec}) to bridges (cyan) is then found by maximising this quantity over all possible choices of the yellow curve (vortex line).

The type of 3D `slipping' reconnection that occurs within each individual vortex tube is illustrated schematically in Figure~\ref{fig:intrectype3drec}(b, c). Consider a single vortex tube, within which exists a localised region of $(\nabla\times\vort)\cdot\vort\neq 0$ (marked as a grey sphere). The existence of this non-ideal region implies a rotational `slipping' of field lines that are integrated from either side of the non-ideal region \citep{1988JGR....93.5559H,hornig2003,2005ApJ...631.1227H}. By a similar argument to above the reconnection acts to change the flux through the green and blue surfaces in the figure -- despite there being no relative rotation between the two end planes -- since $\int\nabla\times\vort \cdot {\bf dl}\neq 0$ along the (yellow) axis. A key feature of all 3D reconnection is that field line connectivity change no longer occurs at a single point or line, but throughout the region in which $(\nabla\times\vort)\cdot\vort\neq 0$ \citep{2003JGRA..108.1285P}.  

\begin{acknowledgments}
The authors gratefully acknowledge financial support from the Leverhulme Trust and EPSRC. The authors are grateful for the use of the computing cluster (Magneto) of the School of Science and Engineering.
\end{acknowledgments}


\begin{thebibliography}{38}%
\makeatletter
\providecommand \@ifxundefined [1]{%
 \@ifx{#1\undefined}
}%
\providecommand \@ifnum [1]{%
 \ifnum #1\expandafter \@firstoftwo
 \else \expandafter \@secondoftwo
 \fi
}%
\providecommand \@ifx [1]{%
 \ifx #1\expandafter \@firstoftwo
 \else \expandafter \@secondoftwo
 \fi
}%
\providecommand \natexlab [1]{#1}%
\providecommand \enquote  [1]{``#1''}%
\providecommand \bibnamefont  [1]{#1}%
\providecommand \bibfnamefont [1]{#1}%
\providecommand \citenamefont [1]{#1}%
\providecommand \href@noop [0]{\@secondoftwo}%
\providecommand \href [0]{\begingroup \@sanitize@url \@href}%
\providecommand \@href[1]{\@@startlink{#1}\@@href}%
\providecommand \@@href[1]{\endgroup#1\@@endlink}%
\providecommand \@sanitize@url [0]{\catcode `\\12\catcode `\$12\catcode
  `\&12\catcode `\#12\catcode `\^12\catcode `\_12\catcode `\%12\relax}%
\providecommand \@@startlink[1]{}%
\providecommand \@@endlink[0]{}%
\providecommand \url  [0]{\begingroup\@sanitize@url \@url }%
\providecommand \@url [1]{\endgroup\@href {#1}{\urlprefix }}%
\providecommand \urlprefix  [0]{URL }%
\providecommand \Eprint [0]{\href }%
\providecommand \doibase [0]{http://dx.doi.org/}%
\providecommand \selectlanguage [0]{\@gobble}%
\providecommand \bibinfo  [0]{\@secondoftwo}%
\providecommand \bibfield  [0]{\@secondoftwo}%
\providecommand \translation [1]{[#1]}%
\providecommand \BibitemOpen [0]{}%
\providecommand \bibitemStop [0]{}%
\providecommand \bibitemNoStop [0]{.\EOS\space}%
\providecommand \EOS [0]{\spacefactor3000\relax}%
\providecommand \BibitemShut  [1]{\csname bibitem#1\endcsname}%
\let\auto@bib@innerbib\@empty
\bibitem [{\citenamefont {{Crow}}(1970)}]{1970AIAAJ...8.2172C}%
  \BibitemOpen
  \bibfield  {author} {\bibinfo {author} {\bibfnamefont {S.~C.}\ \bibnamefont
  {{Crow}}},\ }\bibfield  {title} {\enquote {\bibinfo {title} {{Stability
  theory for a pair of trailing vortices}},}\ }\href {\doibase 10.2514/3.6083}
  {\bibfield  {journal} {\bibinfo  {journal} {AIAA Journal}\ }\textbf {\bibinfo
  {volume} {8}},\ \bibinfo {pages} {2172--2179} (\bibinfo {year}
  {1970})}\BibitemShut {NoStop}%
\bibitem [{\citenamefont {{Hussain}}(1983)}]{1983PhFl...26.2816H}%
  \BibitemOpen
  \bibfield  {author} {\bibinfo {author} {\bibfnamefont {A.~K.~M.~F.}\
  \bibnamefont {{Hussain}}},\ }\bibfield  {title} {\enquote {\bibinfo {title}
  {{Coherent structures - Reality and myth}},}\ }\href {\doibase
  10.1063/1.864048} {\bibfield  {journal} {\bibinfo  {journal} {Physics of
  Fluids}\ }\textbf {\bibinfo {volume} {26}},\ \bibinfo {pages} {2816--2850}
  (\bibinfo {year} {1983})}\BibitemShut {NoStop}%
\bibitem [{\citenamefont {{Hussain}}\ and\ \citenamefont
  {{Duraisamy}}(2011)}]{2011PhFl...23b1701H}%
  \BibitemOpen
  \bibfield  {author} {\bibinfo {author} {\bibfnamefont {F.}~\bibnamefont
  {{Hussain}}}\ and\ \bibinfo {author} {\bibfnamefont {K.}~\bibnamefont
  {{Duraisamy}}},\ }\bibfield  {title} {\enquote {\bibinfo {title} {{Mechanics
  of viscous vortex reconnection}},}\ }\href {\doibase 10.1063/1.3532039}
  {\bibfield  {journal} {\bibinfo  {journal} {Physics of Fluids}\ }\textbf
  {\bibinfo {volume} {23}},\ \bibinfo {pages} {021701--021701} (\bibinfo {year}
  {2011})}\BibitemShut {NoStop}%
\bibitem [{\citenamefont {Saffman}(1993)}]{saffman1993}%
  \BibitemOpen
  \bibfield  {author} {\bibinfo {author} {\bibfnamefont {P.~G.}\ \bibnamefont
  {Saffman}},\ }\href {\doibase 10.1017/CBO9780511624063} {\emph {\bibinfo
  {title} {Vortex Dynamics:}}}\ (\bibinfo  {publisher} {Cambridge University
  Press},\ \bibinfo {address} {Cambridge},\ \bibinfo {year} {1993})\BibitemShut
  {NoStop}%
\bibitem [{\citenamefont {{Pumir}}\ and\ \citenamefont
  {{Kerr}}(1987)}]{1987PhRvL..58.1636P}%
  \BibitemOpen
  \bibfield  {author} {\bibinfo {author} {\bibfnamefont {A.}~\bibnamefont
  {{Pumir}}}\ and\ \bibinfo {author} {\bibfnamefont {R.~M.}\ \bibnamefont
  {{Kerr}}},\ }\bibfield  {title} {\enquote {\bibinfo {title} {{Numerical
  simulation of interacting vortex tubes}},}\ }\href {\doibase
  10.1103/PhysRevLett.58.1636} {\bibfield  {journal} {\bibinfo  {journal}
  {Physical Review Letters}\ }\textbf {\bibinfo {volume} {58}},\ \bibinfo
  {pages} {1636--1639} (\bibinfo {year} {1987})}\BibitemShut {NoStop}%
\bibitem [{\citenamefont {{Melander}}\ and\ \citenamefont
  {{Hussain}}(1989{\natexlab{a}})}]{1989PhFl....1..633M}%
  \BibitemOpen
  \bibfield  {author} {\bibinfo {author} {\bibfnamefont {M.~V.}\ \bibnamefont
  {{Melander}}}\ and\ \bibinfo {author} {\bibfnamefont {F.}~\bibnamefont
  {{Hussain}}},\ }\bibfield  {title} {\enquote {\bibinfo {title}
  {{Cross-linking of two antiparallel vortex tubes}},}\ }\href {\doibase
  10.1063/1.857437} {\bibfield  {journal} {\bibinfo  {journal} {Physics of
  Fluids}\ }\textbf {\bibinfo {volume} {1}},\ \bibinfo {pages} {633--636}
  (\bibinfo {year} {1989}{\natexlab{a}})}\BibitemShut {NoStop}%
\bibitem [{\citenamefont {{Melander}}\ and\ \citenamefont
  {{Hussain}}(1989{\natexlab{b}})}]{iutam1989melander}%
  \BibitemOpen
  \bibfield  {author} {\bibinfo {author} {\bibfnamefont {M.~V.}\ \bibnamefont
  {{Melander}}}\ and\ \bibinfo {author} {\bibfnamefont {F.}~\bibnamefont
  {{Hussain}}},\ }\bibfield  {title} {\enquote {\bibinfo {title} {{Topological
  Aspects of Vortex Reconnection}},}\ }in\ \href@noop {} {\emph {\bibinfo
  {booktitle} {Topological Fluid Mechanics: Proceedings of the IUTAM
  Symposium}}},\ \bibinfo {editor} {edited by\ \bibinfo {editor} {\bibfnamefont
  {H.K.}\ \bibnamefont {Moffatt}}\ and\ \bibinfo {editor} {\bibfnamefont
  {A.}~\bibnamefont {Tsinober}}}\ (\bibinfo {year} {1989})\ pp.\ \bibinfo
  {pages} {485--499}\BibitemShut {NoStop}%
\bibitem [{\citenamefont {{van Rees}}\ \emph {et~al.}(2012)\citenamefont {{van
  Rees}}, \citenamefont {{Hussain}},\ and\ \citenamefont
  {{Koumoutsakos}}}]{2012PhFl...24g5105V}%
  \BibitemOpen
  \bibfield  {author} {\bibinfo {author} {\bibfnamefont {W.~M.}\ \bibnamefont
  {{van Rees}}}, \bibinfo {author} {\bibfnamefont {F.}~\bibnamefont
  {{Hussain}}}, \ and\ \bibinfo {author} {\bibfnamefont {P.}~\bibnamefont
  {{Koumoutsakos}}},\ }\bibfield  {title} {\enquote {\bibinfo {title} {{Vortex
  tube reconnection at Re = 10$^{4}$}},}\ }\href {\doibase 10.1063/1.4731809}
  {\bibfield  {journal} {\bibinfo  {journal} {Physics of Fluids}\ }\textbf
  {\bibinfo {volume} {24}},\ \bibinfo {pages} {075105--075105} (\bibinfo {year}
  {2012})}\BibitemShut {NoStop}%
\bibitem [{\citenamefont {{Ashurst}}\ and\ \citenamefont
  {{Meiron}}(1987)}]{1987PhRvL..58.1632A}%
  \BibitemOpen
  \bibfield  {author} {\bibinfo {author} {\bibfnamefont {W.~T.}\ \bibnamefont
  {{Ashurst}}}\ and\ \bibinfo {author} {\bibfnamefont {D.~I.}\ \bibnamefont
  {{Meiron}}},\ }\bibfield  {title} {\enquote {\bibinfo {title} {{Numerical
  study of vortex reconnection}},}\ }\href {\doibase
  10.1103/PhysRevLett.58.1632} {\bibfield  {journal} {\bibinfo  {journal}
  {Physical Review Letters}\ }\textbf {\bibinfo {volume} {58}},\ \bibinfo
  {pages} {1632--1635} (\bibinfo {year} {1987})}\BibitemShut {NoStop}%
\bibitem [{\citenamefont {Kida}\ \emph {et~al.}(1991)\citenamefont {Kida},
  \citenamefont {Takaoka},\ and\ \citenamefont
  {Hussain}}]{1991JFM...230..583K}%
  \BibitemOpen
  \bibfield  {author} {\bibinfo {author} {\bibfnamefont {S}~\bibnamefont
  {Kida}}, \bibinfo {author} {\bibfnamefont {M}~\bibnamefont {Takaoka}}, \ and\
  \bibinfo {author} {\bibfnamefont {F}~\bibnamefont {Hussain}},\ }\bibfield
  {title} {\enquote {\bibinfo {title} {{Collision of two vortex rings}},}\
  }\href@noop {} {\bibfield  {journal} {\bibinfo  {journal} {Journal of Fluid
  Mechanics (ISSN 0022-1120)}\ }\textbf {\bibinfo {volume} {230}},\ \bibinfo
  {pages} {583--646} (\bibinfo {year} {1991})}\BibitemShut {NoStop}%
\bibitem [{\citenamefont {Kida}\ and\ \citenamefont
  {Takaoka}(1987)}]{kida1987}%
  \BibitemOpen
  \bibfield  {author} {\bibinfo {author} {\bibfnamefont {S}~\bibnamefont
  {Kida}}\ and\ \bibinfo {author} {\bibfnamefont {M}~\bibnamefont {Takaoka}},\
  }\bibfield  {title} {\enquote {\bibinfo {title} {{Bridging in vortex
  reconnection}},}\ }\href@noop {} {\bibfield  {journal} {\bibinfo  {journal}
  {Phys Fluids}\ }\textbf {\bibinfo {volume} {30}},\ \bibinfo {pages} {2911--5}
  (\bibinfo {year} {1987})}\BibitemShut {NoStop}%
\bibitem [{\citenamefont {{Kleckner}}\ and\ \citenamefont
  {{Irvine}}(2013)}]{2013NatPh...9..253K}%
  \BibitemOpen
  \bibfield  {author} {\bibinfo {author} {\bibfnamefont {D.}~\bibnamefont
  {{Kleckner}}}\ and\ \bibinfo {author} {\bibfnamefont {W.~T.~M.}\ \bibnamefont
  {{Irvine}}},\ }\bibfield  {title} {\enquote {\bibinfo {title} {{Creation and
  dynamics of knotted vortices}},}\ }\href {\doibase 10.1038/nphys2560}
  {\bibfield  {journal} {\bibinfo  {journal} {Nature Physics}\ }\textbf
  {\bibinfo {volume} {9}},\ \bibinfo {pages} {253--258} (\bibinfo {year}
  {2013})}\BibitemShut {NoStop}%
\bibitem [{\citenamefont {Moore}\ and\ \citenamefont
  {Saffman}(1973)}]{Moore491}%
  \BibitemOpen
  \bibfield  {author} {\bibinfo {author} {\bibfnamefont {D.~W.}\ \bibnamefont
  {Moore}}\ and\ \bibinfo {author} {\bibfnamefont {P.~G.}\ \bibnamefont
  {Saffman}},\ }\bibfield  {title} {\enquote {\bibinfo {title} {Axial flow in
  laminar trailing vortices},}\ }\href {\doibase 10.1098/rspa.1973.0075}
  {\bibfield  {journal} {\bibinfo  {journal} {Proceedings of the Royal Society
  of London A: Mathematical, Physical and Engineering Sciences}\ }\textbf
  {\bibinfo {volume} {333}},\ \bibinfo {pages} {491--508} (\bibinfo {year}
  {1973})},\ \Eprint
  {http://arxiv.org/abs/http://rspa.royalsocietypublishing.org/content/333/1595/491.full.pdf}
  {http://rspa.royalsocietypublishing.org/content/333/1595/491.full.pdf}
  \BibitemShut {NoStop}%
\bibitem [{\citenamefont {{Feys}}\ and\ \citenamefont
  {{Maslowe}}(2016)}]{2016JFM...803..556F}%
  \BibitemOpen
  \bibfield  {author} {\bibinfo {author} {\bibfnamefont {J.}~\bibnamefont
  {{Feys}}}\ and\ \bibinfo {author} {\bibfnamefont {S.~A.}\ \bibnamefont
  {{Maslowe}}},\ }\bibfield  {title} {\enquote {\bibinfo {title} {{Elliptical
  instability of the Moore-Saffman model for a trailing wingtip vortex}},}\
  }\href {\doibase 10.1017/jfm.2016.512} {\bibfield  {journal} {\bibinfo
  {journal} {Journal of Fluid Mechanics}\ }\textbf {\bibinfo {volume} {803}},\
  \bibinfo {pages} {556--590} (\bibinfo {year} {2016})}\BibitemShut {NoStop}%
\bibitem [{\citenamefont {Kida}\ and\ \citenamefont
  {Takaoka}(1991)}]{kida1991}%
  \BibitemOpen
  \bibfield  {author} {\bibinfo {author} {\bibfnamefont {Shigeo}\ \bibnamefont
  {Kida}}\ and\ \bibinfo {author} {\bibfnamefont {Masanori}\ \bibnamefont
  {Takaoka}},\ }\bibfield  {title} {\enquote {\bibinfo {title} {{Breakdown of
  Frozen Motion of Vorticity Fieldand Vorticity Reconnection}},}\ }\href@noop
  {} {\bibfield  {journal} {\bibinfo  {journal} {Journal of the Physical
  Society of Japan}\ }\textbf {\bibinfo {volume} {60}},\ \bibinfo {pages}
  {2184--2196} (\bibinfo {year} {1991})}\BibitemShut {NoStop}%
\bibitem [{\citenamefont {Takaoka}(1996)}]{takaoka1996}%
  \BibitemOpen
  \bibfield  {author} {\bibinfo {author} {\bibfnamefont {M}~\bibnamefont
  {Takaoka}},\ }\bibfield  {title} {\enquote {\bibinfo {title} {{Helicity
  generation and vorticity dynamics in helically symmetric flow}},}\
  }\href@noop {} {\bibfield  {journal} {\bibinfo  {journal} {Journal of Fluid
  Mechanics}\ }\textbf {\bibinfo {volume} {319}},\ \bibinfo {pages} {125--149}
  (\bibinfo {year} {1996})}\BibitemShut {NoStop}%
\bibitem [{\citenamefont {{Schindler}}\ \emph {et~al.}(1988)\citenamefont
  {{Schindler}}, \citenamefont {{Hesse}},\ and\ \citenamefont
  {{Birn}}}]{1988JGR....93.5547S}%
  \BibitemOpen
  \bibfield  {author} {\bibinfo {author} {\bibfnamefont {K.}~\bibnamefont
  {{Schindler}}}, \bibinfo {author} {\bibfnamefont {M.}~\bibnamefont
  {{Hesse}}}, \ and\ \bibinfo {author} {\bibfnamefont {J.}~\bibnamefont
  {{Birn}}},\ }\bibfield  {title} {\enquote {\bibinfo {title} {{General
  magnetic reconnection, parallel electric fields, and helicity}},}\ }\href
  {\doibase 10.1029/JA093iA06p05547} {\bibfield  {journal} {\bibinfo  {journal}
  {Journal of Geophysics Research}\ }\textbf {\bibinfo {volume} {93}},\
  \bibinfo {pages} {5547--5557} (\bibinfo {year} {1988})}\BibitemShut {NoStop}%
\bibitem [{\citenamefont {Hornig}\ and\ \citenamefont
  {Schindler}(1996)}]{hornig1996}%
  \BibitemOpen
  \bibfield  {author} {\bibinfo {author} {\bibfnamefont {G.}~\bibnamefont
  {Hornig}}\ and\ \bibinfo {author} {\bibfnamefont {K.}~\bibnamefont
  {Schindler}},\ }\bibfield  {title} {\enquote {\bibinfo {title} {Magnetic
  topology and the problem of its invariant definition},}\ }\href@noop {}
  {\bibfield  {journal} {\bibinfo  {journal} {Phys. Plasmas}\ }\textbf
  {\bibinfo {volume} {3}},\ \bibinfo {pages} {781--791} (\bibinfo {year}
  {1996})}\BibitemShut {NoStop}%
\bibitem [{\citenamefont {{McGavin}}\ and\ \citenamefont
  {{Pontin}}(2018)}]{mcgavin2018a}%
  \BibitemOpen
  \bibfield  {author} {\bibinfo {author} {\bibfnamefont {P.}~\bibnamefont
  {{McGavin}}}\ and\ \bibinfo {author} {\bibfnamefont {D.~I.}\ \bibnamefont
  {{Pontin}}},\ }\bibfield  {title} {\enquote {\bibinfo {title} {{Vortex line
  topology during vortex tube reconnection}},}\ }\href {\doibase
  10.1103/PhysRevFluids.3.054701} {\bibfield  {journal} {\bibinfo  {journal}
  {Physical Review Fluids}\ }\textbf {\bibinfo {volume} {3}},\ \bibinfo {eid}
  {054701} (\bibinfo {year} {2018})},\ \Eprint
  {http://arxiv.org/abs/1805.01202} {arXiv:1805.01202 [physics.flu-dyn]}
  \BibitemShut {NoStop}%
\bibitem [{\citenamefont {{Greene}}(1993)}]{1993PhFlB...5.2355G}%
  \BibitemOpen
  \bibfield  {author} {\bibinfo {author} {\bibfnamefont {J.~M.}\ \bibnamefont
  {{Greene}}},\ }\bibfield  {title} {\enquote {\bibinfo {title} {{Reconnection
  of vorticity lines and magnetic lines}},}\ }\href {\doibase 10.1063/1.860718}
  {\bibfield  {journal} {\bibinfo  {journal} {Physics of Fluids B}\ }\textbf
  {\bibinfo {volume} {5}},\ \bibinfo {pages} {2355--2362} (\bibinfo {year}
  {1993})}\BibitemShut {NoStop}%
\bibitem [{\citenamefont {{Hornig}}(2001)}]{2001LNP...571..373H}%
  \BibitemOpen
  \bibfield  {author} {\bibinfo {author} {\bibfnamefont {G.}~\bibnamefont
  {{Hornig}}},\ }\bibfield  {title} {\enquote {\bibinfo {title} {{The Geometry
  of Magnetic and Vortex Reconnection}},}\ }in\ \href@noop {} {\emph {\bibinfo
  {booktitle} {Quantized Vortex Dynamics and Superfluid Turbulence}}},\
  \bibinfo {series} {Lecture Notes in Physics, Berlin Springer Verlag}, Vol.\
  \bibinfo {volume} {571},\ \bibinfo {editor} {edited by\ \bibinfo {editor}
  {\bibfnamefont {C.~F.}\ \bibnamefont {{Barenghi}}}, \bibinfo {editor}
  {\bibfnamefont {R.~J.}\ \bibnamefont {{Donnelly}}}, \ and\ \bibinfo {editor}
  {\bibfnamefont {W.~F.}\ \bibnamefont {{Vinen}}}}\ (\bibinfo {year} {2001})\
  p.\ \bibinfo {pages} {373}\BibitemShut {NoStop}%
\bibitem [{\citenamefont {Frankel}(2004)}]{frankel2004geometry}%
  \BibitemOpen
  \bibfield  {author} {\bibinfo {author} {\bibfnamefont {T.}~\bibnamefont
  {Frankel}},\ }\href {https://books.google.co.uk/books?id=2iq2EGNgX24C} {\emph
  {\bibinfo {title} {The Geometry of Physics: An Introduction}}}\ (\bibinfo
  {publisher} {Cambridge University Press},\ \bibinfo {year}
  {2004})\BibitemShut {NoStop}%
\bibitem [{\citenamefont {{Bustamante}}\ and\ \citenamefont
  {{Kerr}}(2008)}]{2008PhyD..237.1912B}%
  \BibitemOpen
  \bibfield  {author} {\bibinfo {author} {\bibfnamefont {M.~D.}\ \bibnamefont
  {{Bustamante}}}\ and\ \bibinfo {author} {\bibfnamefont {R.~M.}\ \bibnamefont
  {{Kerr}}},\ }\bibfield  {title} {\enquote {\bibinfo {title} {{3D Euler about
  a 2D symmetry plane}},}\ }\href {\doibase 10.1016/j.physd.2008.02.007}
  {\bibfield  {journal} {\bibinfo  {journal} {Physica D Nonlinear Phenomena}\
  }\textbf {\bibinfo {volume} {237}},\ \bibinfo {pages} {1912--1920} (\bibinfo
  {year} {2008})},\ \Eprint {http://arxiv.org/abs/0802.3369} {arXiv:0802.3369
  [physics.flu-dyn]} \BibitemShut {NoStop}%
\bibitem [{\citenamefont {{Nordlund}}\ and\ \citenamefont
  {{Galsgaard}}(1995)}]{nordlund19953d}%
  \BibitemOpen
  \bibfield  {author} {\bibinfo {author} {\bibfnamefont {{\AA}.}~\bibnamefont
  {{Nordlund}}}\ and\ \bibinfo {author} {\bibfnamefont {K.}~\bibnamefont
  {{Galsgaard}}},\ }\href@noop {} {\emph {\bibinfo {title} {{A 3D MHD code for
  Parallel Computers}}}},\ \bibinfo {type} {Tech. Rep.}\ (\bibinfo
  {institution} {Niels Bohr Institute for Astronomy, Physics and Geophysics},\
  \bibinfo {year} {1995})\BibitemShut {NoStop}%
\bibitem [{\citenamefont {Galsgaard}\ and\ \citenamefont
  {Nordlund}(1996)}]{galsgaard1996}%
  \BibitemOpen
  \bibfield  {author} {\bibinfo {author} {\bibfnamefont {K.}~\bibnamefont
  {Galsgaard}}\ and\ \bibinfo {author} {\bibfnamefont {A.}~\bibnamefont
  {Nordlund}},\ }\bibfield  {title} {\enquote {\bibinfo {title} {{Heating and
  activity of the solar corona: 1. Boundary shearing of an initially
  homogeneous magnetic field}},}\ }\href@noop {} {\bibfield  {journal}
  {\bibinfo  {journal} {J. Geophys. Res.}\ }\textbf {\bibinfo {volume} {101}},\
  \bibinfo {pages} {13445--13460} (\bibinfo {year} {1996})}\BibitemShut
  {NoStop}%
\bibitem [{\citenamefont {Mayer}\ and\ \citenamefont
  {Powell}(1992)}]{mayer1992}%
  \BibitemOpen
  \bibfield  {author} {\bibinfo {author} {\bibfnamefont {Ernst~W.}\
  \bibnamefont {Mayer}}\ and\ \bibinfo {author} {\bibfnamefont {Kenneth~G.}\
  \bibnamefont {Powell}},\ }\bibfield  {title} {\enquote {\bibinfo {title}
  {Viscous and inviscid instabilities of a trailing vortex},}\ }\href {\doibase
  10.1017/S0022112092000363} {\bibfield  {journal} {\bibinfo  {journal}
  {Journal of Fluid Mechanics}\ }\textbf {\bibinfo {volume} {245}},\ \bibinfo
  {pages} {91--114} (\bibinfo {year} {1992})}\BibitemShut {NoStop}%
\bibitem [{\citenamefont {McGavin}(2017)}]{mcgavin2017}%
  \BibitemOpen
  \bibfield  {author} {\bibinfo {author} {\bibfnamefont {P.}~\bibnamefont
  {McGavin}},\ }\emph {\bibinfo {title} {Understanding vortex reconnection in
  complex fluid flows}},\ \href@noop {} {Ph.D. thesis},\ \bibinfo  {school}
  {University of Dundee} (\bibinfo {year} {2017})\BibitemShut {NoStop}%
\bibitem [{\citenamefont {Scheeler}\ \emph {et~al.}(2017)\citenamefont
  {Scheeler}, \citenamefont {van Rees}, \citenamefont {Kedia}, \citenamefont
  {Kleckner},\ and\ \citenamefont {Irvine}}]{scheeler2017}%
  \BibitemOpen
  \bibfield  {author} {\bibinfo {author} {\bibfnamefont {Martin~W.}\
  \bibnamefont {Scheeler}}, \bibinfo {author} {\bibfnamefont {Wim~M.}\
  \bibnamefont {van Rees}}, \bibinfo {author} {\bibfnamefont {Hridesh}\
  \bibnamefont {Kedia}}, \bibinfo {author} {\bibfnamefont {Dustin}\
  \bibnamefont {Kleckner}}, \ and\ \bibinfo {author} {\bibfnamefont {William
  T.~M.}\ \bibnamefont {Irvine}},\ }\bibfield  {title} {\enquote {\bibinfo
  {title} {Complete measurement of helicity and its dynamics in vortex
  tubes},}\ }\href {\doibase 10.1126/science.aam6897} {\bibfield  {journal}
  {\bibinfo  {journal} {Science}\ }\textbf {\bibinfo {volume} {357}},\ \bibinfo
  {pages} {487--491} (\bibinfo {year} {2017})}\BibitemShut {NoStop}%
\bibitem [{\citenamefont {{Haynes}}\ and\ \citenamefont
  {{Parnell}}(2007)}]{haynes2007}%
  \BibitemOpen
  \bibfield  {author} {\bibinfo {author} {\bibfnamefont {A.~L.}\ \bibnamefont
  {{Haynes}}}\ and\ \bibinfo {author} {\bibfnamefont {C.~E.}\ \bibnamefont
  {{Parnell}}},\ }\bibfield  {title} {\enquote {\bibinfo {title} {{A trilinear
  method for finding null points in a three-dimensional vector space}},}\
  }\href {\doibase 10.1063/1.2756751} {\bibfield  {journal} {\bibinfo
  {journal} {Phys.~Plasmas}\ }\textbf {\bibinfo {volume} {14}},\ \bibinfo
  {pages} {082107--082107} (\bibinfo {year} {2007})}\BibitemShut {NoStop}%
\bibitem [{\citenamefont {Parnell}\ \emph {et~al.}(1996)\citenamefont
  {Parnell}, \citenamefont {Smith}, \citenamefont {Neukirch},\ and\
  \citenamefont {Priest}}]{parnell1996}%
  \BibitemOpen
  \bibfield  {author} {\bibinfo {author} {\bibfnamefont {C.~E.}\ \bibnamefont
  {Parnell}}, \bibinfo {author} {\bibfnamefont {J.~M.}\ \bibnamefont {Smith}},
  \bibinfo {author} {\bibfnamefont {T.}~\bibnamefont {Neukirch}}, \ and\
  \bibinfo {author} {\bibfnamefont {E.~R.}\ \bibnamefont {Priest}},\ }\bibfield
   {title} {\enquote {\bibinfo {title} {The structure of three-dimensional
  magnetic neutral points},}\ }\href@noop {} {\bibfield  {journal} {\bibinfo
  {journal} {Phys.~Plasmas}\ }\textbf {\bibinfo {volume} {3}},\ \bibinfo
  {pages} {759--770} (\bibinfo {year} {1996})}\BibitemShut {NoStop}%
\bibitem [{\citenamefont {Bhattacharjee}\ and\ \citenamefont
  {Wang}(1992)}]{bhattacharjee1992}%
  \BibitemOpen
  \bibfield  {author} {\bibinfo {author} {\bibfnamefont {A}~\bibnamefont
  {Bhattacharjee}}\ and\ \bibinfo {author} {\bibfnamefont {X}~\bibnamefont
  {Wang}},\ }\bibfield  {title} {\enquote {\bibinfo {title} {{Finite-time
  vortex singularity in a Model of Three-dimensional Euler flows}},}\
  }\href@noop {} {\bibfield  {journal} {\bibinfo  {journal} {Phys Rev. Lett.}\
  }\textbf {\bibinfo {volume} {69}},\ \bibinfo {pages} {2196--2199} (\bibinfo
  {year} {1992})}\BibitemShut {NoStop}%
\bibitem [{\citenamefont {Pelz}(2002)}]{pelz2002}%
  \BibitemOpen
  \bibfield  {author} {\bibinfo {author} {\bibfnamefont {Richard~B.}\
  \bibnamefont {Pelz}},\ }\enquote {\bibinfo {title} {Discrete groups,
  symmetric flows and hydrodynamic blowup},}\ in\ \href {\doibase
  10.1007/0-306-48420-X_34} {\emph {\bibinfo {booktitle} {Tubes, Sheets and
  Singularities in Fluid Dynamics: Proceedings of the NATO RAW held in
  Zakopane, Poland, 2--7 September 2001, Sponsored as an IUTAM Symposium by the
  International Union of Theoretical and Applied Mechanics}}},\ \bibinfo
  {editor} {edited by\ \bibinfo {editor} {\bibfnamefont {K.}~\bibnamefont
  {Bajer}}\ and\ \bibinfo {editor} {\bibfnamefont {H.~K.}\ \bibnamefont
  {Moffatt}}}\ (\bibinfo  {publisher} {Springer Netherlands},\ \bibinfo
  {address} {Dordrecht},\ \bibinfo {year} {2002})\ pp.\ \bibinfo {pages}
  {269--283}\BibitemShut {NoStop}%
\bibitem [{\citenamefont {Moffatt}(1978)}]{1978magnetic}%
  \BibitemOpen
  \bibfield  {author} {\bibinfo {author} {\bibfnamefont {K.}~\bibnamefont
  {Moffatt}},\ }\href {https://books.google.co.uk/books?id=cAo4AAAAIAAJ} {\emph
  {\bibinfo {title} {Magnetic Field Generation in Electrically Conducting
  Fluids}}},\ Cambridge Monographs on Mechanics\ (\bibinfo  {publisher}
  {Cambridge University Press},\ \bibinfo {year} {1978})\BibitemShut {NoStop}%
\bibitem [{\citenamefont {{Kida}}\ and\ \citenamefont
  {{Takaoka}}(1994)}]{1994AnRFM..26..169K}%
  \BibitemOpen
  \bibfield  {author} {\bibinfo {author} {\bibfnamefont {S.}~\bibnamefont
  {{Kida}}}\ and\ \bibinfo {author} {\bibfnamefont {M.}~\bibnamefont
  {{Takaoka}}},\ }\bibfield  {title} {\enquote {\bibinfo {title} {{Vortex
  reconnection}},}\ }\href {\doibase 10.1146/annurev.fl.26.010194.001125}
  {\bibfield  {journal} {\bibinfo  {journal} {Annual Review of Fluid
  Mechanics}\ }\textbf {\bibinfo {volume} {26}},\ \bibinfo {pages} {169--189}
  (\bibinfo {year} {1994})}\BibitemShut {NoStop}%
\bibitem [{\citenamefont {{Hesse}}\ and\ \citenamefont
  {{Schindler}}(1988)}]{1988JGR....93.5559H}%
  \BibitemOpen
  \bibfield  {author} {\bibinfo {author} {\bibfnamefont {M.}~\bibnamefont
  {{Hesse}}}\ and\ \bibinfo {author} {\bibfnamefont {K.}~\bibnamefont
  {{Schindler}}},\ }\bibfield  {title} {\enquote {\bibinfo {title} {{A
  theoretical foundation of general magnetic reconnection}},}\ }\href {\doibase
  10.1029/JA093iA06p05559} {\bibfield  {journal} {\bibinfo  {journal} {Journal
  of Geophysics Research}\ }\textbf {\bibinfo {volume} {93}},\ \bibinfo {pages}
  {5559--5567} (\bibinfo {year} {1988})}\BibitemShut {NoStop}%
\bibitem [{\citenamefont {{Hornig}}\ and\ \citenamefont
  {{Priest}}(2003)}]{hornig2003}%
  \BibitemOpen
  \bibfield  {author} {\bibinfo {author} {\bibfnamefont {G.}~\bibnamefont
  {{Hornig}}}\ and\ \bibinfo {author} {\bibfnamefont {E.}~\bibnamefont
  {{Priest}}},\ }\bibfield  {title} {\enquote {\bibinfo {title} {{Evolution of
  magnetic flux in an isolated reconnection process}},}\ }\href {\doibase
  10.1063/1.1580120} {\bibfield  {journal} {\bibinfo  {journal} {Physics of
  Plasmas}\ }\textbf {\bibinfo {volume} {10}},\ \bibinfo {pages} {2712--2721}
  (\bibinfo {year} {2003})}\BibitemShut {NoStop}%
\bibitem [{\citenamefont {{Hesse}}\ \emph {et~al.}(2005)\citenamefont
  {{Hesse}}, \citenamefont {{Forbes}},\ and\ \citenamefont
  {{Birn}}}]{2005ApJ...631.1227H}%
  \BibitemOpen
  \bibfield  {author} {\bibinfo {author} {\bibfnamefont {M.}~\bibnamefont
  {{Hesse}}}, \bibinfo {author} {\bibfnamefont {T.~G.}\ \bibnamefont
  {{Forbes}}}, \ and\ \bibinfo {author} {\bibfnamefont {J.}~\bibnamefont
  {{Birn}}},\ }\bibfield  {title} {\enquote {\bibinfo {title} {{On the Relation
  between Reconnected Magnetic Flux and Parallel Electric Fields in the Solar
  Corona}},}\ }\href {\doibase 10.1086/432677} {\bibfield  {journal} {\bibinfo
  {journal} {The Astrophysical Journal}\ }\textbf {\bibinfo {volume} {631}},\
  \bibinfo {pages} {1227--1238} (\bibinfo {year} {2005})}\BibitemShut {NoStop}%
\bibitem [{\citenamefont {{Priest}}\ \emph {et~al.}(2003)\citenamefont
  {{Priest}}, \citenamefont {{Hornig}},\ and\ \citenamefont
  {{Pontin}}}]{2003JGRA..108.1285P}%
  \BibitemOpen
  \bibfield  {author} {\bibinfo {author} {\bibfnamefont {E.~R.}\ \bibnamefont
  {{Priest}}}, \bibinfo {author} {\bibfnamefont {G.}~\bibnamefont {{Hornig}}},
  \ and\ \bibinfo {author} {\bibfnamefont {D.~I.}\ \bibnamefont {{Pontin}}},\
  }\bibfield  {title} {\enquote {\bibinfo {title} {{On the nature of
  three-dimensional magnetic reconnection}},}\ }\href {\doibase
  10.1029/2002JA009812} {\bibfield  {journal} {\bibinfo  {journal} {Journal of
  Geophysical Research (Space Physics)}\ }\textbf {\bibinfo {volume} {108}},\
  \bibinfo {eid} {1285} (\bibinfo {year} {2003})}\BibitemShut {NoStop}%
\end{thebibliography}

%

\end{document}